Titre :
« Traduction commentée de *la Conférence Bakerienne sur la Théorie de la Lumière et des Couleurs* de Thomas Young ».

Auteur :
MORIZOT Olivier,
Aix-Marseille Univ, CNRS, Centre Gilles Gaston Granger, Aix-en-Provence, France

Coordonnées :
Olivier.morizot@univ-amu.fr

Date de création :
11 juin 2020

# Traduction commentée de *la Conférence Bakerienne sur la Théorie de la Lumière et des Couleurs* de Thomas Young

Le 12 novembre 1801, Thomas Young – alors âgé de vingt-huit ans – expose à la Royal Society sa *Théorie de la Lumière et des Couleurs*, pour laquelle il reçoit la médaille bakerienne pour la deuxième année consécutive. Dans ce texte d'une quarantaine de pages dans sa version originale, Young défend une conception ondulatoire de la lumière face à un parterre acquis au système corpusculaire hérité de Newton. L'idée d'un modèle ondulatoire de la lumière n'a rien d'original en soi et Young se reconnait de nombreux et illustres prédécesseurs, tels que Huygens, Euler, Hooke ou Malebranche. Cependant, une attention particulière portée aux conséquences du croisement de deux ondulations lumineuses lui permet de proposer dans ce court texte une première version du principe d'interférence lumineuse. Et d'en déduire la première justification correcte de la couleur des lames minces – telles que celles que l'on observe sur les bulles de savon – grâce à laquelle il fournit même la toute première série de mesures des longueurs d'onde de la lumière – concept qu'il introduit à cette occasion. Sur la base de ce même principe il présente et interprète aussi la première expérience de diffraction par un réseau, lui permettant d'anticiper sur l'origine structurelle des couleurs de certains insectes et papillons. Enfin, la défense d'un modèle ondulatoire de la lumière est ici l'occasion pour lui de formuler l'hypothèse visionnaire de la structure triple des photorécepteurs de la rétine humaine justifiant des sensations colorées. La densité de concepts et de résultats novateurs que ce court texte a présenté à l'époque de sa parution nous a donc semblé justifier à elle seule de l'intérêt de mettre aujourd'hui à disposition du public français une traduction de *la Théorie de la Lumière et des Couleurs* de Thomas Young.

Sans que l'anglais du début du XIX[e] siècle employé par son auteur soit fondamentalement inaccessible, il nous a en effet paru qu'une traduction française de ce texte permettrait au lecteur d'entrer plus directement en contact avec le fourmillement d'idées de l'auteur que s'il avait dû le faire au travers du filtre de la langue. A ce titre, et autant qu'il l'était raisonnable, nous avons fait le choix d'une traduction fidèle aux termes et formulations employées par l'auteur, soutenue par des notes de bas de page proposant une reformulation moderne des quelques archaïsmes pouvant rendre certains passages difficilement compréhensibles. La traduction proposée est par ailleurs augmentée d'un abondant commentaire, superflu pour qui voudrait se contenter d'une lecture superficielle, mais nécessaire pour clarifier les inévitables ambiguïtés et lourdeurs qu'un texte écrit il y a plus de deux siècles peut comporter pour nous. Surtout, ce commentaire doit permettre de lever le voile que les ans ont lentement déposé sur les préoccupations et connaissances d'une époque, et plus particulièrement sur celles de son auteur ; le génie et l'audace de ses propositions ne pouvant être véritablement sensibles que si elles sont replacées dans le contexte des problèmes qu'il cherchait à résoudre et des outils dont il disposait pour les élaborer.

La présente traduction commentée a donc une ambition triple. Celle de proposer à un large public de langue française un accès facilité à un texte fondamental pour l'histoire des théories de l'optique ; celle de mettre à disposition d'étudiants, enseignants et chercheurs en sciences physiques un texte fondateur de la discipline[1] enrichi non seulement d'une analyse physico-mathématique moderne, mais aussi d'une mise en lumière par l'exemple de certains mécanismes d'élaboration d'une théorie scientifique en cours de construction ; celle enfin de mettre à disposition des historiens des sciences un texte fidèle à l'intention de son auteur, et contextualisé dans son rapport aux conceptions du monde qui le portaient, aux sources auxquelles il a emprunté, aux exigences des lecteurs auxquels il savait s'adresser et à la manière dont la communauté scientifique de la première moitié du XIX[e] siècle l'a diversement reçu et intégré.

---

[1] Breno Arsioli Moura et Sergio Luiz Bragatto Boss, *Thomas Young e o resgate da teoria ondulatória da luz : Uma tradução comentada de sua Teoria Sobre Luz e Cores*. Revista Brasileira de Ensino de Física, vol. 37, n° 4, 2015, 4203/1-24.

**L'auteur**

Thomas Young est né en 1773 dans une famille quaker de Milverton, dans le Sud-Ouest de l'Angleterre. De son incroyable personnalité, richement décrite par son ami d'enfance Hudson Gurney[2], ou plus récemment par Alexander Wood[3], on se contentera de présenter ici quelques éléments susceptibles d'éclairer une partie du cheminement intellectuel l'ayant mené à l'élaboration de *la Théorie de la Lumière et des Couleurs.*

Thomas Young est nommé membre de la Royal Society de Londres à l'âge de vingt-et-un ans, suite à la publication d'un premier travail dédié au mécanisme de la vision, exposant particulièrement sa découverte que le cristallin peut se contracter tel un muscle, et que cette contraction est à l'origine de la capacité d'accommodation de l'œil humain[4]. Son intérêt particulier pour le fonctionnement de l'œil aura l'occasion de s'approfondir au cours des quelques huit années d'études de la médecine qui suivront pour lui à Edinburgh, Göttingen et Cambridge et qui lui permettront de s'installer comme docteur en médecine à Londres en 1799.

Ces années d'études de la médecine sont aussi l'occasion pour lui d'étudier plus précisément les mécanismes de perception des sons et de production de la voix humaine, lesquels finissent même par constituer le sujet de sa thèse de doctorat[5]. Mais il semble qu'il ne put développer ces considérations sur la perception et la production des sons, sans se plonger également dans une réflexion sur la nature même des sons et les mécanismes de leur propagation par l'air, les tubes ou les cordes[6].

Ce travail à la fois expérimental et théorique se matérialise en 1800 par sa seconde publication dans les Transactions de la Royal Society, intitulée *Brefs exposés d'expériences et d'investigations relatives au son et à la lumière*[7], consacrée presque exclusivement aux phénomènes sonores, mais dont la dixième section intitulée « *de l'analogie entre lumière et son* » (p.125-130) est consacrée à un plaidoyer en faveur d'une application des résultats et connaissances relatives au son pour une meilleure compréhension des phénomènes lumineux encore inexpliqués. Young y suggère donc d'envisager la propagation de la lumière sur le modèle d'une onde longitudinale se propageant dans un milieu matériel, de manière analogue à celle du son.

---

[2] Hudson Gurney, *Memoir of the life of Thomas Young, M.D., F.R.S. With a catalogue of his works and essays*, John & Arthur Arch, Cornhill, 1831.
[3] Alexander Wood, *Thomas Young Natural Philosopher*, Cambridge University Press, 1954.
[4] Thomas Young, *Observations on vision*, Philosophical Transactions of the Royal Society of London, vol. 83, 1793, 169-181.
[5] Thomas Young, *De Corporis Humani Viribus Conservatricibus,* Thèse de doctorat, Gottingae : typis J.C. Dieterich, 1796.
[6] Dans une lettre de 1797 à Andrew Dalzel, Thomas Young écrit : « *Je suis en ce moment passablement occupé par le léger schéma synoptique pour la fin de ma thèse, la définition et la classification des divers sons de toutes les langues dont je peux avoir connaissance ; et j'ai dernièrement divergé quelque peu vers la théorie physique et mathématique du son en général. Je me plais à croire que j'ai fait quelques observations singulières sur les cordes vibrantes, et j'ai l'intention de poursuivre mes expériences.* » (Andrew Dalzel, *History of the Universisty of Edimburgh from its foundation*, Edmonston and Douglas, vol. I, 1862, 144).
[7] Thomas Young, *Outlines of Experiments and Inquiries respecting Sound and Light*, Philosophical Transactions of the Royal Society of London, vol. 90, 1800, 106-150.

La reconnaissance de ses talents par la communauté scientifique londonienne de l'époque se manifeste par sa nomination comme tout premier professeur de philosophie naturelle à la naissante Royal Institution[8], où il donnera de 1801 à 1803 un cours de physique générale portant sur une série de sujets aussi vaste[9] que la mécanique (allant des lois de la statique et du mouvement, à l'élasticité des solides, l'ingénierie, ou l'architecture), l'hydrodynamique (dans laquelle il inclue l'étude des liquides statiques ou en mouvement, mais aussi des sons et, de manière significative, de la lumière) et les sciences de la nature (où il traite d'astronomie, autant que de météorologie, d'électricité ou de chaleur). Cours foisonnant donc, dont le contenu sera publié[10] postérieurement à la lecture de *la Théorie de la Lumière et des Couleurs*.

L'estime que la communauté scientifique britannique lui porte se traduit aussi par l'obtention à trois reprises au tournant du siècle de la médaille bakerienne[11] de la Royal Society. Thomas Young est en effet mandaté une première fois en 1800 pour donner une conférence bakerienne *Sur le Mécanisme de l'œil*[12]. Cette conférence est l'occasion pour lui de revenir sur sa découverte du rôle du cristallin dans l'accommodation qu'il avait présentée en 1793, mais qui avait été si vigoureusement contestée dans l'intervalle qu'il avait dû la rétracter. De nouvelles expériences, appuyées par une série de calculs basés sur des modèles simples de l'œil et du cristallin, et menées dans le cadre d'une conception corpusculaire de la lumière, lui permettent de confirmer la valeur et l'originalité de ses premières découvertes, ainsi que de découvrir et interpréter pour la première fois le potentiel défaut d'astigmatisme de l'œil humain.

Et d'être à nouveau distingué par ce prix l'année suivante, où il expose enfin *la Théorie de la lumière et des couleurs*[13] présentée ici. Il ne sera donc pas surprenant de voir se développer au fil de ce texte une interprétation des phénomènes de la lumière et des couleurs fortement influencée par une longue réflexion conjointe sur les phénomènes du son et de la musique, par une connaissance mécanique et anatomique très fine des systèmes visuel et auditif humains et par une conception unificatrice des phénomènes physiques exaltée au cours de ces années d'enseignement généraliste à la Royal Institution.

---

[8] Geoffrey N. Cantor, *Thomas Young's lectures at the Royal Institution*, Notes and Records of the Royal Society of London, vol. 25, n° 1, 1970, 87-112.

[9] Au-delà de sa maîtrise de tous les champs de la physique de son époque, Thomas Young est un véritable génie polymathe : on a déjà évoqué son doctorat en médecine et ses travaux fondamentaux sur le fonctionnement de l'œil ou de la voix humaine. Mais pour être complet il faudrait aussi signaler ses travaux originaux en mathématiques, en ingénierie, en langues anciennes (son article « Languages » dans l'*Encyclopedia Britannica* compare et classe les grammaires de plus de 400 langues, il a travaillé à la restauration de papyrus extraits à Herculanum et fut le premier à déchiffrer les cartouches présents sur la pierre de Rosette), sa production d'almanach nautiques pour le Bureau des Longitudes, sa rédaction de soixante-et-un articles ou biographies pour l'*Encyclopedia Britannica*, ou ses nombreux calculs relatifs aux assurances vie. Pour plus d'information sur ces sujets, voir Alexander Wood, *Thomas Young Natural Philosopher*, Cambridge University Press, 1954.

[10] Thomas Young, *Syllabus of a course of Lectures on Natural and Experimental Philosophy*, Royal Institution, London, Art. 289, 1802. Et Thomas Young, *A Course of Lectures in Natural Philosophy and Mechanical Arts*, vol. I et II, London, 1807.

[11] Henry Baker (1698-1774) fonde à sa mort le prix et le cycle des conférences portant son nom en léguant une dotation de £100 par an à la Royal Society pour la mise en place d'une conférence annuelle donnée par un membre de la Royal Society sur quelque sujet d'histoire naturelle ou de philosophie expérimentale que la société jugera d'importance majeure.

[12] Thomas Young, *On the Mechanism of the Eye*, Philosophical Transactions of the Royal Society of London, vol. 91, 1801, 23–88.

[13] Thomas Young, *On the Theory of Light and Colours*, Philosophical Transactions of the Royal Society of London, vol. 92, 1802, 12-48.

Thomas Young présente une dernière conférence bakerienne en 1803, qu'il dédie à une nouvelle série d'*Expériences et Calculs relatifs à l'Optique Naturelle*[14]. Il nous arrivera donc de faire référence à cet article quand il permettra de compléter ou de souligner un contraste avec les propositions développées dans la version originale de *la Théorie de la Lumière et des Couleurs*.

**Aperçu des théories de la lumière et des couleurs circulant à l'époque**

A l'aube du XVIII$^e$ siècle, au Royaume-Uni comme en France, la théorie optique la plus répandue dans la communauté scientifique est très certainement celle qui a été développée par Newton dans son célèbre *Opticks*[15]. Ou plus précisément, puisque près d'un siècle s'est écoulé depuis sa première publication, on retrouve alors une multitude de propositions relatives à l'optique se présentant essentiellement comme filles de l'*Opticks*, et développées comme un approfondissement des questions pendantes à la fin de cet ouvrage. Notamment en proposant des modèles mathématiques de forces mécaniques à appliquer à la lumière, afin de justifier de son comportement selon les lois de la dynamique newtonienne[16]. Car conformément à la position explicitement adoptée par Newton en conclusion de l'*Opticks*[17], la totalité de ces propositions envisage strictement la lumière sous la forme d'un flux de corpuscules matériels émanant des corps lumineux.

Parmi les quelques théories encore opposées à l'optique newtonienne se trouvent probablement en tête quelques propositions descendant en ligne plus ou moins directe de la *Dioptrique* de Descartes[18]. Mais dans celles-ci aussi, conformément à la théorie cartésienne, la lumière est fondamentalement composée de corpuscules, répondant aux lois de la mécanique[19].

La seule hypothèse crédible opposée au système corpusculaire à l'époque étant celle selon laquelle la lumière serait le fruit d'une vibration, ou d'une onde de pression, se propageant dans un milieu extrêmement rare et subtil, communément appelé éther. Les plus prestigieux défenseurs de ce modèle ondulatoire de la lumière ayant été Huygens[20], Hooke[21], Euler[22] ou

---

[14] Thomas Young, *Experiments and Calculations relative to Physical Optics*, Philosophical Transactions of the Royal Society of London, vol. 94, 1804, 1–16. On a traduit « Physical Optics » par « Optique Naturelle » pour éviter un malentendu trop facile avec ce que l'on appelle aujourd'hui communément « Optique Physique » et qui correspond justement à l'optique ondulatoire. De fait, cet article parle d'optique au sens de *science de l'œil* et le mot « physical » renvoie à l'idée de nature.

[15] Isaac Newton, *Opticks: or a treatise of the reflections, refractions, inflections and colors of light*. L'ouvrage a donné lieu à quatre éditions de 1704 à 1730, présentant de légères évolutions du texte, et l'ajout progressif de *queries* (ou questions ouvertes) à la fin de l'ouvrage.

[16] Geoffrey N. Cantor, *Optics after Newton : Theories of light in Britain and Ireland, 1704-1840,*, Manchester University Press, Manchester, 1983. Ou André Chappert, *Histoire de l'optique ondulatoire de Fresnel à Maxwell*, premier chapitre, Belin, Paris, 2007.

[17] Voir notamment les questions 29 à 31.

[18] Charles Adam et Paul Tannery, *Œuvres de René Descartes*, La Dioptrique, vol. VI, Paris, 1897-1909.

[19] Voir notamment le Premier Discours *de la Dioptrique*, intitulé « De la Lumière ».

[20] Christian Huygens (1629-1695) développe dans son *Traité de la Lumière,* 1690, une théorie ondulatoire de la lumière, ouvertement opposée aux théories newtoniennes circulant depuis 1672. Il semble néanmoins que

Malebranche[23]. Mais bien qu'ayant tous avancé des arguments puissants en faveur du système ondulatoire, leurs propositions sont globalement estimées au XVIIIe comme ayant été complètement décrédibilisées par Newton lui-même, ou ne pouvant aucunement prétendre à sa remise en cause pour celles qui lui sont postérieures.

On a évoqué plus haut quelques pistes qui permettraient de justifier l'intérêt de Thomas Young pour le système ondulatoire. Néanmoins la séduction de l'analogie entre lumière et son, ou simplement entre couleur et musique, ne peut tout justifier ; de nombreux auteurs, dont Newton lui-même[24], l'ayant notée et exploitée avant lui.

Viennent certainement s'y ajouter deux difficultés que Young juge insurmontables dans le cadre corpusculaire. Il s'agit en premier lieu de l'égalité de la vitesse de la lumière quelle que soit sa source d'émission ; Young estime effectivement inconcevable que les potentiels corpuscules lumineux projetés par le plus petit des vers luisants puissent se propager à la même vitesse exactement que ceux émis par le Soleil. En second lieu, le phénomène de réflexion partielle qui accompagne systématiquement toute réfraction est selon lui incompatible avec la description de la lumière comme un flux de corpuscules : un faisceau lumineux partiellement réfléchi et partiellement transmis au niveau d'un dioptre transparent, dans des proportions toujours identiques pourvu que les milieux et l'angle d'incidence soient inchangés, révèlerait en effet pour lui un comportement inexplicablement différent pour ces corpuscules supposément identiques[25].

Young, déjà convaincu de l'existence d'un milieu éthéré plus subtil que l'air et porteur notamment des phénomènes magnétiques, électriques[26] et gravitationnels, bascule donc en faveur de l'idée unificatrice que ce même éther, imprégnant intégralement l'univers, est le vecteur par ses ondulations de la propagation de la lumière et de la chaleur. C'est donc aussi dans le cadre d'une recherche d'un principe unique commun à l'ensemble des phénomènes physiques que Young s'engage dans la défense d'un système ondulatoire de la lumière.

---

rapidement après sa mort cette théorie fut déjà massivement oubliée ou rejetée en Europe (Casper Hakfoort, *Optics in the Age of Euler*, Cambridge University Press, 1995, 56).
1695, his theories were widely forgotten or rejected in Europe

[21] Robert Hooke (1635-1703), observe dès 1660 les irisations des lames minces, les anneaux colorés apparaissant au contact d'une lentille convexe et d'un plan, ainsi que la lumière colorée à la sortie d'un prisme. Pour expliquer ces phénomènes il propose un modèle ondulatoire de la lumière débouchant sur la première de nombreuses confrontations avec Newton - alors tout récent membre de la Royal Society – suite à la publication de *Robert's Hooke critique of Newton's theory of light and colours* (envoyé en 1672, et publié dans Thomas Birch, The History of the Royal Society, vol. 3, 1757, 10-15).

[22] Leonhard Euler (1707-1783), a exprimé son désaccord avec la théorie corpusculaire de la lumière de Newton dans *Nova Theoria Lucis et Colorum*, 1746. Texte dans lequel il associe la propagation de la lumière à un mouvement vibratoire de l'éther.

[23] Nicolas Malebranche (1638-1715) est d'après Pierre Duhem (*L'optique de Malebranche*, Revue de Métaphysique et de Morale, tome 23, n° 1, 1916) le premier à élaborer une théorie selon laquelle la lumière est due à des vibrations de l'éther. Il affirme notamment dans son *Mémoire*, 1699, 260 : « La force ou l'éclat des couleurs vient donc aussi du plus et du moins de force des vibrations, non de l'air, mais de la matière subtile ; et les différentes espèces de couleurs du plus ou moins de promptitude de ces mêmes vibrations ».

[24] Peter Pesic, *Isaac Newton and the mystery of the major sixth : a transcription of his manuscript "Of Musick" with commentary*, Interdisciplinary science reviews, vol. 31, n° 4, 2006, 291-306.

[25] Thomas Young, *Outlines of Experiments and Inquiries respecting Sound and Light*, Philosophical Transactions of the Royal Society of London, vol. 90, 1800, 125-126.

[26] Inspiré probablement par Benjamin Franklin, *Experiments and Observations on Electricity, made at Philadelphia in America,* London, 1769. D'ailleurs cette édition de l'ouvrage contient une lettre de 1752 dans laquelle Franklin affirme être insatisfait par la doctrine supposant l'existence de particules de lumière et où il développe une brève interprétation vibrationnelle des phénomènes optiques (p. 264).

Pour lui, cette vision globale du rôle de l'éther implique en particulier l'idée que sa densité doit être plus grande dans les corps matériels que dans l'air, et que ces corps sont entourés d'une atmosphère d'éther dont la densité décroît continûment avec la distance[27]. Cette idée se matérialise dans l'Hypothèse IV de *la Théorie de la Lumière et des Couleurs*. Cependant, l'abandon de cette hypothèse dès la fin de l'année 1803 sur la base de résultats expérimentaux qui l'invalident[28], forcera non seulement Young à modifier ou censurer plusieurs passages de *la Théorie de la Lumière et des Couleurs* – dont l'Hypothèse IV elle-même – dans ses éditions ultérieures, mais fera aussi partie des raisons qui le pousseront à renoncer à l'ambition de découvrir le substrat unique de tous les phénomènes physiques[29].

**Structure du texte**

La structure du texte proposée par Thomas Young est simple et directe. Elle est introduite par quatre Hypothèses relatives à la nature de la lumière, affirmant l'existence d'un éther comme support de sa propagation. De ces Hypothèses découlent huit Propositions décrivant les propriétés mécaniques de l'éther permettant d'expliquer le comportement communément observé de la lumière ; la huitième Proposition consistant en la description des effets de la combinaison de deux ondulations distinctes de même fréquence de vibration. De ce qui se révèle être la première formulation d'un principe d'interférences des ondes lumineuses, Young déduit huit corollaires, développés dans le dernier tiers du texte comme autant d'exemples distincts d'applications de ce principe à l'interprétation des phénomènes lumineux. Une neuvième et dernière proposition vient alors conclure le texte en proposant une extrapolation de l'interprétation ondulatoire de la lumière à celle de la chaleur, dans une tentative d'emporter l'adhésion définitive du lecteur par la démonstration de la validité encore plus large de la solution ondulatoire.
Bâtissant son raisonnement sur une série d'hypothèses et affirmant d'entrée de jeu qu'il ne s'agira pas pour lui de produire la moindre expérience nouvelle, Young sait qu'il prend le risque de s'opposer non seulement sur le fond, mais aussi sur la forme, aux préceptes méthodologiques baconiens et newtoniens qui ont orienté si fortement le développement de la Royal Society. Ainsi tient-il à préciser dès la première phrase que s'il propose bien une série d'hypothèses dans ce texte, celles-ci n'ont rien à voir avec ces spéculations chimériques, « *indépendantes de toute connexion avec les observations expérimentales* » dont Newton affirme s'être affranchi. Mais que sa méthode consiste au contraire à proposer quelques principes simples, « *par lesquels un grand nombre de phénomènes apparemment hétérogènes sont réduits à des lois cohérentes et universelles* ». En l'occurrence il s'agira de montrer qu'en supposant uniquement l'existence d'un milieu éthéré extrêmement rare et élastique dont les ondulations produisent la lumière, il est possible d'extrapoler l'explication de l'ensemble des phénomènes sonores à quantité de phénomènes similaires ou spécifiques de la lumière, et d'en déduire même les propriétés de la chaleur.

---

[27] Geoffrey N. Cantor, *The changing role of Young's ether,* The British Journal for the History of Science, vol. 3, 17, 1970, 44-62.
[28] Young évoque dans des carnets de notes qui suivent de peu la publication de *la Théorie de la Lumière et des Couleurs* le résultat d'une expérience qui se révèlera cruciale, consistant à observer la manière dont l'inflexion de la lumière passant à proximité des corps est affectée par la nature du corps infléchissant. Si ceux-ci sont entourés d'une atmosphère d'un gradient d'atmosphère d'éther, ce gradient doit non seulement participer au phénomène d'inflexion, mais aussi dépendre selon toute probabilité de la nature (masse, densité…) du corps qu'il enrobe. L'invariance de l'inflexion avec la nature du corps matériel la produisant est mentionnée par Young fin 1801 – début 1802, et semble marquer le début de la fin de son hypothèse d'un gradient d'éther.
[29] Geoffrey N. Cantor, *Ibid.*, 56-61.

En cela Young anticipe de manière exemplaire le concept de « consilience d'induction », introduit un demi-siècle plus tard par William Whewell[30] pour désigner ces situations dans lesquelles une théorie explicative causale est conjecturée avec succès pour expliquer une série connue de phénomènes, et se trouve alors capable, sans ajustement supplémentaire, de fournir une explication causale toute aussi concluante d'une autre série de phénomènes, en apparence très différents de ceux qui ont été exploités pour la formulation de la théorie. Cet accord imprévu dans lequel « *des règles émergeant de domaines éloignés et déconnectés bondissent jusqu'à un même point* » (vol. II, p.65) était même pour Whewell le critère de confiance le plus fort que pouvait fournir une théorie scientifique. Et c'est certainement un sentiment que Young partage lorsqu'il insiste sur le fait qu'il ne sera pas nécessaire pour lui de produire de nouvelles expériences, mais seulement de montrer comment certains principes déjà proposés peuvent être réorganisés pour expliquer « *un grand nombre de faits très divers qui sont jusqu'ici restés enterrés dans l'obscurité* », dont un grand nombre d'expériences « *d'autant plus indiscutables qu'elles ont dû être conduites sans la moindre partialité pour le système par lequel elles vont être expliquées* », et même encore « *certains faits jusque-là inobservés* ». C'est l'une des raisons pour lesquelles on verra Young se référer systématiquement dans ce texte aux résultats d'auteurs antérieurs tels que Huygens, Euler ou Lagrange, mais aussi et surtout à ceux de Newton.

Néanmoins, ces références répétées aux théories antérieures sont également dues au fait que l'autre risque majeur que prend Young – en pleine conscience – est celui de proposer une interprétation de la lumière diamétralement opposée à ce qu'il sait être non seulement l'opinion de la majorité, mais aussi celle du plus grand monument de la science anglaise qu'est Isaac Newton. C'est donc avec plus ou moins d'habileté et de malice qu'on le verra tout au long du texte citer de longs extraits de textes dans lequel Newton lui-même peut sembler défendre des positions similaires à la sienne. Mais si la proximité des positions citées avec celles de Young semble assez convaincante au premier abord, il ne peut échapper à un lecteur attentif de l'*Opticks*, qu'elles sont pour l'essentiel extraites de la partie finale du livre dédiée à des questions spéculatives. Et que lorsqu'elles évoquent la possibilité d'existence d'un éther, il s'agit d'un milieu plus subtil que l'air avec lequel les corpuscules de lumière pourraient interagir, mais qui ne constituerait aucunement la matière de celle-ci, ni même le support de sa propagation[31].

La structure de *la Théorie de la Lumière et des Couleurs* répond donc à une logique clairement explicitée par l'auteur, pouvant faire figure d'exemple méthodologique pour l'élaboration et la présentation d'une théorie scientifique. Néanmoins la forme iconoclaste de cette théorie reposant sur une série d'hypothèses et ne proposant pas d'expériences nouvelles, tout autant que son fond diamétralement opposé aux idées héritées de Newton, font qu'elle émerge dans un contexte scientifique qui n'est probablement pas prêt à les recevoir, et qu'ils alimenteront l'un comme l'autre une violente polémique autour de cette publication.

**Réception immédiate et postérité de ce texte**

---

[30] William Whewell, *The Philosophy of the Inductive Sciences Founded on Their History*, vol. I et II, London, 1847.
[31] Comme le confirme explicitement la question 28 : « *Toutes les hypothèses dans lesquelles la lumière est supposer consister en une pression ou un mouvement propagé dans un milieu fluide ne sont-elles pas erronées ?* » (Isaac Newton, *Opticks*, Livre III, Question, 1730, 336).

Malgré son format et son contenu non standards, on peut raisonnablement projeter que l'attribution à Young de la médaille bakerienne de la Royal Society en 1801 pour ce texte, puis à nouveau deux années plus tard pour d'autres travaux relatifs à l'optique, sont le signe d'un accueil favorable de *la Théorie de la Lumière et des Couleurs* par la communauté scientifique britannique. On s'étonnera donc peut-être de voir le faible écho qu'auront ses propositions révolutionnaires dans les années qui suivront sa publication.

L'une des explications vient peut-être de la très violente critique de ce texte qui sera publiée anonymement dans l'*Edinburgh Review*[32]. Son auteur est rapidement identifié comme étant Henry Brougham, jeune membre de la Royal Society et co-fondateur en 1802 du journal en question. Dans celui-ci, il publiera d'ailleurs deux articles supplémentaires à charge contre les théories mais aussi la personnalité de Thomas Young : le premier dédié à la critique d'un article relatif aux couleurs publié en 1802[33] ; et le dernier consacré à la conférence bakerienne donnée par Young en 1803[34]. Si la virulence de ces critiques est en partie expliquée par une amertume développée par Brougham à l'égard de Young, il semble clair que leur fond idéologique est lié non seulement à une défense vigoureuse du système corpusculaire newtonien de la lumière, mais aussi du courant de la tradition méthodologique écossaise dont se réclame Henry Brougham. Laquelle repose sur une forte suspicion vis-à-vis de tout type d'hypothèse, particulièrement celles postulant l'existence d'un éther[35]. Il y a donc un fond dogmatique à la critique adressée à Young par Brougham, dû à une vision radicale de la méthode scientifique qui ne lui semble pas être respectée dans ces textes. Reste que le ton de l'article, ridiculisant la proposition de Young par des formules aussi acides que « *ce papier ne contient rien qui mérite le nom d'expérience ou de découverte* », qu'il est « *dépourvu de la moindre valeur* » ou que son auteur « *semble avoir systématisé dans une sorte de théorie la méthode pour gaspiller du temps* » empêchera de poursuivre le débat dans le champ des idées et nuira indubitablement à la réception des théories de Young par ses contemporains.

Il est manifeste cependant que le style même de l'œuvre de Young a lui aussi participé à ce que la communauté scientifique ne s'empare pas véritablement des découvertes qui y pullulent pourtant. Il semble en effet clair lorsque l'on compile ses textes, que Young rencontrait souvent des difficultés à transposer les problèmes physiques qu'il rencontrait en termes mathématiques, et qu'il exprimait par conséquent le plus souvent ses solutions sous la forme d'une prose parfois trop elliptique, parfois trop verbeuse car dénuée de l'efficacité du formalisme mathématique, et que nombre de ses contemporains devaient échouer à comprendre parfaitement. On laissera à ce titre le lecteur se faire son propre avis au fil des pages qui suivent. Mais il est certain que le développement indépendant d'une nouvelle théorie ondulatoire moins de quinze ans plus tard par Augustin Fresnel[36], reposant sur une approche expérimentale et calculatoire beaucoup plus robuste et rigoureuse, et qui emportera l'assentiment général en quelques années seulement, tend à appuyer cette interprétation de l'histoire[37].

---

[32] Edinburgh Review, vol. 1, 450.
[33] Edinburgh Review, vol. 1, 459.
[34] Edinburgh Review, vol. 5, 97.
[35] Brougham semble en effet emprunter beaucoup à la tradition radicalement anti-hypothétique introduite par Thomas Reid dans la science britannique à la fin du XVIII[e] siècle. Voir Geoffrey N. Cantor, *Henry Brougham and the Scottish methodological tradition*, Studies in History and Philosophy of Science Part A, vol. 2, n° 1, 1971, 69-89.
[36] *Œuvres complètes d'Augustin Fresnel*, Eds : H. de Senarmont, E. Verdet, L. Fresnel, tomes I et II, 1866.

Fresnel reconnait d'ailleurs rapidement l'antériorité de la théorie de Young, mais pas la paternité. Puisqu'il semble confirmé qu'il n'en avait pas connaissance au début de ses travaux[38] ; signe supplémentaire que celle-ci avait été relativement peu diffusée.

C'est donc peut-être sur le point sur lequel Young s'y serait le moins attendu que le contenu de *la Théorie de la Lumière et des Couleurs* a le plus significativement irrigué la science. A savoir l'Hypothèse III relative au mécanisme de la perception des couleurs[39], supposant la présence en chaque point de la rétine de trois types de récepteurs, sensibles chacun préférentiellement à une gamme de fréquences d'ondulation ; l'une correspondant à la sensation du rouge, la deuxième du jaune, la troisième du bleu. La théorie psychophysique de la couleur prend son essor à partir du milieu du XIX[e] siècle, en particulier suite aux travaux exceptionnels de James Clerk Maxwell[40] et Hermann von Helmholtz[41]. Helmholtz, probablement sensibilisé aux travaux de Young via ses publications relatives au fonctionnement de l'œil, envisage très tôt sa théorie de la perception des couleurs, pour la rejeter néanmoins en 1852 du fait d'une interprétation trop stricte de celle-ci. Cependant dès 1855 Maxwell, inspiré par les travaux de Grassmann, affirme la véracité de l'hypothèse des trois récepteurs de Young, forçant Helmholtz à la reconsidérer sous une version amendée[42] qui s'avérera l'une des clés de leur théorie de couleurs à tous deux, dite aujourd'hui « de Young-Helmholtz ». Helmholtz rend d'ailleurs régulièrement hommage à Young dans son *Optique Physiologique*, notamment en témoignant que : « *La théorie des sensations colorées de Th. Young était restée inaperçue comme tant d'autres choses que ce merveilleux chercheur a trouvées, en devançant son époque, jusqu'à ce que mes recherches et celles de Maxwell attirassent l'attention sur cette théorie.* »[43]

Cette remarque fait douloureusement écho à une phrase que l'on retrouve dans une lettre du 3 janvier 1821 adressée par Young à son ami Hudson Gurney[44] : Mon éditeur « *a dit que la série de conférences <à la Royal Institution> étaient un livre trop bon, et que c'était la raison pour laquelle il ne s'était pas vendu. En bref, je pense que s'il m'arrive de vivre jusqu'à*

---

[37] Il n'est pas inintéressant de mettre en perspective la place très différente que tiennent les mathématiques dans le traitement que font Fresnel et Young de problèmes similaires, avec le fait que les théories essentiellement qualitatives de la physique expérimentale (thermique, électricité, magnétisme, optique) ne se sont réellement et massivement ouvertes aux mathématiques que dans les vingt premières années du XVIII[e] siècle, et sous l'impulsion d'auteurs presque exclusivement français (Laplace, Carnot, Fourier, Ampère, Poisson, Fresnel…) ; cette tradition inaugurée en France dans le premier quart du siècle n'étant reprise en Allemagne et en Angleterre qu'à partir du milieu des années 1840. Thomas S. Kuhn, *Tradition mathématique et tradition expérimentale dans le développement de la physique*, Annales. Économies, sociétés, civilisations. 30[e] année, n° 5, 1975, 993-994.

[38] Alexander Wood, *Thomas Young Natural Philosopher*, Cambridge University Press, 1954, 179-205.

[39] David Hargreave, *Thomas Young' s Theory of Color Vision: Its Roots, Development, and Acceptance by the British Scientific Community*, Thèse de doctorat, University of Wisconsin, 1973.

[40] James Clerk Maxwell, *On the Theory of Colours in Relation to Colour-Blindness*, Transactions of the Royal Scottish Society of Arts, 1856, 420-428.

[41] Hermann von Helmholtz, *Handbuch der physiologischen Optik*, 1867. Traduction française : *Optique Physiologique*, Trad. : Émile Javal et N. Th. Klein, Paris, Victor Masson, 1867.

[42] R. Steven Turner, *The origins of colorimetry : what did Helmholtz and Maxwell learn from Grassmann ?*, In: Schubring G. (eds) Hermann Günther Graßmann (1809–1877): Visionary Mathematician, Scientist and Neohumanist Scholar, Boston Studies in the Philosophy of Science, vol. 187, 1996.

[43] Hermann von Helmholtz, *Optique Physiologique*, Deuxième partie, Chapitre 20, Trad. : Émile Javal et N. Th. Klein, Paris, Victor Masson, 1867, 408.

[44] Alexander Wood, *Thomas Young Natural Philosopher*, Cambridge University Press, 1954, 91.

*mes deux cents ans, je commencerai enfin à récolter quelque crédit et rétribution pour les travaux de ma jeunesse – mais qu'à cent ans je ne serai toujours qu'un petit garçon pour l'opinion publique.* » Ainsi était-il plus que temps qu'une traduction française de sa *Théorie de la Lumière et des Couleurs* vienne participer à sa manière au rayonnement du dernier des génies polymathes qu'était Thomas Young.

{P.12}[45] *Conférence Bakerienne. Sur la Théorie de la Lumière et des Couleurs. Par* Thomas Young, *M. D. F. R. S.*[46] *Professeur de Philosophie Naturelle à la Royal Institution.*

Lu le 12 Novembre 1801[47].

Bien que l'invention d'hypothèses plausibles, indépendantes de toute connexion avec les observations expérimentales, soit d'une utilité très médiocre pour la promotion de la connaissance de la nature ; la découverte de principes simples et uniformes, par lesquels un grand nombre de phénomènes apparemment hétérogènes sont réduits à des lois cohérentes et universelles, doit cependant toujours être admise comme étant d'une importance considérable pour le progrès de l'intellect humain[48].

    L'objet de la présente dissertation n'est pas tant de proposer des opinions absolument nouvelles que de réattribuer certaines théories déjà avancées à leurs inventeurs originaux, de les appuyer par des preuves nouvelles et de les appliquer à un grand nombre de faits très divers qui sont jusqu'ici restés enterrés dans l'obscurité. Ainsi n'est-il pas absolument nécessaire à cette occasion de produire la moindre expérience nouvelle[49] ; car d'expériences nous avons déjà une large provision, qui sont d'autant plus indiscutables qu'elles ont dû être conduites sans la moindre partialité pour le système par lequel elles vont être expliquées ; cependant certains faits jusque-là inobservés seront mis en avant, afin de montrer l'accord parfait de ce système avec les très divers phénomènes de la nature.

    {P.13} Les observations en optique de Newton restent à ce jour inégalées ; et, exception faite de quelques imprécisions occasionnelles, elles ne font que monter dans notre estime lorsqu'on les compare aux tentatives ultérieures de les améliorer. Une réflexion approfondie sur les couleurs des lames minces, telles qu'elles sont décrites dans le second livre de l'optique de Newton, a converti cette opinion préconçue que j'avais déjà du

---

[45] On indiquera entre accolades le numéro de la page sur laquelle se trouve la partie du texte qui les suit, dans la version originale de la conférence publiée dans les Philosophical Transactions of the Royal Society of London, vol. 92, 1802, 12–48.

[46] Thomas Young est docteur en médecine (Doctor of Medicine : M. D.) depuis 1796 et membre de la Royal Society (Fellow of the Royal Society : F. R. S.) depuis 1794 déjà.

[47] Ce texte est écrit en juillet 1801, lu le 12 novembre de la même année devant la Royal Society, et publié dans les Philosophical Transactions of the Royal Society of London de 1802. C'est cette version qui sera traduite et commentée ici. Cette conférence est de nouveau publiée, en deux parties, dans le Nicholson's Journal of Natural Philosophy, Chemistry and the Arts, en juin et juillet 1802, avec des ajustements mineurs et dans Thomas Young, *A Course of Lectures in Natural Philosophy and Mechanical Arts*, vol. II, London, 1807, avec quelques modifications très significatives dont on signalera les principales.

[48] Thomas Young introduit sa *Théorie de la Lumière et des Couleurs* par quatre Hypothèses sur lesquelles repose toute sa réinterprétation des phénomènes lumineux et colorés. Ce choix méthodologique, bien que justifié, le place immédiatement en porte-à-faux avec le fameux « hypotheses non fingo » de Newton, ou plutôt avec l'interprétation que ses successeurs ont pu en faire en s'opposant farouchement à tout type d'hypothèse (comme peut le faire par exemple Thomas Reid dans *Essays on the Intellectual Powers of Man*, première édition, Edinburgh, 1785). C'est la raison pour laquelle Young prend soin dès ce premier paragraphe de distinguer les mauvaises des bonnes hypothèses.

[49] Thomas Young n'a jamais été un expérimentateur passionné. Il confie à son ami d'enfance et biographe Hudson Gurney « *qu'à aucune période de sa vie il n'affectionna particulièrement de répéter des expériences, ni même de fréquemment tenter d'en produire de nouvelles : considérant que, bien que nécessaires au progrès de la science, elles demandaient un important sacrifice de temps, et que quand le fait était établi une bonne fois, ce temps était mieux employé à considérer les objectifs auxquels il pourrait être appliqué, et les principes qu'il pourrait tendre à élucider.* » (Hudson Gurney, *Memoir of the life of Thomas Young, M.D.F.R.S.*, 1831, 12-13).

système ondulatoire de la lumière en une très forte conviction de sa vérité et de sa suffisance ; conviction qui a depuis été confirmée d'une manière des plus frappantes par l'analyse des couleurs des corps striés. Les phénomènes des lames minces sont de fait si singuliers que leur aspect général n'est que très difficilement conciliable avec quelque théorie, aussi compliquée soit-elle, qui leur ait été jusqu'ici appliquée ; et quelques-unes des principales observations n'ont jamais été expliquées que par les hypothèses les plus gratuites ; mais il apparaitra que les plus infimes détails de ces phénomènes sont non seulement parfaitement compatibles avec la théorie qui va maintenant être détaillée, mais qu'ils sont tous les conséquences nécessaires de cette théorie, sans aucune supposition auxiliaire ; et ce par des inférences si simples qu'ils deviennent des corollaires particuliers, ne méritant que rarement une énumération distincte.

  Un examen plus vaste des divers écrits de NEWTON m'a montré qu'il était en réalité le premier à avoir suggéré une théorie telle que celle que je vais faire tout mon possible pour soutenir ; que ses propres opinions différaient moins de cette théorie qu'on le suppose aujourd'hui presque universellement ; et qu'une variété d'arguments ont été avancés, comme pour le réfuter, qui peuvent être trouvés sous une forme presque similaire dans ses propres travaux ; et ce par un mathématicien non moindre que LEONARD EULER, dont le système de la lumière[50], autant qu'il vaille la peine de le remarquer, était, ou aurait pu être, {P.14} intégralement emprunté à NEWTON, HOOKE[51], HUYGENS[52], et MALEBRANCHE[53].

  Ceux qui sont attachés, comme ils peuvent l'être très justement, à toute doctrine estampillée de l'approbation Newtonienne, seront probablement disposés à prêter à ces considérations d'autant plus d'attention qu'elles semblent coïncider au plus près avec les propres opinions de NEWTON. Pour cette raison, après avoir brièvement exposé chaque position particulière de ma théorie, je recueillerai dans les divers écrits de NEWTON des passages qui semblent les plus favorables à son acceptation ; et, bien que je puisse citer certains papiers que l'on peut penser avoir été partiellement rétractés à la publication de l'optique[54], je ne leur emprunterai cependant rien que l'on puisse supposer militer contre son jugement à maturité.

**Hypothèse I**

*Un Éther[55] luminifère imprègne l'Univers, rare[56] et élastique[57] à très haut degré.*

---

[50] Leonhard Euler, *Nova Theoria Lucis et Colorum*, 1746.
[51] Robert Hooke, *Robert's Hooke critique of Newton's theory of light and colours* (envoyé en 1672, et publié dans Thomas Birch, The History of the Royal Society, vol. 3, 1757, 10-15).
[52] Christian Huygens, *Traité de la Lumière,* 1690.
[53] Nicolas Malebranche, *Mémoire*, 1699.
[54] Newton a écrit de nombreux textes sur la lumière parmi lesquels Young va aussi puiser, depuis la lettre où il présente pour la première fois sa théorie des couleurs, appuyée par l'expérience cruciale du prisme en 1672 (*A Letter of Mr. Isaac Newton, Professor of the Mathematicks in the University of Cambridge ; Containing His New Theory about Light and Colors*, Philosophical Transactions of the Royal Society of London, vol. 6, 1672, 3075–3087) jusqu'à la dernière édition de l'*Opticks* en 1730. Pour les citations tirées de l'*Opticks* c'est à la pagination de cette dernière édition de 1730 que nous renverrons par la suite.
[55] Le terme *éther* semble avoir existé de tout temps dans les théories du Monde pour désigner un élément insaisissable et imperceptible du système, sans lequel celui-ci ne peut pourtant fonctionner. Dans chaque théorie l'éther joue un rôle différent, pourtant dans la majorité des cas il est imaginé comme un milieu permettant de transmettre des effets entre des corps distants. Ainsi selon les époques et les théoriciens il a pu être le support (ou la matière même) de la lumière, de l'action à distance, de l'électricité, de la chaleur (Edmund T. Whittaker, *A History of the theories of aether and electricity,* Dublin : Longman, Green and Co.,

*Passages de* NEWTON.

« L'hypothèse a certainement une bien plus grande affinité avec la sienne, » c'est-à-dire, celle du Dr. HOOKE « qu'il ne semble en être conscient ; les vibrations de l'éther étant aussi utiles et nécessaires dans celle-ci, que dans la sienne. »[58] (*Phil. Trans*. Vol. VII. p. 5087. *Abr*. Vol. I. p. 145. Nov. 1672.[59])

« Pour en venir à l'hypothèse : premièrement il faut supposer dans celle-ci, qu'il y a un milieu éthéré, d'une constitution très similaire à celle de l'air, mais bien plus rare, plus subtil et plus fortement élastique. […][60] Il n'est pas à supposer, que ce milieu est une matière uniforme, mais qu'il est composé, en partie du corps flegmatique[61] principal de l'éther, en partie d'autres esprits éthérés, {P.15} de manière très similaire à celle dont l'air est composé du corps flegmatique de l'air, entremêlé avec diverses vapeurs et exhalaisons : car les

---

1910). Il est intéressant de remarquer la versatilité du concept d'éther qui, comme on s'en rendra compte en lisant les passages de Newton ci-après, a la particularité d'être décliné dans la majorité des théories corpusculaires, comme des théories ondulatoires de la lumière développées du XVIIe au XIXe siècle.

[56] Tout au long de ce texte, le mot « rare » reviendra pour qualifier l'éther. Il devra être entendu au sens de *subtil*, ou *très peu dense*. L'éther de Young doit nécessairement être *rare* car il doit, entre autres choses, pouvoir pénétrer les corps solides, permettre la propagation de la lumière dans des milieux vidés de leur air, et ne pas freiner la révolution des planètes.

[57] C'est-à-dire qu'il a la propriété de retrouver sa forme initiale après avoir été contraint ou déformé. La propriété d'élasticité, et donc d'extrême dureté des particules d'éther s'avère une nécessité dans les théories ondulatoires de Huygens (*Traité de la lumière*, 1690, Chapitre I) et Young, où la propagation de la lumière est envisagée sous le mode d'une onde longitudinale (comme dans le cas du son). Dans cette analogie, la lumière est propagée de proche en proche, par le choc des particules d'éther avec leurs voisines ; et pour justifier à la fois de l'extrême rapidité de la propagation de ces chocs, et du cheminement possible de cette onde de choc sur d'immenses distances (depuis les étoiles jusqu'à nous) sans amortissement, il est nécessaire d'affirmer l'extrême élasticité du milieu de propagation.

[58] Newton poursuit : « *Car, en supposant que les rayons de lumière sont de petits corps émis dans toutes les directions depuis les substances brillantes, ceux-ci, lorsqu'ils parviennent à quelque surface réfringente ou réfléchissante que ce soit, doivent tout aussi nécessairement exciter des vibrations dans l'éther, comme les pierres le font dans l'eau quand elles y sont lancées.* »

[59] Isaac Newton, *Mr. Isaac Newton's Answer to Some Considerations upon His Doctrine of Light and Colors; Which Doctrine Was Printed in Numb. 80. of These Tracts,* Philosophical Transactions of the Royal Society of London, vol. 7, 1672, 5084-5103. Lettre contenant les contre-objections de Newton aux objections émises par Robert Hooke (*Robert's Hooke critique of Newton's theory of light and colours, op.cit.*) à sa nouvelle théorie de la lumière et des couleurs, elle-même publiée dans Philosophical Transactions of the Royal Society of London, vol. 6, 1672, 3075–3087. La même réponse de Newton est rééditée dans les Philosophical Transactions Abriged 2, 1672, 13-29 ; donc dans une notation différente de celle indiquée par Young.

[60] Ici comme en de nombreux endroits, Young coupe certaines phrases du texte original de Newton ; souvent pour alléger le texte, toujours en mentionnant ces coupes par un « – » dans son texte original, que nous avons remplacé dans cette traduction par le symbole moins ambigu « […] ». Ici, dans la phrase coupée Newton ajoute : « *De l'existence de ce milieu, le mouvement d'un pendule dans une cellule de verre vidée d'air aussi rapide qu'à l'air libre, n'est pas un argument négligeable.* » Phrase dans laquelle on pressent que manquent des mots tels que « atténuation du » devant le mot « mouvement ». Où Newton défend donc l'existence d'un éther sur la base d'une expérience observant l'atténuation du mouvement d'un pendule oscillant dans une chambre vidée de son air, et constatant qu'elle est à peu près aussi rapide qu'à l'air libre. Très certainement du fait que l'atténuation de ce mouvement est généralement dominée par les frottements au niveau du point d'ancrage du pendule plutôt que par les frottements de l'air.

[61] C'est-à-dire *humide*.

effluves électriques et magnétiques, et le principe gravitationnel[62], semblent prouver une telle variété. » (*Birch. Hist. of R. S.* Vol. III. p. 249. Dec. 1675.)[63]

« La chaleur (de la chambre chaude)[64] n'est-elle pas transmise à travers le vide par les vibrations d'un milieu bien plus subtil que l'air ? […][65] Et ce milieu n'est-il pas le même qui réfracte et réfléchit la lumière, et par les vibrations duquel la lumière communique la chaleur aux corps, et qui la met dans des accès de facile réflexion et facile transmission ?[66] Et les vibrations de ce milieu à l'intérieur des corps chauds, ne contribuent-elles pas à l'intensité et la durée de leur chaleur ? Et les corps chauds ne communiquent-ils pas leur chaleur aux corps froids contigus par les vibrations de ce milieu propagées depuis eux vers les corps froids ? Et ce milieu n'est-il pas excessivement plus rare et subtil que l'air, et excessivement plus élastique et actif ? Et n'imprègne-t-il pas aisément tous les corps ? Et n'est-il pas, de par sa force élastique, répandu à travers tous les cieux ?[67] […] Ne se peut-il pas que les planètes et les comètes, et tous les corps grossiers, accomplissent leurs mouvements[68] dans ce milieu éthéré ? […][69] Et sa résistance ne peut-elle être si faible qu'elle soit négligeable ? Pour exemple, si cet éther (car c'est ainsi que je l'appellerai) était supposé 700.000 fois plus élastique que notre air, et plus de 700.000 fois plus rare, sa résistance serait environ 600.000.000 de fois plus faible que celle de l'eau. Et une résistance si faible saurait difficilement provoquer quelconque altération des mouvements des planètes, en dix mille ans même. Si qui que ce soit demandait comment un milieu peut être aussi rare, qu'il me dise […] comment un corps électrique peut-il par friction émettre une exhalaison si rare et subtile, et pourtant si puissante ? […] Et comment les {P.16} effluves d'un aimant peuvent passer à travers une plaque de verre sans résistance, et encore faire tourner une aiguille magnétique située au-delà du verre ? » (*Optics*, Qu. 18, 22)[70]

**Hypothèse II**
*Des ondulations sont excitées dans cet Éther chaque fois qu'un Corps devient lumineux.[71]*

---

[62] Pour plus d'éléments sur la théorie de Newton de l'éther produisant les phénomènes lumineux, mais aussi électriques, magnétiques, calorifiques et gravitationnels, voir par exemple Edmund T. Whittaker, *A History of the theories of aether and electricity,* Dublin : Longman, Green and Co., 1910, 17-20.

[63] Isaac Newton, *An Hypothesis explaining the properties of light, discoursed of in my several papers*, lu en 1675 et publié dans Thomas Birch, The History of the Royal Society, vol. 3, London, 1757, 248-260.

[64] « (de la chambre chaude) » ajouté par Young. En effet Newton décrit ici l'expérience consistant à mettre côte à côte deux vases cylindriques en verre scellés, à suspendre au centre de chacun un thermomètre, à faire le vide dans l'un des deux cylindres, et enfin à déplacer ensemble les deux vases d'une chambre froide vers une chambre plus chaude. Newton observe que le thermomètre indique une augmentation de chaleur aussi haute et aussi rapide dans le vase vide que dans celui rempli d'air. Il développe alors sur la conclusion reprise ici par Young.

[65] Newton : « …, *qui après que l'Air fut retiré restait encore dans le Vide ?* »

[66] Les termes d' « accès de facile réflexion » et d' « accès de facile transmission » ont été introduits par Newton pour expliquer les couleurs des lames fines. On détaille leur mécanisme en note de bas de page du Corollaire I de la Proposition VIII, dédié à ces couleurs.

[67] Ici s'arrête la question 18 de l'*Opticks*. Par l'ellipse qui suit, Young saute au début de la question 22.

[68] Newton : « …*plus librement et avec moins de résistance…* »

[69] Newton : « …*que dans tout autre Fluide, qui remplit tout l'Espace adéquatement sans laisser le moindre Pore, et qui est par conséquent plus dense que le Vif-argent ou l'Or ?* »

[70] Isaac Newton, *Opticks*, Livre III, Question 18 et Question 22, 1730, 323 et 327.

[71] A partir de cette proposition, Young s'évertue à défendre la nature ondulatoire de la propagation de la lumière. Les ondulations évoquées ici, deviendront au fil de ce texte les ondes lumineuses et même, en conclusion, des ondes de chaleur. Toutefois il faut garder en tête qu'ici, et dans l'intégralité des travaux de Young, la propagation de la lumière est imaginée comme une ondulation *longitudinale*, ayant lieu dans la direction de propagation de la lumière (comme une onde sonore), et non comme une onde *transverse* oscillant

*Scholie*. J'utilise le mot ondulation, préférablement à vibration, car vibration est généralement compris comme impliquant un mouvement qui se poursuit alternativement d'arrière en avant, par une combinaison du moment[72] du corps avec une force accélératrice, et qui est naturellement plus ou moins permanent ; mais une ondulation est supposée consister en un mouvement vibratoire transmis successivement par différents parties d'un milieu sans la moindre tendance de chaque particule à poursuivre son mouvement, excepté en conséquence de la transmission d'ondulations successives depuis un corps vibrant distinct ; comme, dans l'air, les vibrations d'une corde produisent les ondulations constituant le son.

*Passages de* NEWTON.

« Si je devais faire une hypothèse, ce serait celle-ci, avancée plus généralement de manière à ne pas déterminer ce qu'est la lumière au-delà du fait qu'il s'agit d'une chose ou d'une autre capable d'exciter des vibrations dans l'éther ; car ainsi elle deviendra si générale et inclusive des autres hypothèses qu'elle laissera peu de place pour que d'autres soient inventées. » (*Birch*. Vol. III. p. 249. Dec. 1675)[73]

« En second lieu, il est à supposer que l'éther est un milieu vibrant comme l'air, bien que de vibrations bien plus rapides et minuscules ; celles de l'air, provoquées par la voix ordinaire d'un homme, se succédant les unes aux autres à plus d'un demi-pied ou un pied[74] {P.17} de distance ; mais celles de l'éther à une distance inférieure à la cent-millième partie d'un pouce[75]. Et, comme dans l'air certaines vibrations sont plus grandes que d'autres mais toutes également rapides, (puisque dans un carillon de cloches, le son de chaque ton est audible à deux ou trois miles de distance dans le même ordre où les cloches sont frappées) ainsi, je suppose, les vibrations éthérées diffèrent en grandeur mais non en rapidité. A présent ces vibrations, au-delà de leur utilité pour la réflexion et la réfraction, peuvent être supposées comme le principal moyen par lequel les parties des substances en fermentation

---

perpendiculairement à la propagation (comme les ondes à la surface de l'eau). Difficile en effet d'admettre qu'un milieu aussi *rare* que doit l'être l'éther puisse osciller transversalement, aucune force ne pouvant raisonnablement lier les particules le composant ; le modèle de l'onde de choc s'impose donc assez naturellement par analogie avec le son. Ce n'est qu'après les démonstrations publiées par Fresnel en 1821 (Augustin Fresnel, *Note sur le calcul des teintes que la polarisation développe dans les lames cristallisées*, Annales de chimie et de physique, tome XVII, 1821, p.102, 167 et 312) que la nature *transverse* de la propagation de la lumière sera globalement acceptée.

[72] « momentum » : *moment*, ou *quantité de mouvement*. C'est-à-dire le produit de la masse par le vecteur vitesse du corps. Le second principe de la mécanique newtonienne nous dit que la somme des forces extérieures appliquées au corps est égale à la dérivée par rapport au temps de sa quantité de mouvement.

[73] Isaac Newton, *An Hypothesis explaining the properties of light, discoursed of in my several papers*, lu en 1675 et publié dans Thomas Birch, The History of the Royal Society, vol. 3, 1757, 248-260. La position de Newton dans ce texte est bien moins tranchée en faveur de la nature ondulatoire de la lumière que l'extrait choisi par Young ne le laisse supposer. Dans cette réponse aux objections de Hooke, déjà évoquées plus haut, Newton insiste sur le fait qu'une hypothèse corpusculaire n'est pas incompatible avec la présence d'ondulations dans l'éther. Puisque le mouvement des corpuscules est très certainement capable d'exciter des vibrations. Ainsi malgré une concession superficielle, il y a peu de doute dans ce texte quant au fait que pour Newton la lumière n'est pas constituée de vibrations, mais qu'elle est bien « *une chose ou une autre capable d'exciter des vibrations dans l'éther* », donc qu'elle est très certainement composée de corpuscules en mouvement.

[74] Bien que les étalons de longueur aient légèrement évolué depuis, on pourra considérer ici : 1 Mile = 1,609 km. 1 pied = 30,48 cm. 1 pouce = 2,54 cm.

[75] 1/100000 * 2,54 cm, soit 254 nm. L'ordre de grandeur est correct.

ou putréfaction, les fluides liquoreux, ou les corps fondus, brûlants, et autres corps chauds poursuivent leur mouvement. » (*Birch*. Vol. III. p. 251. Dec. 1675)[76]

« Lorsqu'un rayon de lumière tombe sur la surface de quelque corps transparent, et qu'il y est réfracté ou réfléchi, des vagues de vibrations, ou des secousses, ne peuvent-elles pas de la sorte être excitées dans le milieu réfringent ou réfléchissant ? […] Et ces vibrations ne sont-elles pas propagées depuis le point d'incidence sur de grandes distances ? Et ne dépassent-elles pas les rayons de lumière, et en les dépassant successivement, ne les mettent-elles pas dans les accès de facile réflexion et facile transmission décrits plus haut. » (*Optics*. Qu. 17.)[77]

« La lumière est dans des accès de facile réflexion et facile transmission, avant son incidence sur les corps transparents. Et probablement est-elle mise dans de tels accès dès sa première émission par les corps lumineux, et persévère dans ceux-ci pendant toute sa propagation. » (*Optics.* Livre Second. Partie III. Prop. 13)[78]

{P.18} **Hypothèse III**
*La Sensation des différentes Couleurs dépend des différentes fréquences de Vibrations, excitées par la Lumière dans la Rétine.*[79]

*Passages de* NEWTON.

« L'hypothèse de la personne faisant objection[80], pour sa partie fondamentale, n'est pas contre moi. Cette supposition fondamentale est que les parties des corps, lorsqu'elles sont vivement agitées, excitent des vibrations dans l'éther qui sont propagées dans toutes les directions depuis ces corps en lignes droites, et causent une sensation de lumière en battant et s'écrasant contre le fond de l'œil, de la même manière que les vibrations dans l'air causent une sensation de son en battant contre les organes de l'ouïe[81]. A présent, l'application la plus libre et naturelle de cette hypothèse à la résolution de phénomènes me semble être celle-ci : que les parties agitées des corps, en fonction de leurs diverses tailles, formes et mouvements, excitent des vibrations de l'éther de différentes profondeurs ou grandeurs qui, étant propagées confusément à travers le milieu jusqu'à nos yeux, provoquent en nous une sensation de lumière de couleur blanche ; mais si par quelque moyen celles de grandeurs inégales sont séparées les unes des autres, les plus grandes engendrent une sensation de couleur rouge, les moindres ou plus courtes d'un violet profond, et les intermédiaires de couleurs intermédiaires ; de la même manière que les corps, selon leurs différentes tailles, formes et mouvements, excitent des vibrations dans l'air de grandeurs variées qui, selon ces grandeurs, font des tons différents dans le son : que les plus grandes vibrations sont plus capables de surmonter la résistance des surfaces réfringentes, et donc à y pénétrer avec une moindre réfraction ; d'où les vibrations {P.19} de différentes grandeurs, c'est-à-dire les rayons de couleurs différentes, qui sont mélangées ensemble dans la lumière, doivent être séparées les unes des autres par réfraction, et ainsi causer les phénomènes des prismes et des autres substances réfringentes ; et qu'il dépend de l'épaisseur d'une lame fine transparente ou d'une bulle si une vibration sera réfléchie à ses surfaces suivantes ou transmise ; de façon que, selon le nombre de vibrations intervenant entre les deux surfaces, elles peuvent être réfléchies ou transmises par de nombreuses surfaces successives. Et, puisque les vibrations qui causent le bleu et le violet sont supposées plus courtes que celles qui causent le rouge et le jaune, elles doivent être réfléchies pour une épaisseur moindre de la lame : ce qui est suffisant pour expliquer tous les phénomènes ordinaires de ces lames ou de ces bulles, ainsi que de tous les corps naturels, dont les parties sont comme de très nombreux fragments de telles lames[82]. Celles-ci semblent être les plus simples, véritables et nécessaires conditions de cette hypothèse. Et elles s'accordent si convenablement avec ma théorie, que si le critique pense qu'il est approprié de les appliquer, il ne lui est pas nécessaire, sur ce compte, de réclamer un divorce d'avec elle. Mais cependant, comment la défendra-t-il d'autres difficultés, je ne le sais point. » (*Phil. Trans.* Vol. VII. p. 5088. *Abr*. Vol. I. p. 145. Nov. 1672)[83].

---

[76] Isaac Newton, *An Hypothesis explaining the properties of light, discoursed of in my several papers*, lu en 1675 et publié dans Thomas Birch, The History of the Royal Society, vol. 3, 1757, 248-260.
[77] Isaac Newton, *Opticks*, Livre III, Question 17, 1730, 322.
[78] Isaac Newton, *Opticks*, Livre II, Partie III, Proposition 13, 1730, 257.
[79] Cette hypothèse visionnaire, justifiant la triplicité des photorécepteurs de la rétine nécessaire à la sensation des couleurs, ne sera pas saisie à sa juste valeur avant les travaux de Helmholtz (Hermann von Helmholtz, *Handbuch der physiologischen Optik*, vol. I, Part. II, 1867, version française : *Optique Physiologique*, Trad. : Émile Javal et N. Th. Klein, Paris, Victor Masson, 1867) et ne sera démontrée expérimentalement qu'au milieu du XX$^e$ siècle (Gunnar Svaetichin, *Spectral response curves from single cones*, Acta Physiologica Scandinavica, vol. 39, suppl. 134, 1956, 17-46). Cette hypothèse, bien que reposant fondamentalement sur un modèle ondulatoire de la lumière, semble avoir une position à part dans *la Théorie de la Lumière et des Couleurs*, tant elle est déconnectée du reste du texte. On l'interprétera volontiers comme une extension de la précédente conférence bakerienne de Young (*On the Mechanism of the Eye*, Philosophical Transactions of the Royal Society of London, vol. 91, 1801, 23–88) dédiée à l'étude de l'œil sans pour autant que la vision colorée y soit traitée, mais dès le début de laquelle on peut lire (III, p.25) que l'oreille : « *est l'unique organe qui peut être strictement comparé* » à l'œil. Car on peut voir effectivement l'hypothèse III projetée ici par Young comme une rencontre entre comparaison subtile de l'anatomie de l'œil (*On the Mechanism of the Eye*, 1801) et de l'oreille (Thomas Young, *De Corporis Humani Viribus Conservatricibus,* Thèse de doctorat, Gottingae : typis J.C. Dieterich, 1796), de la très forte analogie que Young défend entre le son et la lumière (*Outlines of Experiments and Inquiries respecting Sound and Light*, Philosophical Transactions of the Royal Society of London, vol. 90, 1800, 106-150) et de ses réflexions sur l'élasticité des matériaux (*A Course of Lectures on Natural Philosophy and Mechanical Arts*, vol. 2, Part II, Section 9 : « On the equilibrium and strength of elastic substances », London, 1807, 46-51), catalysée par la volonté de Young, dans cette *Théorie de la Lumière et des Couleurs*, de justifier d'un maximum de phénomènes par le modèle ondulatoire.
[80] Il s'agit encore de Robert Hooke.

« Pour expliquer les couleurs, je suppose que comme les corps de tailles, densités, ou sensations variées excitent par percussion ou autre action des sons de tons différents, et conséquemment des vibrations dans l'air de différentes grandeurs ; ainsi les rayons de lumière, en frappant contre les surfaces réfringentes rigides, excitent des vibrations dans l'éther […] de différentes grandeurs ; les rayons les plus grands, les plus forts ou les plus puissants, les vibrations les plus grandes ; et les autres des plus petites, selon leur grandeur, force, ou puissance : ainsi les terminaisons des filaments[84] du nerf optique qui pavent {P.20} ou recouvrent la rétine étant de telles surfaces réfringentes, lorsque les rayons les frappent, ils doivent à cet endroit exciter ces vibrations, lesquelles (comme celles du son dans une trompe ou une trompette) courront le long des pores aqueux ou de la moelle cristalline des filaments, par les nerfs optiques, jusqu'au sensorium ; […][85] et à cet endroit, je suppose, affectent le sens avec différentes couleurs selon leur grandeur et mélange ; les plus grandes avec les couleurs les plus fortes, rouges et jaunes ; les plus petites avec les plus faibles, bleues et violettes ; les moyennes avec le vert ; et une confusion de toutes avec le blanc, de la même manière que dans le sens de l'ouïe la nature fait usage de vibrations de l'air de diverses tailles pour générer des sons de divers tons ; car l'analogie de la nature doit être respectée. » (*Birch* Vol. III. P. 262. Dec. 1675)[86].

---

[81] Toute cette phrase est écrite en italique dans le texte original de Newton afin de bien faire ressortir ce qu'il appelle la supposition fondamentale de Hooke, laquelle est l'objet de toute l'analyse qui suit.
[82] Ici Newton anticipe l'interprétation des couleurs des lames minces par la théorie des accès déjà évoquée plus haut et qu'il développera dans l'*Opticks*, mais dont nous parlerons plus précisément en note de bas de page au Corollaire I de la Proposition VIII, dédié à cette question.
[83] Isaac Newton, *Mr. Isaac Newton's Answer to Some Considerations upon His Doctrine of Light and Colors; Which Doctrine Was Printed in Numb. 80. of These Tracts,* Philosophical Transactions of the Royal Society of London, vol. 7, 1672, 5084-5103. Le même texte est réédité dans Philosophical Transactions Abriged 2, 1672, 13-29. Newton poursuit en effet : « *Selon moi, la supposition fondamentale elle-même semble impossible ; A savoir, que les ondes ou vibrations de quelque fluide puissent, comme les rayons de lumière, être propagées en lignes droites, sans un étalement et une incurvation continus, les faisant déborder largement de leurs limites dans toutes les directions à l'intérieur du milieu au repos, là où elles sont limitées par celui-ci. Je me trompe, s'il n'y a pas à la fois l'expérience et la démonstration du contraire.* » Objection qui l'éloigne radicalement de l'hypothèse ondulatoire défendue par Young, mais que ce-dernier reprendra *in extenso* pour la contredire dans la Proposition III.

« Considérant la durée des mouvements excités dans le fond de l'œil par la lumière, ne sont-ils pas de nature vibratoire ?[87] […] Les rayons les plus réfrangibles n'excitent-ils pas les vibrations les plus courtes, […] les moins réfrangibles les plus courtes ?[88] L'harmonie et la discordance des couleurs ne naitrait-elle pas des proportions des vibrations propagées par les fibres du nerf optique jusqu'au cerveau, comme l'harmonie et la discordance des sons nait des proportions des vibrations de l'air. » (*Optics*, Qu. 16, 13, 14)[89]

*Scholie.* Puisque, pour la raison assignée ici par Newton, il est probable que le mouvement de la rétine est plus de nature vibratoire qu'ondulatoire, la fréquence des vibrations doit être dépendante de la constitution de cette substance. Maintenant, puisqu'il est pratiquement impossible de concevoir que chaque point sensible de la rétine contienne un nombre infini de particules, chacune capable de vibrer en parfait unisson avec toute ondulation possible[90], il {P.21} devient nécessaire de supposer leur nombre limité, par exemple aux trois principales couleurs, rouge, jaune, et bleu, auxquelles les ondulations sont liées en magnitude environ comme les nombres 8, 7 et 6 [91]; et que chacune des particules est capable d'être mise en mouvement de manière plus ou moins forcée, par des ondulations différant plus ou moins d'un parfait unisson ; par exemple, les ondulations d'une lumière verte étant environ dans le rapport $6_{1/2}$, affecteront également les particules qui sont à l'unisson avec le jaune et le bleu, et produiront le même effet qu'une lumière composée de ces deux espèces : et chaque filament sensible du nerf est probablement constitué de trois portions, une pour chaque couleur principale. En admettant cette proposition, il apparait que toute tentative de produire un effet musical à partir des couleurs doit être infructueuse, ou tout du moins que rien mieux qu'une mélodie très simple pourrait être imitée avec elles ; car la période, qui en fait constitue l'harmonie de tout accord, étant un multiple des périodes des ondulations individuelles, serait dans ce cas intégralement en-dehors des limites de sympathie de la rétine et perdrait son effet ; de la même manière que l'harmonie d'une tierce ou d'une quarte est détruite en la diminuant aux notes les plus basses de l'échelle audible. Dans l'ouïe, il ne semble pas y avoir de vibration permanente de quelque partie de l'organe que ce soit.

---

[84] Le mot utilisé ici par Newton est « capillamenta » dont le dictionnaire de Johnson (1755) propose la définition suivante : « *Petits fils ou poils qui croissent au milieu d'une fleur. Quincy.* »
[85] Newton : « *(chose que la lumière elle-même ne peut pas faire)* ».
[86] Isaac Newton, *An Hypothesis explaining the properties of light, discoursed of in my several papers*, lu en 1675 et publié dans Thomas Birch, The History of the Royal Society, vol. 3, 1757, 248-260.
[87] Ici s'arrête la question 16 de l'*Opticks*. Par l'ellipse qui suit, Young saute à la question 13.
[88] Ici Young quitte la question 13 de l'*Opticks* pour passer à la question 14.
[89] Isaac Newton, *Opticks*, Livre III, Question 16, Question 13 et Question 14, 1730, p.321 et 320.
[90] Young sait qu'il s'agit de l'exploit que réalise la cochlée dans l'oreille ; mais il estime qu'il s'agit d'un organe trop complexe pour être miniaturisé et reproduit en chaque point de la rétine.
[91] Ces valeurs, affichées ici de manière apparemment arbitraire, seront justifiées plus loin par les mesures exploitées dans le Corollaire II de la Proposition VIII. Il deviendra alors clair que le mot *magnitude* employé ici par Young correspond en quelque façon à ce que l'on appelle aujourd'hui *longueur d'onde*. On pourra noter, sans en tenir compte pour autant dans la lecture de la suite du texte, que dans une publication ultérieure (*An account of some Cases of the production of Colours not hitherto described*, Philosophical Transactions of the Royal Society of London, 1802, 395), Young explique que les observations plus précises du spectre prismatique par Wollaston lui imposent de remplacer les couleurs « *rouge, jaune et bleu* » par les couleurs « *rouge, vert et violet* », et la série de rapports « *8, 7 et 6* » par la série « *7, 6 et 5* ».

**Hypothèse IV**
*Tous les Corps matériels ont une Attraction pour le Milieu éthéré, par le moyen de laquelle celui-ci est accumulé dans leur Substance et à une courte Distance autour d'eux, dans un État de Densité plus grande mais pas de plus grande Élasticité.*[92]

Il a été montré, que les trois hypothèses précédentes, qui peuvent être qualifiées d'essentielles, font littéralement partie du plus complexe système Newtonien. Peut-être que cette quatrième hypothèse diffère {P.22} à quelque degré de toute autre qui aurait été proposée par des auteurs antérieurs, et qu'elle est diamétralement opposée à celle de NEWTON[93] ; mais, les deux options étant par elles-mêmes également probables, l'opposition est seulement accidentelle ; et il s'agit seulement de demander laquelle est la mieux capable d'expliquer les phénomènes. Peut-être d'autres suppositions pourraient-elles être substituées à celle-ci, et par conséquent je ne la considère pas comme fondamentale, cependant elle parait être la plus simple et la meilleure de celles qui se sont présentées à moi.

---

[92] Il est intéressant de remarquer que dans la compilation des travaux de Young réunie à la fin de son cours à la Royal Institution publié en 1807 (Thomas Young, *A Course of Lectures in Natural Philosophy and Mechanical Arts*, vol. II, London, 1807) figure une version de *La Théorie de la Lumière et des Couleurs* dont l'hypothèse IV a été complètement réécrite par Young, sans pour autant que le texte l'illustrant ait été modifié. L'hypothèse IV devient alors : « *Tous les Corps matériels doivent être considérés, respectivement aux Phénomènes de la Lumière, comme constitués de Particules suffisamment éloignées les unes des autres pour permettre au Milieu éthéré de les pénétrer parfaitement librement, et soit de le retenir dans un état de plus grande densité et d'élasticité égale, soit de constituer, avec le Milieu lui-même, un Agrégat, pouvant être considéré plus dense mais pas plus élastique* » (p. 618). Un tel revirement en un temps si court est révélateur d'un changement radical dans la conception que Young se fait de l'éther, remarquablement renseigné dans Geoffrey N. Cantor, *The changing role of Young's ether,* The British Journal for the History of Science, vol. 3, 17, 1970, 44-62. Cantor montre là que l'abandon de cette « *hypothèse de distribution de l'éther* » par Young aura des répercussions épistémologiques lourdes sur la suite de ses travaux, localement manifestées par la réécriture (Proposition V Corollaire II, ou Proposition VIII Corollaire V), ou la disparition complète (Proposition VIII Corollaire IV), de passages significatifs de *la Théorie de la Lumière et des Couleurs* dans sa version de 1807.

[93] Les idées développées par Newton dans *An Hypothesis explaining the properties of light*, régulièrement cité jusqu'ici, reposent explicitement sur l'hypothèse opposée d'un éther moins dense dans les corps qu'à l'extérieur de ceux-ci. Newton y écrit par exemple : « the denser æther without the body, and the rarer within it ». C'est donc entre cette dernière hypothèse et son hypothèse IV, toutes deux défendant une conception opposée du mode de répartition de l'éther, que Young demande de trancher. On notera que même s'il viendra à l'abandonner plus tard, cette hypothèse joue un rôle pivot dans ce texte, permettant à Young d'introduire enfin sa propre théorie, et de justifier en particulier la Proposition V et les Corollaire IV et V de la Proposition VIII qui suivent.

Proposition I
*Toutes les Impulsions sont propagées dans un Milieu élastique homogène à Vélocité constante.*[94]

Chaque expérience relative au son coïncide avec l'observation de NEWTON déjà citée, que toutes les ondulations sont propagées dans l'air avec une vélocité égale[95] ; et ceci est plus largement confirmé par les calculs. (LAGRANGE. *Misc. Taur.* Vol. I. p. 91.[96] Mais aussi de manière beaucoup plus concise, dans mon *Programme de Conférences en Philosophie Naturelle et Expérimentale*[97], sur le point d'être publié. Article 289.) Si l'impulsion est suffisamment importante pour perturber matériellement la densité du milieu, celui-ci ne sera plus homogène ; mais autant que nos sens soient concernés, la quantité de mouvement peut être considérée comme infiniment petite. Il est surprenant qu'EULER, bien que conscient de ce fait, ait néanmoins maintenu que les ondulations les plus fréquentes[98] sont propagées plus rapidement. (*Theor. mus.* et *Conject. phys.*)[99] Il est possible que la vitesse réelle des particules de l'éther luminifère corresponde à une proportion de la vélocité des ondulations bien moindre que dans le cas du son ; car la lumière peut être excitée par le mouvement d'un corps se déplaçant à un rythme de seulement un mile dans le temps où la lumière en parcourt cent millions.

{P.23} *Scholie 1*. Il a été démontré que dans différents milieux la vitesse varie dans le rapport un demi de la force directement[100], et de la densité inversement[101]. (*Misc. Taur. Vol. I.* p. 91.[102] *Le programme de YOUNG*. Art. 294.[103])

---

[94] Dans cette proposition Young mêle deux informations relatives à la vitesse des impulsions lumineuses qu'il est difficile de discriminer dans sa formulation de la chose. D'une part il affirme que cette vitesse est *uniforme* ou *constante* pour chaque impulsion au cours de sa propagation dans un milieu homogène (le mot qu'il utilise alors est « equable »). D'autre part il affirme que cette vitesse est *identique* ou *égale* pour toutes les impulsions (le mot qu'il utilise alors est « equal »). Dans les deux cas, les éléments de preuves sont apportés exclusivement par des références à la propagation des ondes sonores ; les références à Lagrange et Young affirmant l'uniformité de la vitesse de propagation, et la référence à Euler militant en faveur d'une vitesse égale pour toutes les impulsions. Cette proposition revêt une importance majeure, car Young affirme dans la partie dédiée à « *l'analogie entre lumière et son* » de l'article *Outlines of Experiments and Inquiries respecting Sound and Light*, (Philosophical Transactions of the Royal Society of London, vol. 90, 1800, 125-130) que l'impossible justification par la théorie newtonienne d'une vitesse de propagation de la lumière toujours identique, quelle que soit la source de son émission, est la première « *difficulté* » pour laquelle il faut se tourner vers le système ondulatoire.

[95] Isaac Newton, *An Hypothesis explaining the properties of light, discoursed of in my several papers*, lu en 1675 et publié dans Thomas Birch, The History of the Royal Society, vol. 3, 1757, 248-260.

[96] Joseph Louis de Lagrange (1736-1813), *Recherches sur la nature et la propagation du son,* Section seconde, Chapitre premier : « De la vitesse du son », Miscellanea Taurinensia, vol. 1, 1759, 86-92.

[97] Thomas Young, *Syllabus of a course of Lectures on Natural and Experimental Philosophy*, Royal Institution, London, Art. 289, 1802, 86.

[98] C'est-à-dire *de fréquences les plus hautes*.

[99] Les titres des textes mentionnés ont été tronqués par Young et la liste des publications de Euler est foisonnante. Mais au vu du contenu évoqué et de la bibliographie sur le sujet rassemblée par Young à la fin de son *Course of Lectures in Natural Philosophy and Mechanical Arts*, vol. II, il s'agit probablement de : Leonhard Euler, *Tentamen Novae Theorae Musicae ex Certissismis Hamoniae Principiis Dilucide Expositae*, Saint Petersbourg, 1729. Et Leonhard Euler, *Conjectura Physica Circa Propagationem Soni ac Luminis*, Berlin, 1750.

[100] La traduction mathématique de cette phrase est $\frac{v_1}{v_2} = \frac{\sqrt{F_1}}{\sqrt{F_2}}$, où $v$ représente la vitesse d'une onde dans un milieu quelconque, évoluant en fonction de la force $F$ de l'onde. Il faut donc comprendre : « La vitesse de l'onde est proportionnelle à la racine carrée de la force ».

*Scholie 2*. Il est évident, de par les phénomènes des corps élastiques et des sons, que les ondulations peuvent se croiser l'une l'autre sans interruption. Mais il n'y a pas de nécessité à ce que les différentes couleurs de la lumière blanche entremêlent leurs ondulations ; car en supposant que les vibrations de la rétine se poursuivent ne serait-ce que cinq centièmes de seconde après leur excitation, un million d'ondulations de chacune du million de couleurs peuvent arriver en succession distincte dans cet intervalle de temps et produire le même effet sensible, comme si toutes les couleurs arrivaient précisément au même instant.[104]

Proposition II

*Une Ondulation conçue comme ayant pour origine la Vibration d'une Particule unique doit se déployer dans un Milieu homogène selon une Forme sphérique, mais avec différentes quantités de Mouvement en différentes Parties.*[105]

Car puisque chaque impulsion considérée comme positive ou négative est propagée avec une vélocité constante, chaque partie de l'ondulation doit dans des temps égaux avoir traversé des distances égales depuis le point vibrant. Et en supposant la particule vibrante, au cours de son mouvement, s'avancer d'une petite distance dans une direction donnée, la vigueur principale de l'ondulation sera naturellement droit devant elle ; derrière elle, le mouvement sera égal dans une direction contraire ; et à angles droits de la ligne de vibration, l'ondulation sera évanescente.

---

[101] Réciproquement, la traduction mathématique de cette phrase est $\frac{v_1}{v_2} = \frac{\sqrt{d_2}}{\sqrt{d_1}}$, où $v$ représente la vitesse d'une onde dans un milieu quelconque et $d$ la densité locale d'éther dans le milieu en question. Il faut donc comprendre ici : « La vitesse de l'onde est inversement proportionnelle à la racine carrée de la densité ». Ce résultat, combiné à l'Hypothèse IV selon laquelle les corps matériels attirent l'éther et l'accumulent dans leur substance, permettra à Young de défendre dans la Proposition V l'idée selon laquelle la lumière se propage plus lentement dans les corps matériels (où l'éther est plus dense) que dans l'air ; idée contraire à celle de Newton.

[102] Joseph Louis de Lagrange, *Recherches sur la nature et la propagation du son,* Miscellanea Taurinensia, vol. 1, 1759, I-112.

[103] Thomas Young, *Syllabus of a course of Lectures on Natural and Experimental Philosophy*, Royal Institution, London, Art. 294, 1802, 91. Le raisonnement réalisé dans ce court article repose sur l'étude de la vitesse du son se propageant dans une colonne remplie d'un fluide élastique de densité variable ; que Young s'autorise ici à extrapoler à la vitesse de propagation de la lumière dans l'éther.

[104] Le fait d'avoir distingué les *vibrations* qui se poursuivent dans le temps des *oscillations* qui sont n'ont lieu qu'une fois dans l'Hypothèse II, puis d'avoir affirmé dans l'Hypothèse III que les *ondulations* de l'éther généraient des *vibrations* des filaments du nerf optique, permet ici à Young de compléter son Hypothèse III relative au fonctionnement de l'œil, en justifiant le mécanisme de persistance rétinienne (la sensation visuelle, fruit d'une *vibration*, persiste quelques centièmes de seconde après que l'*ondulation* ait cessé) et la possibilité de sensation des couleurs complexes, qui en découle.

[105] Young confirme la propagation sphérique de l'ondulation à partir de la particule vibrante déjà affirmée par ses prédécesseurs. Mais il ajoute que l'ondulation ne sera pas aussi vigoureuse dans toutes les directions. Et tout naturellement que l'ondulation sera plus puissante dans la direction de vibration de la particule qui l'a initiée. La Figure I de la planche fournie à la fin du texte illustre bien cette situation : autour du point vibrant A se dessinent les surfaces concentriques de l'onde sphérique, et l'épaisseur irrégulière des deux premières surfaces d'ondes schématise manifestement un excédent de vigueur de l'ondulation dans la direction verticale sur la Figure (qui est la direction de propagation de l'onde) et la faiblesse de l'ondulation dans la direction orthogonale à la propagation. Cette assertion est fort utile à Young pour justifier de la propagation en ligne droite de la lumière. Enfin dans ce schéma mécaniste, on voit clairement se dessiner la nature nécessairement *longitudinale* attribuée par Young à l'onde lumineuse.

A présent, afin qu'une telle ondulation puisse continuer sa progression jusqu'à quelque distance considérable que ce soit, il doit y avoir en chaque partie de celle-ci une tendance à préserver son propre mouvement en ligne droite depuis {P.24} le centre ; car si l'excès de force en quelque partie était communiqué aux particules voisines, il ne peut y avoir de raison pour qu'il ne soit pas très rapidement égalisé tout du long ou, en d'autres termes, ne se retrouve totalement éteint, puisque les mouvements en directions contraires se détruiraient naturellement l'un l'autre. La production d'un son par la vibration d'une corde est manifestement de cette nature ; au contraire, dans une onde circulaire d'eau, chaque partie est au même instant soit élevée soit abaissée. Il peut être difficile de montrer mathématiquement le mode selon lequel cette inégalité de force est préservée ; mais l'inférence depuis les faits parait inévitable ; et alors que la science de l'hydrodynamique est si imparfaite que nous ne pouvons pas même résoudre le simple problème du temps requis pour vider un vase par une ouverture donnée, on ne peut pas s'attendre à ce que l'on soit capable de rendre compte parfaitement d'une série de phénomènes si compliquée, que ceux des fluides élastiques. La théorie de HUYGENS[106] explique effectivement la circonstance de manière tolérablement satisfaisante : il suppose que chaque particule du milieu propage une ondulation distincte dans toutes les directions ; et que l'effet général n'est perceptible que là où une portion de chaque ondulation s'accorde en direction au même instant ; et il est aisé de montrer qu'une telle ondulation générale se propagerait dans tous les cas rectilignement, avec une force proportionnelle ; mais, selon cette proposition, il semble suivre qu'une plus grande quantité de force doit être perdue par la divergence des ondulations partielles que ce qui parait être cohérent avec la propagation de l'effet à quelque distance considérable. Cependant il est évident que l'on doive naturellement s'attendre à ce qu'une telle limitation du mouvement ait lieu ; car si l'intensité du mouvement de quelque partie particulière, au lieu de continuer à être propagé droit vers l'avant, était supposée affecter l'intensité d'une partie de l'ondulation voisine, une {P.25} impulsion devrait alors avoir voyagé depuis un cercle interne vers un externe en direction oblique, en même temps qu'en direction radiale, et conséquemment avec une vélocité plus importante ; ce qui est contre la première proposition. Dans le cas de l'eau, la vélocité n'est en aucun cas si rigoureusement limitée que dans celui d'un milieu élastique. Pourtant il n'est pas nécessaire de supposer, non plus qu'il soit probable, qu'il n'y ait absolument pas la moindre communication latérale de la force de l'ondulation, plutôt que, dans les milieux hautement élastiques, cette communication soit pratiquement imperceptible. Dans l'air, si une corde est parfaitement isolée, de manière à propager exactement les ondulations qui ont été décrites, elles seront de fait beaucoup moins fortes que si la corde était placée au voisinage d'une table d'harmonie, et probablement dans quelque mesure à cause de cette communication latérale de mouvements de tendance opposée. Et la différence d'intensité des différentes parties de la même ondulation circulaire peut être observée, en tenant un diapason à branches commun à bout de bras alors qu'il vibre, et en le tournant d'un plan dirigé vers l'oreille à une position perpendiculaire à ce plan.

Proposition III

---

[106] Christian Huygens, *Traité de la Lumière,* Chapitre I : « Des rayons directement étendus », 1690.

*Une Portion d'Ondulation sphérique, admise par une Ouverture dans un Milieu au repos, va continuer à être propagée rectilignement en Surfaces concentriques, terminées latéralement par des Portions faibles et irrégulières d'Ondulations nouvellement divergentes.*[107]

Au moment de l'admission, il peut être supposé que la circonférence de chacune des ondulations génère une ondulation partielle emplissant l'angle naissant entre les rayons <des surfaces sphériques>[108] et la surface limitant le milieu[109] ; mais aucune augmentation sensible {P.26} de sa force ne sera provoquée par une divergence de mouvement de quelque autre partie de l'ondulation, par défaut d'une coïncidence dans le temps, comme il a déjà été expliqué à l'égard des différentes forces d'une ondulation sphérique. Si en effet l'ouverture ne transmet qu'une petite portion de l'étendue d'une ondulation[110], l'ondulation nouvellement générée doit absorber presque l'intégralité de la force de la portion transmise ; et ceci est le cas considéré par NEWTON dans les *Principia*[111]. Mais aucune expérience ne peut être réalisée dans ces circonstances avec la lumière, du fait de l'extrême petitesse de ses ondulations, et de l'interférence de l'inflexion[112] ; et cependant quelques très faibles radiations divergent effectivement au-delà des limites probables de l'inflexion, rendant le contour de l'ouverture distinctement visible dans toutes les directions ; celles-ci sont attribuées par NEWTON à une raison inconnue, distincte de l'inflexion ; (*Optics*, Livre Troisième, Obs. 5)[113] et elles répondent entièrement à la description de cette proposition.

---

[107] Par cette proposition Young déploie un peu plus son modèle de propagation ondulatoire de la lumière, tout en proposant une justification à l'objection corpusculaire classique selon laquelle une onde ne peut se propager en ligne droite. Ainsi, au passage d'une ouverture, l'onde sphérique se déploie effectivement vers les côtés et contourne l'obstacle à la manière d'une vague. Mais ces « *ondulations nouvellement divergentes* » sont « *faibles et irrégulières* », au point d'être à peine discernables. Par conséquent, en apparence au moins, la lumière « *va continuer à être propagée rectilignement* », conformément à l'observation commune.

[108] La langue anglaise présente cet avantage de lever l'ambiguïté entre les mots « ray », employé pour qualifier le *rayon lumineux* ; et « radius », désignant le *rayon du cercle* ou *de la sphère*. C'est donc par ces groupes nominaux que l'on traduira ces mots lorsqu'on les rencontrera. Exceptionnellement ici, les deux acceptions du mot *rayon* conviendraient pour décrire l'objet indiqué. Cependant, en écrivant ici « radii » plutôt que « rays », Young choisit d'évoquer les *rayons géométriques des surfaces d'ondes sphériques* se propageant depuis un point vibrant A (voir Fig. I, Planche I.) et s'inscrit ainsi d'emblée dans le formalisme ondulatoire. On remarque cependant que ces *rayons géométriques* indiquent précisément la direction de propagation de la lumière dans le système ondulatoire ; et qu'ils sont donc assimilables aux *rayons lumineux* de l'optique corpusculaire. Il est intéressant de voir combien le basculement d'un formalisme à l'autre peut se jouer sur le simple choix d'un terme plutôt qu'un autre dans la description même de l'expérience ; d'autant plus quand ces deux termes se trouvent être confondus dans la langue française.

[109] Young décrit ici l'apparition d'ondes sphériques nouvelles se développant dans les angles morts situés derrière l'écran de la Figure I, de part et d'autre des rayons lumineux ABG et AC limitant l'onde transmise normalement.

[110] Il est en effet fondamental pour cette justification de la propagation rectiligne de l'onde lumineuse que l'ouverture ne laisse passer qu'une « *petite portion de l'étendue d'une ondulation* ». Plus la taille de l'ouverture sera faible par rapport au rayon de l'onde sphérique incidente, plus la portion de l'onde transmise par l'ouverture sera assimilable au plan tangent à l'onde au centre de l'ouverture, et tendra donc vers une onde plane se propageant en ligne droite. Noter que Young emploie ici le mot « breadth » en faisant référence à la *largeur* du front d'onde, telle que schématisée par le segment [BC] sur la Figure I. A ne pas confondre avec la *longueur d'onde*, qualifiée plus tard par le même terme « breadth » dans la Proposition VIII, Corollaires I et II.

[111] Ici, Young fait notamment référence à la Proposition 42 du Livre II des *Principia*, qui sera citée plus bas.

[112] Le mot « inflexion » désigne chez Newton une partie des phénomènes que nous réunissons aujourd'hui sous le terme de *diffraction* ; donc des phénomènes de déviation de la lumière impactant les objets de petite taille ou rasant les corps opaques. Newton justifie ce phénomène de déviation par l'action d'une force exercée sur les corpuscules de lumière par les corpuscules situés à la surface du corps opaque, et sensible uniquement à très courte distance de celui-ci. Par ailleurs, il ne faut pas se méprendre ici sur le sens du mot « interférence » qui indique seulement que le phénomène décrit risque d'être masqué par l'inflexion qui se produit conjointement, comme développé dans le passage de Isaac Newton, *Opticks*, Livre III, Observation 5, mentionné plus loin.

Soient les lignes concentriques de la Fig. 1 (Planche I.) représentant la position simultanée de parties similaires d'un nombre d'ondulations successives divergeant depuis le point A ; elles représenteront aussi les positions successives de chaque ondulation individuelle : soit la force de chaque ondulation représentée par l'épaisseur de la ligne, et soit le cône de lumière ABC admis par l'ouverture BC ; alors les ondulations principales se propageront en direction rectiligne vers GH, et les très faibles radiations de chaque côté divergeront avec B et C pour centres, sans recevoir quelque force additionnelle que ce soit d'un quelconque point intermédiaire D de l'ondulation pour cause de l'inégalité des lignes DE et DF. Mais, si l'on admet une légère divergence latérale depuis les extrémités des ondulations, celle-ci doit diminuer leur force sans augmenter matériellement celle de la lumière dissipée ; et leur {P.27} extrémité, au lieu de la ligne droite BG, prendra la forme CH ; car la déperdition de force doit être plus considérable à proximité de C qu'à de plus grandes distances. Cette ligne correspond à la limite de l'ombre dans la première observation de NEWTON[114], Fig. 1 ; et il est beaucoup plus probable qu'une telle dissipation de lumière ait été la cause de l'augmentation de l'ombre dans cette observation, plutôt que celle-ci ait été due à l'action de l'atmosphère infléchissante, qui aurait dû s'étendre jusqu'à un trentième de pouce[115] dans chaque direction afin de la produire ; spécialement lorsque l'on considère que la lumière n'était pas diminuée en entourant le cheveu d'un milieu plus dense que l'air, qui selon toute probabilité devrait avoir affaibli et contracté son atmosphère infléchissante. En d'autres circonstances, la divergence latérale pourrait bien paraître augmenter, plutôt que diminuer, la largeur du faisceau.

---

[113] Isaac Newton, *Opticks*, Livre III, Observation 5, 1730, 300. Le passage en question mentionne une expérience d'inflexion de la lumière du soleil par la lame d'un couteau, ou par un cheveu. Expérience au cours de laquelle Newton remarque une brillance particulière du fil de la lame, très faible mais notable quelle que soit la position de son œil. Or il attribue cette brillance à une cause différente de celle produisant les effets d'inflexion, eux-mêmes visibles dans certaines directions seulement. Young tente donc de donner ici du poids à l'idée que l'onde lumineuse - comme la vague - s'ouvre après passage par une ouverture, en se référant à une observation de Newton que ce dernier n'explique pas et différencie de sa théorie de l'inflexion. L'emprunt est audacieux, tant Newton a vigoureusement nié, et certainement jamais observé, le fait que la lumière puisse se courber en direction de l'ombre des objets (Robert H. Stuewer, *A Critical Analysis of Newton's Work on Diffraction*, Isis 61, n° 2, 1970, 188-205). Et l'on remarque en effet que dans l'expérience schématisée sur la Fig. I. empruntée à Newton, la lumière est déviée dans le sens opposé à celui de la déviation décrite ici par Young et schématisée sur les Figures 1 et 4.

[114] Young évoque ici l'Observation I du Livre III de l'*Opticks* de Newton et la Figure 1 qui lui est associée :

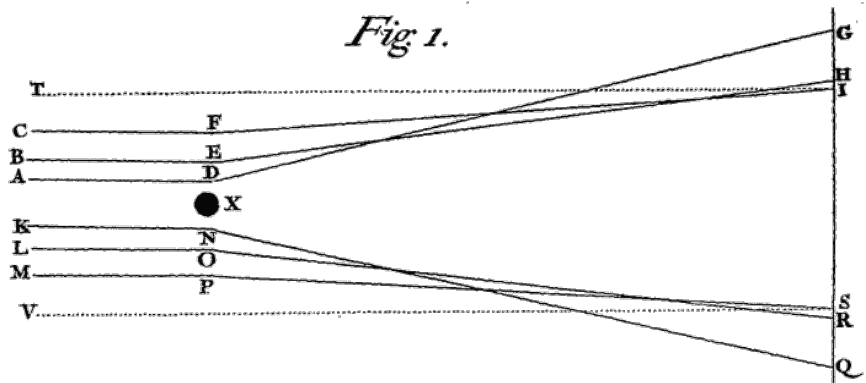

[115] 1/30 * 2,54 cm soit environ 850 µm.

Comme on a toujours estimé que le sujet de cette proposition était la partie la plus difficile du système ondulatoire, il sera approprié d'examiner ici les objections que Newton a établies contre lui.

« Selon moi, la supposition fondamentale elle-même semble impossible ; A savoir, que les ondes ou vibrations de quelque fluide puissent, comme les rayons de lumière, être propagées en lignes droites, sans un étalement et une incurvation continus, les faisant déborder largement de leurs limites dans toutes les directions à l'intérieur du milieu au repos, là où elles sont limitées par celui-ci. Je me trompe, s'il n'y a pas à la fois l'expérience et la démonstration du contraire. » (*Phil. Trans.* VII. 5089, *Abr*. I. 146. Nov. 1672.)[116]

« Tout mouvement propagé dans un fluide diverge de la transmission rectiligne dans les espaces immobiles. »[117]

« Puisque le milieu ici, » au centre d'une ondulation {P.28} qui se propage « est plus dense, que dans les espaces d'un côté et de l'autre, il se dilatera tant en direction de ces espaces qu'en direction des intervalles raréfiés entre les impulsions ; et ainsi […] les impulsions se dilateront à la même vitesse *environ* d'un côté et de l'autre dans les parties du milieu qui sont au repos ; […] et ainsi occuperont tout l'espace. […] Comme nous l'expérimentons pour les sons. »[118] (*Princ.* Livre II. Prop. 42.)[119]

« Toutes les hypothèses dans lesquelles la lumière est supposée consister en une pression ou un mouvement propagé à travers un milieu fluide ne sont-elles pas erronées ? […] Si elle consistait en une pression ou un mouvement se propageant soit en un instant, soit dans le temps, elle s'infléchirait vers l'intérieur de l'ombre. Car la pression ou le mouvement ne peuvent pas être propagés dans un fluide en lignes droites au-delà d'un obstacle qui interrompt une partie du mouvement, mais s'infléchiront et se répandront dans toutes les directions dans le milieu au repos situé au-delà de l'obstacle. […] Les ondes à la surface de l'eau stagnante, passant sur les côtés d'un large obstacle qui les interrompt en partie, se courbent ensuite et se dilatent graduellement dans l'eau calme derrière l'obstacle. Les ondes, impulsions ou vibrations de l'air, dont le son est composé, se courbent manifestement, bien que pas autant que les ondes de l'eau. Car une cloche ou un canon peuvent être entendus au-delà d'une colline qui intercepte la vue du corps sonore ; et les sons sont propagés aussi promptement par les tubes tortueux que par les droits. Mais autant que l'on sache, la lumière ne suit jamais de passages tortueux, ni ne se courbe vers l'intérieur de l'ombre. Car les étoiles fixes, par interposition de n'importe laquelle des planètes, cessent d'être vues. Tout comme les parties du soleil par l'interposition de la lune, de Mercure ou de Venus. Les rayons qui passent très près des bords de quelque corps que ce soit sont un peu courbés par l'action du corps […] ; mais cette courbure ne se fait pas vers l'ombre, mais depuis celle-ci {P.29} et n'est produite qu'au passage du rayon à proximité du

---

[116] Isaac Newton, *Mr. Isaac Newton's Answer to Some Considerations upon His Doctrine of Light and Colors; Which Doctrine Was Printed in Numb. 80. of These Tracts,* Philosophical Transactions of the Royal Society of London, vol. 7, 1672, 5084-5103. Le même texte est réédité dans Philosophical Transactions Abriged 2, 1672, 13-29.

[117] Titre de la proposition, traduit du latin : « Motus omnis per fluidum propagatus divergit a recto transmite in spatia immota. »

[118] Traduit du latin : « Quoniam medium ibi, densius est, quam in spatiis hinc inde, dilatabit ses tam versus spatia utrinque sita, quam versus pulsuum rariora intervalla ; eoque pacto – pulsus eadem *fere* celeritate ses in medii partes quiescentes hinc inde relaxare debent ; – ideoque spatium totum occupabunt. – Hoc experimur in sonis. »

[119] Isaac Newton, *Philosophiae Naturalis Principia Mathematica*, Livre II, Proposition 42, Théorème 23.

corps, et à très petite distance de celui-ci. Aussitôt que le rayon a passé le corps, il continue tout droit. » (*Optics*, Qu. 28)[120]

Voyez que la proposition citée des *Principia* ne contredit pas directement cette proposition[121] ; car elle ne fait pas l'assertion qu'un tel mouvement doit diverger également[122] dans toutes les directions ; non plus qu'il ne peut être maintenu en vérité, que les parties d'un milieu élastique communicant un mouvement quelconque doivent propager ce mouvement également dans toutes les directions. (*Phil. Trans. for 1800*. p. 109-112.)[123] Tout ce que l'on peut inférer par le raisonnement est que les parties marginales de l'ondulation doivent être de quelque façon affaiblies, et qu'il doit y avoir une très légère divergence dans chaque direction ; mais le fait que l'un ou l'autre de ces effets puisse être d'ampleur suffisante pour être sensible ne pourrait avoir été inféré par un raisonnement si une réponse affirmative n'avait été rendue probable par l'expérience.

Quant à l'analogie avec d'autres fluides, l'inférence la plus naturelle de celle-ci est que : « Les ondes de l'air, dont le son est composé, se courbent manifestement, bien que pas autant que les ondes de l'eau ; » l'eau étant un milieu non-élastique, et l'air modérément élastique ; mais l'éther étant des plus hautement élastique, ses ondes se courbent beaucoup moins que celles de l'air, et donc presque imperceptiblement. Les sons sont propagés dans les passages tortueux car leurs côtés sont capables de réfléchir le son, tout comme la lumière serait propagée dans un tube courbé, s'il était parfaitement poli à l'intérieur[124].

La lumière d'une étoile est bien trop faible pour produire, par sa très faible divergence, quelque illumination visible à la marge d'une planète l'éclipsant ; et l'interception de la lumière du soleil par la lune est aussi étrangère à cette question que la déclaration à propos de l'inflexion est inexacte.

A l'argument fourni par Huygens en faveur de la {P.30} propagation rectilinéaire des ondulations, Newton n'a pas fait de réponse ; peut-être à cause de l'idée fausse qu'il se faisait de la nature des mouvements des milieux élastiques, comme dépendant d'une loi particulière de la vibration qui a été corrigée par les mathématiciens ultérieurs. (*Phil. Trans. for 1800*, p. 116)[125] En fin de compte, il est présumé que cette proposition peut être admise en toute tranquillité comme étant parfaitement consistante avec l'analogie et avec l'expérience.

Proposition IV
*Lorsqu'une Ondulation atteint une Surface qui est la Limite entre des Milieux de différentes Densités, une Réflexion partielle a lieu, proportionnelle en Force à la Différence des Densités.*[126]

---

[120] Isaac Newton, *Opticks*, Livre III, Question 28, 1730, 336.
[121] Comprendre : « la proposition citée des *Principia* ne contredit pas directement *la Proposition III* de Young. »
[122] Comprendre : « uniformément »
[123] Thomas Young, *Outlines of Experiments and Inquiries respecting Sound and Light*, Philosophical Transactions of the Royal Society of London, vol. 90, 1800, 106-150.
[124] De fait, les fibres optiques les plus simples, dites à saut d'indice, fonctionnent aujourd'hui sur le principe d'une réflexion totale de la lumière à l'interface entre leur cœur et leur gaine, permettant un guidage de la lumière même le long de chemins tortueux.
[125] Thomas Young, *Outlines of Experiments and Inquiries respecting Sound and Light*, Philosophical Transactions of the Royal Society of London, vol. 90, 1800, 106-150.
[126] Cette proposition envisage une explication mélangeant le modèle ondulatoire de la lumière et la théorie des chocs pour justifier de l'existence d'une réflexion partielle toujours plus ou moins prononcée de la lumière à l'interface entre deux milieux. L'enjeu est de taille, car Young affirme dans son article précédent (Thomas

Ceci peut être illustré, sinon démontré, par l'analogie des corps élastiques de différentes tailles. « Si un corps élastique plus petit heurte un plus grand, il est bien connu que le plus petit est réfléchi plus ou moins puissamment, en fonction de la différence de leurs tailles respectives : ainsi, il y a toujours une réflexion lorsque les rayons de lumière passent d'une strate d'éther plus rare à une plus dense ; et fréquemment un écho lorsqu'un son heurte un nuage. Un plus grand corps heurtant un plus petit le propulse, sans perdre tout son mouvement : ainsi, les particules d'une strate d'éther plus dense ne transmettent-elles pas l'intégralité de leur mouvement à une plus rare mais, dans leur effort à aller de l'avant, elles sont rappelées par l'attraction de la surface réfringente avec une force égale ; et par conséquent une réflexion est toujours produite dans un second temps lorsque les rayons de lumière passent d'une strate plus dense à une plus rare. » (*Phil. Trans. for 1800*. p. 127.)[127] Mais il n'est pas absolument nécessaire de supposer une attraction dans ce dernier cas, puisque l'effort à aller de l'avant serait propagé vers l'arrière sans celle-ci, et l'ondulation serait inversée, une raréfaction {P.31} revenant à la place d'une condensation ; et l'on trouvera peut-être ceci plus consistant avec les phénomènes.

Proposition V
*Lorsqu'une Ondulation est transmise par une Surface qui est la limite entre des Milieux différents, elle poursuit dans une Direction telle que les Sinus[128] des Angles d'Incidence et de Réfraction sont dans le Rapport constant de la Vélocité de Propagation dans les deux Milieux.*[129]

---

Young, *Outlines of Experiments and Inquiries respecting Sound and Light*, *Ibid.*, 125-126) que l'inaptitude « *totale* » de la théorie newtonienne à expliquer la réflexion partielle est la deuxième « *difficulté* » pour laquelle il faut se tourner vers le système ondulatoire.

[127] Thomas Young, *Outlines of Experiments and Inquiries respecting Sound and Light, Ibid*, 127.

[128] Les définitions du sinus et du cosinus d'un angle employées par Young diffèrent légèrement de celles qui sont devenues les nôtres. Il les rappelle dans la partie intitulée « *Eléments mathématiques de philosophie naturelle* » de Thomas Young, *A Course of Lectures in Natural Philosophy and Mechanical Arts*, vol. II, London, 1807, 15. Ces définitions sont illustrées par la figure suivante :

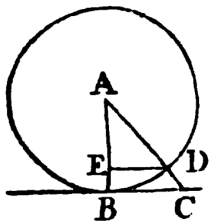

Et sont formulées de la sorte « *DE perpendiculaire à AB est le sinus de BD ou BAD* » et « *AE est le cosinus de BD ou BAD* ». Les mots « sinus » et « cosinus » employés par Young dans ses textes correspondent donc à des longueurs, conformément à leur acception première de « longueur de la demi-corde DE de l'arc double de BD » et de « portion du rayon AB coupée par la demi-corde ». S'il est simple à partir de cette définition de remonter à notre sinus moderne de norme 1 (qui n'est que le rapport du sinus utilisé par Young par le rayon du cercle considéré AD) et si le choix de la définition du sinus n'a aucun impact sur la loi de la réfraction exprimée ici comme un rapport, la confusion entre ces définitions pourrait mener à des erreurs d'interprétation dans la suite du texte. Ainsi, afin de s'assurer de la vigilance du lecteur sur ce point, nous écrirons systématiquement « Sinus » et « Cosinus » avec une majuscule lorsqu'ils désignent des longueurs de segments. Young, les écrivant évidemment en lettres minuscules dans le texte original.

[129] La référence à la bien connue loi de la réfraction et à la justification de sa direction par le changement de vitesse de la lumière est moins anodine qu'il n'y parait. Car si cette loi - stipulant que « *les Sinus des angles d'incidence et de réfraction sont dans [un] rapport constant* » - et que la valeur numérique de ce rapport sont reconnus par tous à l'époque, la justification théorique de la valeur de ce rapport diffère d'un auteur à l'autre.

(BARROW, *Lect. Opt.* II. P. 4.[130] HUYGENS, *de la Lum.* cap. 3.[131] EULER, *Conj. Phys.*[132] *Phil. Trans. for 1800*, p. 128.[133] *Le programme de YOUNG*. Art. 382.[134])

*Corollaire 1.* La même démonstration prouve l'égalité des angles de réflexion et d'incidence.

*Corollaire 2.* Il apparaît d'après les expériences sur la réfraction de l'air condensé que le rapport de la différence des Sinus varie simplement comme la densité.[135] D'où il suit, d'après la Schol. I. Prop. I. que l'excès de densité du milieu éthéré est dans le rapport double de la densité de l'air[136] ; chaque particule collaborant avec ses voisines pour attirer une plus grande portion de celui-ci[137].

Proposition VI

---

En particulier, le choix d'un modèle corpusculaire de la lumière impose, comme conséquence de l'analogie mécaniste qu'il charrie, d'identifier le rapport des Sinus au rapport inverse des vitesses dans les milieux, soit $\frac{\sin i_1}{\sin i_2} = \frac{v_2}{v_1}$. Ce qui est équivalent à soutenir que la propagation de la lumière est plus rapide, ou plus facile, dans un milieu plus dense. C'est ce que l'on trouve chez Descartes (*La Dioptrique*, premier discours, vol. VI, 1637), ou chez Newton (*Opticks*, Livre I, Proposition VI, Théorème VI, Expérience XV, 1730). Cependant, la justification de la loi de la réfraction dans le modèle ondulatoire (voir Christian Huygens, *Traité de la Lumière,* Chapitre 3 : « De la réfraction », 1690) ou par le principe de moindre temps (voir Pierre de Fermat, *Synthèse pour les réfractions*, 1662) repose pour sa part sur l'hypothèse d'un ralentissement de la lumière au passage du milieu moins dense au plus dense. Qui se traduit alors nécessairement par une loi de la forme $\frac{\sin i_1}{\sin i_2} = \frac{v_1}{v_2}$ , identique à celle proposée par Young ci-dessus. La mesure de la vitesse de la lumière dans des milieux différents n'ayant pas été réalisée avant 1850 (Léon Foucault, *Thèse de physique. Sur les vitesses relatives de la lumière dans l'air et dans l'eau, présentée à la Faculté des sciences de Paris, pour obtenir le grade de docteur ès-sciences physiques*, Bachelier, Paris, 1853), il est impossible de trancher par l'expérience entre ces formules à l'époque. Ainsi, le choix de l'une ou de l'autre révèle de fait (ou impose en retour) un choix théorique entre le système ondulatoire ou corpusculaire. La formulation de la loi de réfraction dans les termes proposés ici par Young traduit donc discrètement, mais sans ambiguïté possible, le choix d'une description ondulatoire de la lumière. Pour s'en convaincre on peut d'ailleurs la comparer à la forme beaucoup plus neutre que Young propose de cette loi dans sa conférence bakerienne de l'année précédente (*On the Mechanism of the Eye*, Proposition I, Philosophical Transactions of the Royal Society of London, vol. 91, 1801, 27) : « *Dans toute réfraction, le rapport du Sinus de l'angle d'incidence au Sinus de l'angle de réfraction est constant.* »

[130] Isaac Barrow, *Lectiones Opticae*, Partie II, 1669, 4.

[131] Christian Huygens, *Traité de la Lumière,* Chapitre 3 : « De la réfraction », 1690.

[132] Leonhard Euler, *Conjectura Physica Circa Propagationem Soni ac Luminis*, Berlin, 1750.

[133] Thomas Young, *Outlines of Experiments and Inquiries respecting Sound and Light,* Philosophical Transactions of the Royal Society of London, vol. 90, 1800, 106-150.

[134] Thomas Young, *Syllabus of a course of Lectures on Natural and Experimental Philosophy*, Royal Institution, London, 1802, Art. 382, 116.

[135] Après l'abandon de l'Hypothèse IV d'existence d'une atmosphère éthérée de densité décroissante entourant les corps matériels, Young réédite une version de *la Théorie de la Lumière et des Couleurs* dans laquelle ce Corollaire se retrouve sensiblement modifié (Thomas Young, *A Course of Lectures in Natural Philosophy and Mechanical Arts*, vol. II, London, 1807, 623). D'une part il perd son statut de corollaire pour devenir une scholie. Ensuite il ne parle plus de « *rapport de la différence des Sinus* » mais seulement de « *différence des Sinus* ». Enfin, la suite du corollaire disparait pour être remplacée par la phrase : « *Et la même chose est probablement vraie dans d'autres cas similaires* », justifiant une généralisation du cas de l'air condensé aux autres matériaux.

[136] L'excès de densité d'éther dans le milieu est proportionnel au carré de la densité de l'air. Il ne faudra donc pas dans la suite confondre la « densité de l'air », avec la « densité d'éther », même si ces deux grandeurs sont liées.

[137] Sous-entendu « d'éther ».

*Lorsqu'une Ondulation tombe à la Surface d'un Milieu plus rare si obliquement qu'elle ne peut être réfractée régulièrement, elle est totalement réfléchie, à un Angle égal à celui de son Incidence.*

(*Phil. Trans. for 1800*, p. 128.)[138]

*Corollaire*. Ce phénomène tend à prouver l'augmentation et la diminution graduelles de densité à la surface séparant deux milieux, comme supposé dans l'Hypothèse IV[139] ; bien que Huygens ait tenté de l'expliquer un peu différemment.

{P.32} Proposition VII
*Si l'on suppose des Ondulations équidistantes passant à travers un Milieu dont les Parties sont susceptibles de Vibrations permanentes quelque peu plus lentes que les Ondulations, leur Vélocité sera quelque peu diminuée par cette Tendance vibratoire ; et, dans le même Milieu, d'autant plus que les Ondulations seront plus fréquentes.*[140]

Car, aussi souvent que l'état de l'ondulation requiert un changement dans le mouvement réel de la particule qui la transmet, ce changement sera retardé par la propension de la particule à continuer son mouvement un peu plus longtemps : et ce retard sera d'autant plus fréquent et plus considérable que la différence entre les périodes de l'ondulation et la vibration naturelle sera plus grande.

*Corollaire*. Ce fut longtemps une opinion établie, que la chaleur consiste en des vibrations des particules des corps, et qu'elle est capable d'être transmise par des ondulations à travers un vide apparent. (Newt. *Opt.* Qu. 18.)[141] Dernièrement cette opinion a été largement abandonnée[142]. Le Comte Rumford[143], le Professeur Pictet[144], et M. Davy[145] sont à peu près les seuls auteurs qui ont paru l'approuver ; mais elle semble avoir été rejetée sans aucune bonne raison, et recouvrera probablement très bientôt sa popularité.

---

[138] Thomas Young, *Outlines of Experiments and Inquiries respecting Sound and Light*, Philosophical Transactions of the Royal Society of London, vol. 90, 1800, 106-150.

[139] On notera que l'Hypothèse IV, ne dit pas explicitement si la densité d'éther change abruptement ou continûment à l'interface entre deux milieux. La Proposition VI nous permet donc d'éclaircir le contenu de l'Hypothèse IV et de comprendre mieux ce en quoi cette hypothèse est utile à la théorie optique de Young, notamment pour justifier le Corollaire V de la Proposition VIII sur les couleurs par inflexion.

[140] Cette proposition est une tentative originale d'interprétation par la théorie ondulatoire de la dispersion des différentes composantes colorées de la lumière blanche. Puisqu'il est acquis depuis Newton que les rayons lumineux associés aux différentes couleurs du spectre émergent du prisme dans des directions différentes, il est aussi acquis (et confirmé par la Proposition V ci-dessus) que les composantes colorées de la lumière se propagent à des vitesses différentes dans le prisme. C'est donc ces différences de vitesse, que peuvent présenter parfois les ondes de fréquences différentes, que Young propose d'expliquer ici par la propension des particules de certains milieux à vibrer naturellement, affectant alors la vitesse de propagation des ondes différemment selon la proximité de la fréquence de l'onde avec la fréquence de vibration naturelle des particules du corps.

[141] Isaac Newton, *Opticks*, Livre III, Question 18, 1730, 323.

[142] A l'époque où Young écrit, la nature de la chaleur est encore vivement débattue. Et l'hypothèse selon laquelle il s'agirait d'un fluide impondérable, nommé calorique, est très majoritaire en Angleterre et en France.

[143] Benjamin Thomson, Comte Rumford (1753-1814) est un physicien américain qui fonda en 1799 la Royal Institution à Londres. Rumford observe que le travail exercé par les chevaux activant les machines de forage des canons qu'il supervise, produit une intense chaleur, et en développe l'idée que travail et chaleur sont deux manifestations équivalentes d'une même chose appelée énergie. Il propose alors une interprétation des travaux de Pictet sur le rayonnement thermique en termes d'ondulations de l'éther produites par les corps solides échauffés.

[144] Marc-Auguste Pictet (1752-1827) est un physicien, astronome et météorologue suisse qui publie notamment l'ouvrage *Essais de Physique, tome 1 : Essai sur le feu,* 1790, dont il sera question un peu plus loin dans le texte et dans lequel il développe une théorie du rayonnement thermique.

Supposons que ces vibrations sont moins fréquentes[146] que celles de la lumière ; tous les corps par conséquent sont susceptibles de vibrations plus lentes que celles de la lumière ; et de fait presque tous sont susceptibles de vibrations lumineuses, soit lorsqu'ils sont dans un état d'inflammation, soit dans le cas des phosphores solaires[147] ; mais beaucoup moins facilement, et à degré bien moindre, qu'elles ne sont susceptibles de vibrations de chaleur. Il suivra de ces propositions, que les ondulations lumineuses les plus fréquentes[148] seront plus retardées que les moins fréquentes ; et {P.33} par conséquent, que la lumière bleue sera plus réfrangible que la rouge, et rayonnera la chaleur la plus faible de toutes ; une conséquence qui coïncide exactement avec les expériences hautement intéressantes du Dr. HERSCHEL (*Phil. Trans. for 1800*. p. 284.)[149] Il peut aussi être facilement conçu que l'existence même d'un état de vibration plus lente peut avoir tendance à retarder encore plus les vibrations les plus fréquentes, et que le pouvoir réfringent des corps solides peut être sensiblement augmenté par une augmentation de température, comme cela semble en effet avoir été le cas dans les expériences d'EULER (*Acad. de Berlin.* 1762. p. 328.)[150]

*Scholie*. Si néanmoins cette proposition devait sembler être insuffisamment démontrée, il doit être concédé qu'elle est au moins autant explicative des phénomènes que toute autre chose qui peut être avancée par l'autre bord, depuis la doctrine des projectiles ; puisqu'une supposée force accélératrice doit agir dans une proportion différente de celle du nombre de particules ; et, si nous appelons ceci une attraction élective, ce n'est que voiler sous un terme chimique notre incapacité à assigner une cause mécanique. M. SHORT, lorsqu'il découvrit par l'observation l'égalité de la vitesse de la lumière de toutes les couleurs, prit si fortement conscience de l'objection qu'il en tira immédiatement une inférence en faveur du système ondulatoire[151]. Il est supposé dans la proposition que lorsque la lumière est

---

[145] Humphry Davy (1778-1829) est un physicien et chimiste anglais, nommé professeur de Chimie à la Royal Institution juste avant que Young y soit recruté. Parmi ses nombreux travaux, l'*Essay On Heat, Light, and the Combinations of Light*, publié en 1799, rejette la théorie du calorique.

[146] Comprendre : « ont des fréquences *plus basses* ».

[147] Soit des corps qui, comme le phosphore, sont capables de s'enflammer ou d'émettre une lumière quand ils sont exposés à l'air.

[148] Comprendre : « de plus *hautes fréquences* ».

[149] William Herschel, *Experiments on the refrangibility of the invisible rays of the sun*, Philosophical Transactions of the Royal Society of London, vol. 90, 1800, 284-292. Herschel (1738-1822) observe que lorsque la lumière blanche est dispersée par un prisme, un thermomètre placé au-delà de la partie rouge du spectre (dans la partie que nous associons aujourd'hui au proche infra-rouge) mesure une élévation de température. Young trouve ici l'occasion d'établir un pont entre la théorie vibratoire de la chaleur et sa théorie ondulatoire de la lumière, qu'il utilise comme justification analogique de son système.

[150] Johann Albrecht Euler (1734-1800) est le premier fils de Leonhard Euler. C'est certainement à son mémoire, *De l'influence de la chaleur sur la réfraction des fluides*, Histoire de l'académie royale des sciences et belles-lettres de Berlin, 1762, 328, qu'il est ici fait référence.

[151] James Short (1710-1768) est un astronome et constructeur de télescope faisant à son époque référence à la Royal Society. En réaction aux propositions de Thomas Melvill (*Discourse concerning the cause of the different refrangibility of the rays of light*, Philosophical Transactions of the Royal Society of London, 48, 1753, 261-268) selon qui la dispersion serait due au fait que les particules de lumière de couleurs différentes possèdent différentes vitesses (contrairement à Newton qui leur attribuait différentes tailles), la Royal Society charge James Short d'observations astronomiques visant à vérifier sa théorie. Les observations et conclusions de Short sont publiées à la suite de l'article de Melvill (268-270). Elles relatent l'observation répétée du premier satellite de Jupiter qui, si les particules de lumière rouge allaient plus vite que les autres comme le suppose Melvill, devrait apparaître rouge lors de son émersion de derrière la planète. Short, ne notant pas la moindre altération de la couleur du satellite, conclue à l'invalidation de la théorie, et donc à l'égalité de vitesse des différents

dispersée par réfraction, les corpuscules de la substance réfringente sont véritablement dans un état de mouvement alterné et contribuent à sa transmission ; mais il doit être confessé que nous ne pouvons pour le moment formuler une conception très ferme et précise des forces concernées par le maintien de ces vibrations corpusculaires.

{P.34} Proposition VIII
*Lorsque deux Ondulations, d'Origines différentes, coïncident parfaitement ou presque en Direction, leur effet commun est une Combinaison des Mouvements appartenant à chacune.*[152]

Puisque chaque particule du milieu est affectée par chaque ondulation, partout où les directions coïncident, les ondulations ne peuvent procéder autrement qu'en unissant leurs mouvements, de sorte que le mouvement commun soit la somme ou la différence des mouvements séparés selon que des parties semblables ou dissemblables des ondulations sont coïncidentes.

J'ai insisté, à une précédente occasion, sur l'application de ce principe aux harmoniques ; (*Phil. Trans.*, 1800, p. 130.)[153] et il s'avérera d'une plus vaste utilité encore dans l'explication des phénomènes des couleurs. Les ondulations qui vont maintenant être comparées sont celles de fréquence égale. Lorsque les deux séries coïncident exactement en un point du temps, il est évident que la vitesse unifiée des mouvements particuliers doit être la plus grande et, dans son effet au moins, doubler les vitesses séparées ; et aussi qu'elle doit être la plus faible, et si les ondulations sont de force égale totalement détruite, lorsque l'instant du plus grand mouvement direct appartenant à l'une des ondulations coïncide avec celui du plus grand mouvement rétrograde de l'autre. Dans les états intermédiaires, l'ondulation unifiée sera de force intermédiaire ; mais les lois selon lesquelles cette force intermédiaire doit varier ne peuvent être déterminées sans données supplémentaires. Il est bien connu qu'une cause similaire produit pour le son cet effet que l'on appelle un battement ; deux séries d'ondulations d'amplitudes environ égales coopérant ou se détruisant alternativement, selon qu'elles coïncident {P.35} plus ou moins parfaitement dans les temps d'exécution de leurs mouvements respectifs.

---

lumières colorées dans l'espace - à condition de les considérer comme des projectiles. Car dans la conclusion de sa lettre, James Short, s'autorise en effet à envisager et à justifier la manière dont les différentes lumières colorées pourraient avoir des vitesses et pourtant parcourir la distance de Jupiter à la Terre en un même temps à condition de les assimiler à des vibrations d'un fluide élastique (en se référant aux propositions de Gowin Knight publiées dans *An attempt to demonstrate that all paenomena in nature may be explained by two simple active principles : attraction and repulsion*). Et c'est donc sur l'évocation d'un dilemme entre ces deux possibilités, plutôt qu'en tranchant en faveur du système ondulatoire, que Short conclue son compte rendu.

[152] Cette Proposition VIII est la première formulation de ce qu'on appellera plus tard le Principe d'Interférence des ondes lumineuses, un mécanisme similaire ayant été décrit par Young lui-même dans le cas des ondes sonores (Thomas Young, *Outlines of Experiments and Inquiries respecting Sound and Light*, Philosophical Transactions of the Royal Society of London, vol. 90, 1800, 130). Ce principe d'interférence des ondes lumineuses gagnera par ailleurs en précision et en généralité avec la publication suivante de Young : *An account of some Cases of the production of Colours not hitherto described*, Philosophical Transactions of the Royal Society of London, 1802, 387 : « *Partout où deux portions de la même lumière arrivent à l'œil par différents chemins, exactement, ou presque, dans la même direction, la lumière devient la plus intense lorsque la différence des chemins est un multiple quelconque d'une certaine longueur, et la moins intense dans l'état intermédiaire des portions qui interfèrent ; et cette longueur est différente pour des lumières de longueurs différentes.* »

[153] Thomas Young, *Outlines of Experiments and Inquiries respecting Sound and Light*, Philosophical Transactions of the Royal Society of London, vol. 90, 1800, 106-150.

Corollaire I. *Des Couleurs des Surfaces striées.*

BOYLE semble avoir été le premier à observer les couleurs des rayures sur les surfaces polies[154]. NEWTON ne les a pas remarquées. MAZEAS[155] et M. BROUGHMAN[156] ont réalisé quelques expériences sur le sujet, sans en tirer cependant de conclusion satisfaisante. Mais toutes les variétés de ces couleurs sont très facilement déduites de cette proposition[157].

Soient dans un plan donné deux points réfléchissants très proches l'un de l'autre, et soit le plan situé de telle manière que l'image réfléchie d'un objet lumineux vue dans celui-ci puisse paraître coïncider avec les points ; alors il est évident que la longueur des rayons incident et réfléchi pris ensemble est égale par rapport aux deux points, en les considérant capables de réfléchir dans toutes les directions. Abaissons maintenant l'un des points en-dessous du plan donné ; alors le chemin complet de la lumière réfléchie par lui, sera allongé d'une ligne qui est à l'abaissement du point comme deux fois le rapport du Cosinus d'incidence au rayon <du cercle>. Fig. 2.[158]

Si par conséquent, des ondulations égales de dimensions données sont réfléchies depuis deux points situés suffisamment proches l'un de l'autre pour paraître à l'œil comme un seul <point>, partout où cette ligne[159] sera égale à la moitié de l'étendue[160] d'une ondulation complète, la réflexion du point abaissé interférera avec la réflexion du point <resté> fixe, de façon que le mouvement progressif de l'un coïncidera avec le mouvement rétrograde de l'autre, et ils seront tous deux détruits[161] ; mais, quand cette ligne sera égale à l'étendue complète d'une ondulation, l'effet sera doublé[162] ; et quand à une étendue et demie, détruit à nouveau ; et ainsi de suite pour un nombre considérable d'alternances ; et si les ondulations réfléchies sont de sortes différentes[163], elles {P.36} seront diversement affectées, selon leurs proportions aux diverses longueurs de la ligne qui mesure la différence entre les longueurs de leurs deux chemins, et qui pourrait être dénommée intervalle de retard[164].

---

[154] Robert Boyle (1627-1691). Young fait probablement référence à son ouvrage *Experiments and considerations touching colours*, publié en 1664, dont le premier chapitre analyse la manière dont les aspérités, la rugosité, les irrégularités ou les rayures peuvent affecter la couleur des surfaces.

[155] Guillaume Mazeas (1720-1775), *Observations sur des couleurs engendrées par le frottement des surfaces planes et transparentes*, Histoire de l'académie royale des sciences et belles-lettres de Berlin, 1752, 248-261. Mazeas semble cependant dans ce texte faire plus référence aux couleurs apparaissant à l'interface entre deux surfaces polies que l'on aurait frottées au point qu'elles adhèrent l'une à l'autre, qu'aux couleurs de surfaces striées.

[156] Henry Brougham (1778-1868), *Farther experiments and observations on the affections and properties of light*, Philosophical Transactions of the Royal Society of London, 87, 1797, 352-385. Texte dans lequel sont longuement évoquées les couleurs pouvant apparaître dans les rayures ou aspérités à la surface d'un miroir (365-370). Henry Brougham se révèlera aussi le plus virulent détracteur de *la Théorie de la Lumière et des Couleurs* (Edinburgh Review, vol. 1, 450).

[157] C'est-à-dire « de la Proposition VIII ».

[158] Sur la Figure 2, on retrouve le plan (segment horizontal) sur lequel sont repérés deux points réfléchissants extrêmement proches - que l'on nommera dans cette note A (à gauche) et B (à droite) - l'œil de l'observateur (en haut, à gauche, que l'on appellera O) et un objet lumineux (en haut, à droite, que l'on appellera L). Les longs segments inclinés reliant les deux points réfléchissants à l'œil et à l'objet sont évidemment les rayons lumineux incidents (ceux passant par l'objet) et réfléchis (ceux arrivant à l'œil). Les deux points sont supposés capables de réfléchir tous deux la lumière provenant de l'objet vers l'œil et les trajets de la lumière allant de l'objet à l'œil en passant par chacun des points sont supposés de longueurs égales. Cette hypothèse, reposant sur la grande proximité des deux points et la très grande distance des points O et L à la surface, consiste en pratique à faire l'approximation que les rayons incidents depuis L et réfléchis vers O sont comme parallèles ; approximation très fréquemment utilisée en optique géométrique. Young compare alors ces longueurs à celle du chemin parcouru par la lumière réfléchie depuis un point (que l'on appellera A') situé plus bas que le plan, à la verticale du point A ; donc par exemple un point situé au fond d'une rayure que l'on aurait pratiquée dans le plan au niveau du point A. En utilisant la figure, Young montre que la longueur du chemin LA'O est plus longue que celle du chemin LAO (et donc LBO) d'une quantité égale à $2 * AA' * \cos i$, où $AA'$ est la profondeur de la rayure, $i$ l'angle d'incidence de la lumière en A', et cos le cosinus normalisé à 1 que l'on utilise classiquement aujourd'hui (qui est bien « *le rapport du Cosinus [...] au rayon du cercle* »).

[159] C'est-à-dire « la longueur additionnelle du trajet LA'O ».

[160] Ici Young emploie le mot « breadth » à nouveau (comme dans la Prop. III) et fait néanmoins ici référence à ce que l'on appelle aujourd'hui la *longueur d'onde*, et non plus la largeur du front d'onde.

Afin que l'effet soit le plus perceptible, un nombre de paires de points doit être réuni en deux lignes parallèles ; et, si plusieurs de ces paires de lignes sont placées à côté l'une de l'autre, elles faciliteront l'observation. Si l'une de ces lignes est amenée à tourner autour de l'autre faisant un axe, l'abaissement au-dessous du plan donné sera comme le Sinus de l'inclinaison ; et, tandis que l'œil et l'objet lumineux resteront fixes, la différence de la longueur des chemins variera comme le Sinus[165].

Les meilleurs sujets pour cette expérience sont les excellents micromètres de M. COVENTRY[166] ; ceux d'entre eux qui consistent en des lignes parallèles tracées sur du verre, à la distance d'un cinq-centième de pouce[167], sont les plus pratiques. Chacune de ces lignes apparait sous un microscope comme constituée de deux ou plusieurs lignes plus fines, exactement parallèles, et à une distance d'un peu plus d'un vingtième de celle des lignes adjacentes[168]. J'ai placé l'une d'entre elles de manière à réfléchir la lumière du soleil sous un angle de 45°, et l'ai fixée de telle manière qu'alors qu'elle tournait autour de l'une des lignes utilisée comme un axe je pouvais mesurer son mouvement angulaire ; et j'ai trouvé que la couleur rouge la plus vive apparaissait aux inclinaisons 10¼°, 20¾°, 32° et 45° ; dont les Sinus sont comme les nombres 1, 2, 3 et 4.[169] De plus sous tous les autres angles, lorsque la lumière du soleil était réfléchie par la surface, la couleur s'évanouissait avec l'inclinaison et était égale pour des inclinaisons égales de chaque côté.

---

[161] Young décrit donc ici le cas d'une interférence destructive, où la différence de longueur entre les deux trajets empruntés par la lumière (réfléchie par B et par A') est égale à la moitié de la longueur d'onde d'oscillation.

[162] Et ici est décrit le cas d'une interférence *constructive*, correspondant au cas où la différence de longueur entre les deux trajets empruntés par la lumière est exactement égale de la longueur d'onde d'oscillation.

[163] C'est-à-dire quand la différence des trajets correspond à une longueur quelconque, différente d'un nombre entier de demi-longueurs d'onde.

[164] « *Intervalle de retard* » que l'on nomme aujourd'hui *différence de marche* (ou *différence de chemin optique*), et qui représente dans le cas général non pas la différence de longueur entre les deux trajets empruntés par la lumière, mais la différence des *temps de parcours* de ces deux trajets par la lumière. Par commodité ce temps est néanmoins bel et bien ramené à une distance, en étant divisé systématiquement par la vitesse de propagation de la lumière dans le vide *c*. Ainsi la *différence de marche* $\delta$ correspond précisément à la différence entre les longueurs des trajets multipliées tout au long du parcours par l'indice de réfraction local *n* des milieux traversés : $\delta = (\int_L^O n(s).ds)_{chemin\ 2} - (\int_L^O n(s).ds)_{chemin\ 1}$. Lorsque la lumière parcourt comme ici chaque trajet dans un milieu homogène d'indice 1, le résultat présenté ici par Young est compatible avec le cas général moderne. Et on verra par la suite que Young calcule correctement les *intervalles de retard* en considérant les temps de parcours et non les distances.

[165] Si le point A' n'est pas obtenu par abaissement du point A perpendiculairement au plan, mais par rotation du point A d'un angle $\beta$ autour du point B, alors la distance au plan du point abaissé A' sera égale à $\sin \beta = AB * \sin \beta$, et la différence de chemin optique sera par extension elle aussi proportionnelle à ce Sinus. Si l'on ne comprend pas forcément comment une rayure pourrait être obtenue par rotation autour d'une autre rayure, c'est que ce calcul introduit en réalité l'expérience qui suit, où Young teste son modèle en observant les couleurs apparaissant lorsque l'on fait tourner deux lignes réfléchissantes (deux graduations d'un micromètre) l'une autour de l'autre.

[166] John Coventry (1735-1812). En 1774, John Coventry présente plusieurs micromètres à la Royal Society. Ces plaques de verre ou d'ivoire sur lesquelles était tracées à l'aide d'une fine pointe de diamant une série de traits parallèles (de 50 à 1000 traits pas pouce) sont destinées à être utilisées avec des microscopes, afin d'évaluer la taille des objets observés. Les micromètres de Coventry sont alors réputés pour leur résolution et leur précision. (W. Bicknell, *Account of the late Mr. John Coventry, an ingenious Mechanic,* The Weekly Entertainer, vol. 53, 08 février 1813, 116-118). Un exemple de micromètre de Coventry peut être admiré sur le site du Musée d'Histoire des Sciences d'Oxford : http://www.mhs.ox.ac.uk/collections/imu-search-page/record-details/?TitInventoryNo=59999&querytype=field&thumbnails=on (page consultée le 17 septembre 2019).

[167] 1/500 * 2,54 cm = 50 µm environ.

Cette expérience fournit une confirmation très forte de la théorie. Il est impossible de déduire la moindre explication de celle-ci de quelque hypothèse avancée jusqu'ici ; et je crois qu'il serait {P.37} difficile d'en inventer aucune autre qui saurait en rendre compte. Il y a une analogie frappante entre cette séparation des couleurs et la production d'une note musicale par des échos successifs sur des palissades de fer équidistantes ; que j'ai trouvé correspondre assez précisément avec la vitesse connue du son et les distances des surfaces[170].

Il n'est pas improbable que les couleurs des téguments de certains insectes, et d'autres corps naturels, manifestant sous des lumières différentes la plus belle versatilité, puissent être découvertes comme relevant de cette description, et ne pas être dérivées des lames fines[171]. Dans certains cas, une simple égratignure ou sillon peut produire des effets similaires, par la réflexion depuis ses bords opposés.

Corollaire II. *Des Couleurs des Lames fines.*[172]

---

[168] Donc 1/20 * 1/500 * 2,54 cm = 2,5 µm environ.

[169] Cette expérience correspond à la première description et justification du fonctionnement de ce que l'on appelle aujourd'hui un *réseau de diffraction*. La différence de marche pour les rayons réfléchis par deux lignes successives séparées d'une distance $AB$ = 2,5 µm environ, tournée l'une autour de l'autre d'un angle $\beta$ = 10,25° et éclairées sous une incidence $i$ = 45° et se propageant dans l'air vaut d'après ce qui précède : $\delta = 2 * AB * \sin \beta * \cos i$ = 630 nm environ, ce qui correspond raisonnablement à une interférence constructive pour le rouge (auquel on associe généralement une longueur d'onde $\lambda_r$ comprise entre 625 et 740 nm). La différence de marche calculée pour les autres angles proposés par Young correspondant alors à 2, 3 et 4 fois la longueur d'onde du rouge estimée ici, soit à nouveau à des interférences constructives pour cette couleur et à une augmentation de son intensité relativement aux autres.

[170] Ce problème est développé dans Thomas Young, *Outlines of Experiments and Inquiries respecting Sound and Light*, Philosophical Transactions of the Royal Society of London, vol. 90, 1800, 116-118.

[171] C'est effectivement aux effets d'interférence par réflexion sur leur surface présentant des structures périodiques de tailles micrométriques que sont dues les couleurs irisées, dites *structurelles* ou *physiques*, de certains insectes et végétaux, des ailes de papillons morpho, ou des plumes du paon. D'autres couleurs, comme celles des ailes transparentes de guêpes ou de mouches, étant pour leur part produites par les effets d'interférences que l'on retrouvera dans les lames fines décrites dans le Corollaire II (Serge Berthier, *Iridescences, les couleurs physiques des insectes*, Springer France, 2003 ; Ekaterina Shevtsovaa & al., *Stable structural color patterns displayed on transparent insect wings*, Proceedings of the National Academy of Sciences, vol. 108, n°1, 2011).

[172] Dans les Parties I et II de l'*Opticks*, Newton observe puis interprète les couleurs variables des lames fines. Parmi les systèmes qu'il interprète on trouve notamment les couleurs des bulles de savon. Mais aussi et surtout celles des anneaux colorés dits « de Newton ». Anneaux colorés qui apparaissent organisés autour d'une tache centrale noire lorsque l'on observe la réflexion de la lumière sur une lentille aux surfaces convexes elle-même posée sur une surface plane réfléchissante (et que l'on peut aussi observer en transmission, mais dans des couleurs complémentaires, si la surface plane réfléchissante est aussi partiellement transparente). Newton observe aussi la périodicité régulière de ces anneaux et l'associe rigoureusement à l'augmentation progressive de l'épaisseur de la lame d'air formée entre la face inférieure de la lentille et la surface plane. On reviendra plus loin sur l'interprétation par Newton de ces couleurs des lames minces par la théorie des accès, et sur la manière dont il l'extrapole aux couleurs des autres corps. Mais on souligne dès à présent l'importance

Lorsqu'un faisceau de lumière tombe sur deux surfaces réfringentes parallèles, les réflexions partielles coïncident parfaitement en direction ; et dans ce cas, l'intervalle de retard pris entre les surfaces est à leur distance comme deux fois le rapport du Cosinus de l'angle de réfraction au rayon <du cercle>. Car, sur la Fig. 3,[173] en traçant AB et CD perpendiculairement aux rayons <lumineux>, les temps de parcours de BC et AD seront égaux[174] et DE sera la moitié de l'intervalle de retard ; mais DE est à CE comme le Sinus de DCE est au rayon <du cercle>. D'où, pour que DE soit constant, ou que la même couleur soit réfléchie, l'épaisseur CE doit varier comme la sécante de l'angle de réfraction CED[175] : ce qui s'accorde exactement avec les expériences de Newton ; car la correction est parfaitement négligeable[176].

Supposons que le milieu entre les deux surfaces est plus rare que les milieux l'entourant[177] ; alors l'impulsion réfléchie à la seconde surface, rencontrant à la première une ondulation ultérieure, rendra les particules du milieu le plus rare capables de stopper

---

majeure que revêt pour Young l'explication des couleurs de ces lames minces par la théorie ondulatoire, puisqu'elle remet en question l'essentiel des interprétations fournies dans le Livre II de l'*Opticks*.

[173] A nouveau pour l'explication de cette figure on va nommer L le point d'émission de la lumière (haut, droite) et O l'œil (haut, gauche). Et l'on va supposer comme le fait Young implicitement que les points d'impact A et C de la lumière sur la face supérieure de la lame sont suffisamment proches, et les points O et L suffisamment éloignés de la lame, pour pouvoir faire l'approximation que les rayons incidents depuis L et arrivant en O sont parallèles.

[174] Ce résultat a été établi par Huygens dans le cadre de sa démonstration de la loi de la réfraction (*Traité de la Lumière,* Chapitre 3 : « De la réfraction », 1690). On remarque en effet que par construction, la distance BC est égale à la distance AC multipliée par le sinus de l'angle CAB qui est égal à l'angle d'incidence $i_1$ du rayon LC. D'autre part la distance AD est égale à la même distance AC multipliée par le sinus de l'angle ACD, lequel est égal à l'angle de réfraction $i_2$ du rayon AE. On a donc $\frac{BC}{AD} = \frac{\sin i_1}{\sin i_2}$. Revenant alors à la Proposition V, on se rappelle que pour Young et les tenants de l'optique ondulatoire, la loi de la réfraction manifeste le fait que $\frac{\sin i_1}{\sin i_2} = \frac{v_1}{v_2}$. En conclusion donc $\frac{BC}{AD} = \frac{v_1}{v_2}$, soit $\frac{BC}{v_1} = \frac{AD}{v_2}$. C'est-à-dire que le temps que met la lumière pour parcourir la distance BC à la vitesse $v_1$ est effectivement égal au temps qu'elle met à parcourir la distance AD à la vitesse $v_2$. Par conséquent le retard pris par le rayon LA par rapport au rayon LC correspond effectivement à la portion DE. Et il faudra le multiplier par deux pour connaître le retard pris sur l'ensemble du trajet, du fait de la symétrie de la figure. En réalité ce retard total correspond plutôt au double du temps de parcours de la portion DE, puisque même si Young ne prononce pas le mot « temps » ici ni dans la proposition précédente, c'est bien à partir d'un raisonnement sur les temps et non sur les distances qu'il a conclu à l'égalité des chemins BC et AD.

[175] A l'aide de la Figure associée à la note de bas de page de la Proposition V, Young définit la sécante d'un angle par la phrase « *AC est la sécante de BD, ou BAD.* » (Thomas Young, *A Course of Lectures in Natural Philosophy and Mechanical Arts*, vol. II, London, 1807, 15). La sécante d'un angle est donc égale au rayon du cercle divisé par le cosinus de l'angle en question. Ainsi CE proportionnel à $\frac{1}{\cos(CED)}$ implique effectivement que CE est proportionnel à $\frac{1}{\cos\left(\frac{\pi}{2}-DCE\right)} = \frac{1}{\sin(DCE)}$ et donc que l'intervalle de retard DE est constant.

[176] Isaac Newton, *Opticks*, Livre II, Partie I, Observation XIX, 1730, 191-194. Où Newton associe bien l'uniformité de la couleur d'une couche fine d'eau au fait que son épaisseur sera proportionnelle à la sécante d'un angle, mais d'un angle très légèrement supérieur à l'angle de réfraction (dans le cas d'une lame d'indice supérieur au milieu environnant) et d'une expression si complexe que la solution proposée et démontrée ici par Young doit sembler la plus plausible.

[177] C'est-à-dire, pour Young, dans le cas d'une lame d'indice de réfraction inférieur à celui du milieu environnant, comme dans le cas étudié par Newton d'une fine épaisseur d'air piégée entre deux blocs de verre. L'étude de ce cas précis était préfigurée par le tracé d'un rayon réfracté s'éloignant de la droite normale sur la Figure 3 et justifie même *a posteriori* la possibilité d'assimiler la différence de marche à deux fois la distance DE, puisque l'indice de la lame est ici égal à 1.

intégralement {P.38} le mouvement du plus dense, et de détruire la réflexion, (Prop. IV.)[178] alors qu'elles seront-elles-mêmes propulsées plus fortement que si elles avaient été au repos ; et la lumière transmise sera augmentée. De façon que les couleurs par réflexion seront détruites et celles par transmission rendues plus vives lorsque les doubles épaisseurs, ou intervalles de retard, seront des multiples quelconques de l'étendue totale des ondulations[179] ; et, aux épaisseurs intermédiaires, les effets seront inversés ; conformément aux observations NEWTONIENNES[180].

Si les mêmes proportions se trouvent rester valides pour des lames fines d'un milieu plus dense, ce qui en effet n'est pas improbable, il sera nécessaire d'adopter la démonstration corrigée de la Prop. IV. mais, quoi qu'il en soit, si une lame mince est interposée entre un milieu plus rare et un plus dense, on peut s'attendre à ce que les couleurs par réflexion et transmission changent de places[181].

A partir des mesures des épaisseurs réfléchissant les différentes couleurs effectuées par NEWTON[182], l'étendue[183] et la durée de leurs ondulations respectives peuvent être très précisément déterminées ; bien qu'il ne soit pas improbable, que quand les verres s'approchent très près, l'atmosphère d'éther puisse produire quelque légère irrégularité. L'intégralité du spectre visible parait être compris dans le rapport de trois à cinq, ou une sixte majeure en musique ; et les ondulations de rouge, jaune, et bleu, être liées en magnitude comme les nombres 8, 7 et 6 [184] ; de façon que l'intervalle du rouge au bleu est

---

[178] Proposition IV : « *Lorsqu'une ondulation atteint une surface qui est la limite entre des milieux de différentes densités, une réflexion partielle a lieu, proportionnelle en force à la différence des densités.* »

[179] L'expression employée par Young est à nouveau « breadths of the undulations » pour désigner ce que l'on nomme aujourd'hui *longueur d'onde*. Car ici il décrit le cas d'une interférence destructive de la lumière réfléchie quand la différence de marche entre les rayons réfléchis à la première et à la deuxième surface (en C et en E) est égale à un nombre quelconque de longueurs d'onde *entières* ; ce qui est l'opposé de ce qu'il décrivait au corollaire précédent pour les couleurs des lames striées. La raison pratique est que l'observation expérimentale des anneaux de Newton en réflexion démontre l'existence d'une tache noire au milieu de la figure, soit là où la lentille est en contact avec le bloc de verre et où l'épaisseur de la lame d'air (et donc la différence de marche) est nulle, dont il faut bien rendre compte. La justification théorique *ad hoc* proposée par Young est que lorsque l'impulsion réfléchie par la deuxième face de la lame fine rencontre à la première surface une onde en phase avec elle mais venant dans la direction opposée, les vibrations vont s'annuler dans la direction de la réflexion et vont être dédoublées dans la direction de la transmission. Mais c'est Fresnel qui justifiera complètement en 1823 la nécessité d'ajouter au calcul de la différence de marche un décalage d'une demi-longueur d'onde dû à la différence de nature de la réflexion verre/air du premier rayon réfléchi et de la réflexion air/verre du second. La réflexion d'une onde lumineuse par un milieu d'indice supérieur à celui du milieu dans lequel elle voyage subit effectivement un retard supplémentaire d'une demi-longueur d'onde, quand dans le cas de la réflexion sur un milieu d'indice inférieur le déphasage est nul (Augustin Fresnel, *Mémoire sur la loi des modifications que la réflexion imprime à la lumière polarisée*, Mémoires de l'Académie royale des sciences de l'Institut de France, tome XI, Paris, 1832, 393-433).

[180] Isaac Newton, *Opticks,* Livre II, Parties I et II, 1730, 168-218.

[181] Cette remarque anticipe probablement l'explication des couleurs de fines couches d'oxydes à la surface des métaux (donc situées entre l'air peu dense et le métal dense), dont Young développera l'explication dans *A Course of Lectures in Natural Philosophy and Mechanical Arts*, vol. I, Lecture XXXIX, London, 1807, 457-471.

[182] Au fil des Observations 12 à 16 de la Partie I du Livre II de l'*Opticks*, Newton observe et mesure le diamètre des anneaux colorés obtenus en réflexion lorsque le montage constitué de la lentille convexe posée sur une surface réfléchissante est éclairé successivement par la série des couleurs pures du spectre (1730, 183-187). Et connaissant parfaitement le rayon de courbure de sa lentille, il est capable de remonter à l'épaisseur de la lame d'air à l'endroit où l'on observe le premier anneau brillant pour chacune des couleurs. C'est à cette série de mesure, synthétisée par la Figure 6 du Livre II de l'*Opticks*, que Young fait ici référence.

[183] Ici, le mot « breadth » est à nouveau utilisé pour désigner la *longueur d'onde*.

[184] Où l'on retrouve les proportions avancées dès l'Hypothèse III. Le Tableau illustrant ce corollaire permet de justifier ces valeurs, puisque les longueurs des ondulations associées aux couleurs rouge, jaune et bleu y sont

une quarte. La fréquence absolue exprimée en nombres est trop grande pour être conçue distinctement, mais elle peut être mieux imaginée par une comparaison avec le son. Si une corde produisant le Do tenor[185] pouvait être continûment bissectée 40 fois et devait alors vibrer, elle produirait un jaune-vert clair : celui-ci étant noté Do40 [186], le rouge extrême serait un La39, et le bleu Ré40. {P.39} La longueur[187] et fréquence absolue de chaque vibration est exprimée dans le tableau ; en supposant que la lumière parcourt 500.000.000.000 pieds en $8\frac{1}{8}$ minutes[188].

| Couleurs. | Longueur d'une Ondulation en fractions de Pouce, dans l'Air.[189] | *Longueur d'une ondulation en nm.*[190] | Nombre d'Ondulations en un Pouce. | Nombre d'Ondulations en une Seconde. |
|---|---|---|---|---|

---

estimées en fractions de pouce à respectivement 0,000256 ; 0,0000227 ; 0,000196. Valeurs qui se révèlent être approximativement dans les proportions 8 ; 7 ; 6. Ici, comme dans l'Hypothèse III, c'est le mot « magnitude » qui est utilisé pour désigner ce qui, à la lumière des données du Tableau, correspond manifestement à la *longueur d'onde*.

[185] Do tenor (« tenor $\bar{c}$ ») est un terme de luthier aujourd'hui souvent utilisé pour désigner le Do2 en notation latine (ou $c^3$ en Angleterre), et correspondant à une fréquence de vibration de 130,81 Hz. Cependant le « $c^{41}$ » évoqué dans la suite de la phrase comme produit de l'opération (que nous avons traduit « Do40 »), ainsi que la longueur d'onde associée, semblent indiquer que par l'expression « tenor $\bar{c}$ » Young évoque ici un Do-1 (ou $c^1$ en Angleterre), correspondant à la quatrième touche du piano, et oscillant à 32,70 Hz. C'est en tout cas l'hypothèse que nous retiendrons pour la suite.

[186] La note correspondant à la vibration d'une hypothétique corde obtenue en coupant quarante fois successivement en deux parties égales une corde produisant un Do-1 est aussi un Do, situé 40 octaves au-dessus du Do-1 d'origine ; donc un Do40 (car le Do0 n'existe pas). La fréquence de vibration *f* associée à cette corde sera donc égale à la fréquence de vibration de la corde d'origine multipliée par $2^{40}$, soit : $f = 2^{40} *32,70$ Hz = $5,75.10^{14}$ Hz (ce qui est bien au-delà du spectre audible, évidemment). La longueur d'onde associée à cette fréquence de vibration, pour une onde se propageant à la vitesse *c* de la lumière estimée par Young vaudrait $\lambda$ = *c* / *f* = 544 nm environ, ce qui correspond bien d'après son tableau à un jaune-vert. En associant dans la suite le rouge extrême à un La39 ($a^{40}$ en anglais) et le bleu à un Ré40 ($d^{41}$ en anglais), il attribue à ces deux composantes du spectre des longueurs d'onde respectivement égales à $\lambda_R$ = 647 nm et $\lambda_B$ = 484 nm.

[187] Ici c'est bien le terme *longueur* (« length ») qui est utilisé pour la première fois pour qualifier ce qu'aujourd'hui on appelle communément *longueur d'onde* (ou *wavelength* en anglais). Mais la terminologie « length » pour qualifier la longueur d'onde ne se stabilisera chez Young qu'à partir de sa publication *An account of some Cases of the production of Colours not hitherto described*, Philosophical Transactions of the Royal Society of London, 1802, 387, à l'endroit même où il reformule le Principe d'Interférence (voir note de bas de page de la Proposition VIII).

[188] Un pied correspondant à 30,48 cm, la vitesse de la lumière *c* admise ici par Young vaut dans notre système d'unités *c* = 3,13 * $10^8$ m/s environ. C'est cette valeur que nous utiliserons nous-mêmes dans nos calculs. Cette valeur nous est donnée par Young sous la forme du temps mis pour la lumière pour parcourir la distance du Soleil à la Terre. Et Il attribue ce résultat aux travaux des astronomes Römer sur les occultations des satellites de Jupiter (Ole Christensen Römer, *Démonstration touchant le mouvement de la lumière*, Journal des Sçavans, 7 décembre 1676, 276-279) et Bradley sur l'aberration apparente des étoiles fixes (James Bradley, *A letter giving an Account of a new discovered Motion of the Fixed Stars*, Philosophical Transactions of the Royal Society of London, vol. 35, n° 406, 1728, 637-661).

[189] D'après Young, l'ensemble des valeurs proposées dans ce tableau ont été extrapolées de mesures effectuées par Newton. Dans les faits, dans la partie de l'*Opticks* dédiée aux couleurs des lames minces, Newton exprime très précisément les *rapports* des épaisseurs d'air donnant une réflexion maximale des couleurs situées à la limite entre chacune des sept couleurs qu'il attribue au spectre : « *les épaisseurs de l'air entre les verres là, où les anneaux sont successivement produits par les limites des sept couleurs, rouge, orange, jaune, vert, bleu, indigo et violet, sont l'une à l'autre comme les racines cubiques des carrés des huit longueurs d'un corde, qui font les notes d'une octave, sol, la, fa, sol, la, mi, fa, sol ; c'est-à-dire, comme les racines cubiques des carrés des nombres 1, 8/9, 5/6, ¾, 2/3, 3/5, 9/16, ½.* » (*Opticks,* Livre II, Partie I, Observation 14, 1730, 186). Quant au surprenant enchaînement des notes de l'octave proposé par Newton, on pourra lire

| | | | | |
|---|---|---|---|---|
| Extrême - | .0000266 | *675* | 37640 | 463 millions de millions |
| Rouge - - | .0000256 | *648* | 39180 | 482 |
| Intermédiaire | .0000246 | *624* | 40720 | 501 |
| Orange - - | .0000240 | *610* | 41610 | 512 |
| Intermédiaire | .0000235 | *598* | 42510 | 523 |
| Jaune - - | .0000227 | *577* | 44000 | 542 |
| Intermédiaire | .0000219 | *557* | 45600 | 561  (= $2^{48}$ environ) |
| Vert - - | .0000211 | *535* | 47460 | 584 |
| Intermédiaire | .0000203 | *515* | 49320 | 607 |
| Bleu - - | .0000196 | *497* | 51110 | 629 |
| Intermédiaire | .0000189 | *480* | 52910 | 652 |
| Indigo - - | .0000185 | *470* | 54070 | 665 |
| Intermédiaire | .0000181 | *460* | 55240 | 680 |
| Violet - - | .0000174 | *442* | 57490 | 707 |
| Extrême - | .0000167 | *425* | 59750 | 735 |

*Scholie*. Ce ne fut pas avant de m'être satisfait de tous ces phénomènes, que je trouvai dans la *Micrographie* de HOOKE[191], un passage qui aurait pu me mener plus tôt à une conclusion similaire.

---

Dorothée Devaux et Bernard Maitte, *Newton, les couleurs et la musique*, Alliage, n°59, 2006. Mais quant aux proportions annoncées par Newton, on remarque effectivement que les *longueurs d'ondulations* données dans le présent tableau pour chacune des couleurs *Intermédiaires* (qui sont les limites entre les 7 couleurs) sont précisément dans ces rapports à la *longueur d'ondulation* du Rouge extrême. Car Young identifie l'épaisseur de l'air là où le premier anneau coloré est produit en lumière monochromatique, au quart de la *longueur d'ondulation* correspondant à la couleur considérée : la lumière parcourant deux fois cette épaisseur d'air donnera lieu à une interférence constructive en réflexion, et aura donc parcouru une distance égale à la moitié de la *longueur d'ondulation*. Par ailleurs, la Figure 6 du Livre II de l'*Opticks* et son mode d'emploi donné dans la deuxième partie de ce livre (p. 199-205) sont censés permettre à tout lecteur de déterminer l'épaisseur d'air pour le premier anneau brillant (et donc la *longueur d'ondulation*) pour les couleurs situées au centre des sept zones colorées du spectre déterminées par Newton, donc les longueurs d'ondulation du Rouge, Orange, Jaune, Vert, Bleu, Indigo, Violet données dans le tableau. Cependant si les rapports de longueurs sont très précisément donnés dans l'*Opticks*, il reste nécessaire pour remplir cette colonne de connaître la valeur absolue en pouces d'au moins une épaisseur d'air. Or Newton fournit pour cela une seule valeur, cette fois-ci étonnamment imprécise : une épaisseur d'air de 1/178000 pouce correspondant au premier anneau brillant en lumière blanche, soit d'après son estimation, au maximum du « *jaune citron brillant ou à la limite entre le jaune et l'orange* » (p.204). Or en multipliant cette valeur par quatre, on obtient une longueur de 1/44500 = 0,0000225 inch qui correspond presque, mais pas exactement, à la valeur donnée par Young pour le jaune central, et qui est très éloignée de celle du jaune-orange. Cette incongruité est suffisamment grossière en regard de la précision des valeurs proposées dans le tableau et de leurs rapports pour que l'on puisse raisonnablement projeter que Young a réalisé lui-même l'expérience pour au moins une couleur monochromatique (pour palier l'imprécision de la valeur de référence fournie par Newton) et qu'il l'a extrapolée ensuite aux autres couleurs grâce à la Figure 6 du livre II de l'*Opticks*.

[190] Cette colonne n'existe pas dans le texte de Young. Elle a été rajoutée par le traducteur afin de transposer les données exposées par Young dans une unité plus usuelle pour la mesure de longueurs d'ondes optiques, le nanomètre (nm). Pour cela, les données de la troisième colonne du tableau (plus précises que celles de la deuxième qui ont été arrondies) ont simplement été inversées, puis multipliées par 2,54 pour être converties en cm, puis par $10^7$ pour passer des cm aux nm. Les mesures proposées ici par Young sont les toutes premières mesures des longueurs d'ondes du spectre de la lumière blanche de l'histoire. Et il est frappant de voir comme elles sont proches des valeurs connues actuellement (pas plus de 10% d'écart). En particulier étant donnés la très large gamme de longueurs d'ondes couverte par chacun des noms de couleurs proposés par Newton ; la légère surestimation de la vitesse de la lumière par rapport à la connaissance que nous en avons actuellement ($c$ est surestimée par Young de 5% par rapport à la valeur actuelle) ; et les données parcellaires extraites de l'*Opticks* pour arriver à ces résultats.

[191] Robert Hooke, *Micrographia*, Observation IX, London, 1665, 47-67. Passage dans lequel Robert Hooke décrit les anneaux que l'on appellera « de Newton » et défend une théorie ondulatoire de la lumière.

« Il est des plus évident que la réflexion depuis le côté au-dessous du corps ou le plus éloigné, est la cause principale de la production de ces couleurs. […] Soit le rayon tombant obliquement sur la lame fine, une partie est donc réfléchie vers l'arrière par la première superficie, […] une partie réfractée par la seconde surface, […] d'où elle est réfléchie et réfractée encore. […] De sorte que, après deux réfractions et une {P.40} réflexion, une sorte de rayon plus faible est propagée à cet endroit […], » et, « en raison du temps dépensé à passer et repasser, […] cette impulsion plus faible se retrouve derrière l'impulsion » premièrement réfléchie « de sorte que par cela (les surfaces étant si proches l'une de l'autre que l'œil ne peut les discriminer d'une surface unique) cette impulsion confuse ou dupliquée, dont la partie la plus forte précède et dont la plus faible suit, produit sur la rétine la sensation d'un jaune.[192] Si ces surfaces sont encore plus séparées, l'impulsion la plus faible peut devenir coïncidente avec la » réflexion de la « seconde », ou de l'impulsion la suivant tout juste, à partir de la première surface, « mais aussi se retrouver à la traîne derrière celle-ci, et être coïncidente avec la troisième, quatrième, cinquième, sixième, septième ou huitième […][193] ; de sorte que, s'il y a un corps fin transparent qui de la plus grande finesse requise pour produire des couleurs croit par degrés jusqu'aux plus grandes épaisseurs, […] les couleurs doivent être aussi souvent répétées que l'impulsion la plus faible perd le pas sur sa primaire ou première impulsion, et devient coïncidente avec une impulsion » ultérieure. Et puisque cela « est coïncident, ou que cela suit de la première hypothèse que j'ai prise pour les couleurs, je l'ai trouvé conforme à l'expérience dans une multitude de cas qui semblent le prouver. » (P. 65-67.) Ceci fut imprimé environ sept ans avant qu'aucune des expériences de NEWTON ait été réalisée[194]. Nous sommes informés par NEWTON, du fait que HOOKE fut après coup disposé à adopter sa « suggestion » sur la nature des couleurs ; et cependant il n'apparaît pas que HOOKE ait jamais appliqué cette amélioration à son explication de ces phénomènes, ou ait enquêté sur la nécessaire conséquence d'un changement d'obliquité sur sa supposition originelle, sans quoi il ne pourrait qu'avoir découvert une coïncidence frappante avec les mesures enregistrées par NEWTON d'après expérience. Toutes les tentatives précédentes pour expliquer les couleurs des lames fines ont procédé soit sur la base de suppositions {P.41} qui, comme celles de NEWTON[195], nous

---

[192] Ici Young retire sans le signaler un passage dans lequel Hooke décrit l'effet d'un écartement progressif des surfaces. Si ce passage n'est pas utile à la démonstration de Young, il est intéressant pour ce qu'il illustre bien la théorie des couleurs des lames minces de Robert Hooke : à la surface supérieure de la lame se recombinent un rayon affaibli par la première réflexion, et un second rayon encore plus affaibli (pour avoir été transmis à la première surface, réfléchi à la seconde et transmis à nouveau à la première) et retardé par rapport au premier. C'est de cette combinaison d'une impulsion plus forte et d'une autre plus faible, et surtout de la durée du retard pris par la seconde sur la première que naissent les couleurs des lames (analyse détaillée dans Michel Blay, *Un exemple d'explication mécaniste au XVIIe siècle : l'unité des théories hookiennes de la couleur,* Revue d'histoire des sciences, vol. 34, n°2, 1981, 97-121). Si apparaissent dans ce schéma certains éléments préfigurant la théorie des interférences (conception *ondulatoire* de la lumière, couleur issue de la combinaison de deux impulsions et dépendant de la valeur du retard pris par l'un par rapport à l'autre), on est ici encore loin de la formulation qu'en proposera Young ; du fait notamment d'un mécanisme de recombinaison des impulsions qui ne peut y être envisagé sans l'intervention de l'œil.
[193] Ici par contre Young signale une incise mais il n'y a pas de texte manquant d'après l'original.
[194] D'après ses carnets de notes (*Certain Philosophical questions : Newton's trinity Notebook*, James E. McGuire and Martin Tammy (éds), Cambridge University Press, 1983), les premières expériences de Newton relatives aux anneaux colorés ont été réalisées en 1666, soit juste après la publication de la *Micrographia*, et visaient à déterminer les épaisseurs de lame correspondant à chaque couleur, que Hooke regrettait justement ne pas avoir su obtenir.
[195] Newton développe son interprétation de la couleur des lames fines dans *Opticks*, Livre II, Partie III, Propositions XII à XX, 1730, 253-263. Il suppose qu'au cours de leur trajet, les corpuscules lumineux se trouvent

mèneraient à envisager les plus grandes irrégularités dans la direction des rayons réfractés ; soit, comme celles de M. MICHELL[196], requerraient des effets du changement de l'angle d'incidence qui seraient contraires aux effets observés ; soit qui sont également déficientes par rapport à ces deux circonstances, et sont incohérentes avec l'attention la plus modérée aux phénomènes principaux.

Corollaire III. *Des Couleurs des Lames épaisses.*

Lorsqu'un faisceau de lumière traverse une surface réfringente, spécialement si elle est imparfaitement polie, une portion de celui-ci est irrégulièrement diffusée et rend la surface visible dans toute les directions, mais de façon plus remarquable dans des directions peu distantes de celle de la lumière elle-même : et si une surface réfléchissante est placée parallèlement à la surface réfringente, cette lumière diffusée, comme le faisceau principal, sera réfléchie, et il y aura aussi une nouvelle dissipation de la lumière au retour du faisceau à travers la surface réfringente. Ces deux portions de lumière diffusée coïncideront en direction ; et si les surfaces sont d'une forme telle qu'elles cumulent les effets similaires, <elles> exhiberont des anneaux de couleurs. Ici l'intervalle de retard est la différence entre les chemins du faisceau principal et de la lumière diffusée entre les deux surfaces ; bien sûr, partout où l'inclinaison de la lumière diffusée est égale à celle du faisceau, bien que dans des plans différents, l'intervalle s'annulera et toutes les ondulations conspireront. Aux autres inclinaisons, l'intervalle sera la différence des sécantes de la sécante de l'inclinaison ou angle de réfraction du faisceau principal. De ces causes, toutes les couleurs des miroirs concaves observées par NEWTON et d'autres sont des conséquences nécessaires : et il apparait que leur production, bien que {P.42} quelque peu similaire, n'est en aucun cas, comme NEWTON l'imaginait[197], identique à la production de celles des lames fines.

---

périodiquement dans des « accès de facile transmission » puis de « facile réflexion » ; la longueur de ces accès dépendant de la couleur pure du corpuscule considéré. Un corpuscule arrivant dans un accès de facile transmission à la première surface d'une lame fine la traverse. Et s'il parvient à la seconde dans le même état, il la traverse également ; ainsi l'observateur voit en réflexion une zone obscure à cet endroit. Si par contre le corpuscule arrive à la seconde surface dans un état de facile réflexion, il est réfléchi vers la première surface, où il est transmis (car dans un accès de facile transmission, par symétrie du chemin) ; il parvient donc à l'œil de l'observateur, qui voit une zone lumineuse à l'endroit de la réflexion. Ainsi les anneaux colorés sont expliqués par une évolution périodique de l'état de la lumière, dont la période (la longueur de l'accès) est à envisager en relation à l'épaisseur de la lame fine.

[196] John Michell (1724-1793) est un physicien et géologue britannique, tenant d'une conception corpusculaire de la lumière. Ses théories optiques, sont essentiellement connues grâce aux allusions qu'y fait son ami Joseph Priestley dans *The History and Present State of Discoveries Relating to Vision, Light and Colours*, 1772. Quant aux couleurs des lames fines, Michell admet l'hypothèse newtonienne des accès de faciles transmission et réflexion, qu'il justifie par une « *doctrine des attractions et répulsions* » résumée en ces termes par Priestley (p. 310) : « *chaque particule du milieu a un grand nombre d'intervalles alternés égaux d'attraction et de répulsion, relativement aux particules de lumière, mais ces intervalles sont de différentes magnitudes, comme les particules de lumières sont de différentes couleurs.* » et « *si l'épaisseur de quelque milieu transparent dans lequel les particules de matière sont placées uniformément, est tel, que les intervalles attractifs des particules extrêmes* [NdT : les particules situées à la surface du milieu]*, ainsi que les intervalles répulsifs, coïncident les uns avec les autres, i.e. attractifs avec attractifs, et répulsifs avec répulsifs, à l'égard de chaque sorte de rayon, comme le rouge par exemple, par la force unifiée de ces extrêmes (toutes les particules intermédiaires du milieu détruisant mutuellement leurs effets) ces rayons seront réfléchis.* » Où l'on devine bien comme on est ici éloigné de l'interprétation proposée par Young (pour une analyse plus détaillée, voir John S. Parry, *John Michell's theory of matter and Joseph Priestley's use of it*, Thèse de Master en Philosophie, Imperial College, Londres, 1977, Chapitre 3, 55-86).

[197] Isaac Newton, *Opticks,* Livre II, Partie IV, 1730, 264-291. Cette partie commence par l'observation d'anneaux colorés se dessinant autour de la lumière réfléchie par un miroir concave. Le miroir consistant en un bloc de

Corollaire IV. *De l'Obscurité.*[198]

Dans les trois corollaires précédents, nous avons considéré que les substances réfringentes et réfléchissantes étaient limitées par une surface mathématique ; mais ceci n'est peut-être jamais vrai dans la nature. Les atmosphères éthérées peuvent se prolonger de chaque côté de la surface aussi loin que l'étendue d'une ou plusieurs ondulations ; et si on les suppose varier uniformément en densité en toute partie, les réflexions partielles sur le nombre infini de surfaces où la densité change interféreront beaucoup et détruiront une portion considérable de la lumière réfléchie, de sorte que la substance puisse devenir absolument noire ; et cet effet peut avoir lieu à un degré plus ou moins important selon que la densité de l'atmosphère éthérée varie plus ou moins uniformément ; et dans certains cas, des ondulations particulières étant plus affectées que d'autres, une nuance de couleur peut être produite. Conformément, M. BOUGUER[199] a observé une perte de lumière considérable, et dans certains cas une nuance de couleur, dans des réflexions totales à la surface d'un milieu plus rare.

Corollaire V. *Des Couleurs par Inflexion.*

Quelle que soit la cause de l'inflexion de la lumière passant par une petite ouverture, la lumière la plus proche de son centre doit être la moins déviée, et la plus proche de ses côtés la plus <déviée> : une autre portion de lumière tombant très obliquement sur le bord de l'ouverture sera copieusement réfléchie dans des directions variées ; certaines d'entre elles coïncideront parfaitement ou presque parfaitement en direction avec la lumière non-réfléchie et, ayant pris une {P.43} route indirecte, interféreront avec elle de manière à susciter l'apparition de couleurs[200]. La longueur des deux chemins différera d'autant moins

---

verre épais aux faces sphériques et parallèles (donc concave d'un côté et convexe de l'autre) dont la face extérieure convexe a été recouverte d'une fine couche de mercure réfléchissante, les couleurs sont attribuées à la traversée de l'épaisseur de verre. Le texte se poursuit donc par une explication des couleurs produites par ces *lames épaisses* directement extrapolée de la théorie des accès de faciles transmission et réflexion développée pour expliquer les couleurs des lames fines.

[198] En conséquence de la réécriture de l'hypothèse IV dans sa version de ce texte publiée en 1807 (Thomas Young, *A Course of Lectures in Natural Philosophy and Mechanical Arts*, vol. II, London, 1807) le présent Corollaire IV - reposant en bonne partie sur cette hypothèse - sera supprimé par Young de cette version (p.629).

[199] Pierre Bouguer (1698-1758) est un savant français membre de l'Académie Royale des Sciences, reconnu pour ses contributions en mathématiques, géodésie, architecture navale, hydrographie et optique. En 1760 est édité de manière posthume son *Traité d'optique sur la gradation de la lumière*, dont l'objet est de mesurer les quantités de lumière transmise ou perdue lors de son passage à travers un corps donné (en particulier au travers d'une étendue donnée de l'atmosphère terrestre) et auquel Young fait référence.

[200] Young envisage ici le phénomène des franges colorées apparaissant autour du faisceau lumineux lorsqu'une lumière blanche passe par une ouverture étroite, comme un phénomène d'interférences entre d'une part la lumière plus ou moins infléchie vers l'extérieur du faisceau par la sur-densité d'éther entourant les corps, et d'autre part une lumière réfléchie par les rebords de cette ouverture. Fresnel invalidera ce modèle quelques années plus tard en démontrant l'indépendance du système de franges relativement à la nature du corps diffractant (*Supplément au Deuxième Mémoire sur la diffraction de la lumière, présenté à l'académie des sciences dans la séance du 15 juillet 1816*, Œuvres complètes d'Augustin Fresnel, Eds : H. de Senarmont, E. Verdet, L. Fresnel, tome I, 1866, 129-170). Pour en 1818 expliquer le phénomène à l'aide du principe d'Huygens-Fresnel : « Les vibrations d'une onde lumineuse dans chacun de ses points peuvent être regardées comme la somme des mouvements élémentaires qu'y enverraient au même instant, en agissant isolément, toutes les parties de cette onde considérée dans une quelconque de ses positions antérieures. » (Augustin Fresnel, *Mémoire sur la diffraction de la lumière, couronné par l'Académie des Sciences en 1819*, Œuvres complètes d'Augustin Fresnel, Eds : H. de Senarmont, E. Verdet, L. Fresnel, tome I, 1866, 373-374).

que la direction de la lumière réfléchie aura été moins changée par sa réflexion, soit pour la lumière passant le plus près du bord ; de sorte que les bleus apparaitront dans la lumière la plus proche de l'ombre. L'effet sera augmenté et modifié quand la lumière réfléchie tombera sous l'influence du bord opposé, de sorte à interférer avec la lumière simplement infléchie par celui-là aussi.

Mais afin d'examiner les conséquences plus exactement, il sera commode de supposer l'inflexion causée par une atmosphère éthérée d'une densité variant comme une puissance donnée de la distance à un centre, comme dans la huitième proposition de la dernière Conférence Bakerienne (*Phil. Trans. for 1801*, p. 83.)[201] En posant $r = 3$ et $x = ½$, j'ai construit un diagramme (Fig. 4) qui montre, par les deux paires de courbes, la position relative des portions réfléchie et non-réfléchie d'une ondulation quelconque à deux temps successifs et aussi, par des lignes ombrées tracées au travers, les parties où les intervalles de retard sont en progression arithmétique et où des couleurs similaires se manifesteront à différentes distances de la substance infléchissante[202]. Le résultat s'accorde

---

[201] Thomas Young, *On the Mechanism of the Eye*, Philosophical Transactions of the Royal Society of London, vol. 91, 1801, 83. Young nous renvoie à la toute fin de sa précédente conférence bakerienne, où un addendum détaille le calcul de l'une des propositions qu'il expose au début de son texte sous forme de problème (Proposition VIII, p.33) : « *Déterminer le point de focalisation approximative de rayons parallèles tombant obliquement sur une sphère de densité variable* » ; addendum dans lequel il propose la résolution de ce problème pour : « *une sphère, de densité réfractive variant comme une puissance de la distance à son centre.* » Cette précédente conférence de Young est dédiée à l'étude des mécanismes de la vision, et tout particulièrement à l'accommodation qu'il envisage comme le résultat de la contraction du cristallin. Celui-ci y est décrit comme un muscle transparent d'indice de réfraction décroissant progressivement à mesure que l'on s'éloigne de son centre. C'est donc au calcul théorique de la position foyer du cristallin inhomogène, modélisé par « *une sphère, de densité réfractive variant comme une puissance de la distance à son centre* » que Young dédie cette partie du texte. Calcul et modèle qu'il reprend pour interpréter ici l'inflexion de la lumière passant à proximité d'un corps solide comme la réfraction progressive des rayons à travers une atmosphère éthérée de densité décroissante à mesure que l'on s'éloigne de ce corps. Si dans la page citée du texte *On the Mechanism of the Eye*, Young parle de *densité réfractive*, il évoque ici plutôt une *atmosphère éthérée de densité variable*. Or si ces deux concepts sont corrélés, ils ne sont pas à confondre. La *densité réfractive* mesure la propension des matériaux à dévier la lumière et elle est quantifiée par *l'indice de densité réfractive ou indice de réfraction* (Thomas Young, *A Course of Lectures in Natural Philosophy and Mechanical Arts*, vol. I, London, 1807, 411-412). Elle est donc inversement proportionnelle au rapport des sinus, et à la vitesse de la lumière dans le milieu, donc à $x^{-\frac{1}{r}}$ dans la suite du texte. La *densité d'éther*, d'après la Scholie I de la Proposition I, varie comme l'inverse du carré de la vitesse. Sa loi d'évolution – qui est mentionnée ici comme la raison de l'inflexion sans pourtant être explicitement donnée – doit donc être proportionnelle à $x^{-\frac{2}{r}}$ ; soit comme le carré de la *densité réfractive*.

[202] La Figure 4 représente le trajet de six rayons lumineux parallèles, arrivant verticalement depuis le haut de la figure et passant à proximité d'un corps opaque représenté par le disque uniformément noir de centre C. Le dégradé de gris autour du disque représente l'atmosphère d'éther accumulée autour du corps, de densité continûment décroissante au fur et à mesure que l'on s'éloigne de celui-ci, pour atteindre la valeur de la densité homogène de l'éther dans l'air à l'endroit où le cercle grisé se termine. Des deux rayons les plus à droite de la figure l'un est extérieur, l'autre tangent, à la zone grisée ; par conséquent ils évoluent dans un milieu homogène et cheminent en ligne droite. Le troisième rayon en partant de la droite pénètre dans la zone grisée au point que l'on appellera A en faisant un angle $i_0$ avec la normale à l'atmosphère d'éther, donc avec le rayon (CA) du cercle ; sous l'effet du changement de densité d'éther, la vitesse de ce rayon sera modifiée et il sera réfracté vers le corps opaque, puisque la densité d'éther augmente. Et comme la densité d'éther augmente continûment alors que le rayon se rapproche du corps, le rayon ne cessera d'être dévié au cours de sa traversée de la zone grisée, comme dans les phénomènes de mirages. Il ne reprendra un trajet en ligne droite que lorsqu'il en sortira, au point que l'on appellera B. Et par symétrie du problème et de la loi de la réfraction, le rayon sortant en B fera nécessairement avec la droite (CB) normale au cercle un angle $-i_0$. Le quatrième rayon en partant de la droite est un cas limite du cas précédent, où le rayon lumineux arrive plus

parfaitement avec les observations du troisième livre de NEWTON[203], et avec celles d'auteurs ultérieurs. Mais je ne considèrerai pas comme très certain, avant que d'autres expériences aient été réalisées sur le pouvoir infléchissant de différentes substances, que l'explication du Dr. HOOKE de l'inflexion par la tendance de la lumière à diverger ne puisse avoir quelques prétentions à la vérité.[204] Je suis désolé de devoir revenir ici sur l'assentiment que, à première vue, j'ai été amené à donner à une amélioration supposément proposée par un auteur récent. (*Phil. Trans. for 1800*, p. 128)[205].

---

près encore du corps opaque, donc traversera des couches d'éther plus denses et sera plus dévié encore, au point de frôler le corps opaque, sans le toucher toutefois. Enfin les cinquième et sixième rayons répondent à la condition nécessaire pour qu'il y ait interférence par inflexion selon Young : ces deux rayons sont déviés au point de percuter le corps opaque et d'être réfléchis à sa surface. Ils repartent alors symétriquement selon la loi de la réflexion et continuent leur déviation progressive jusqu'à sortir de la zone grisée pour pouvoir poursuivre en ligne droite sur leur lancée. En bas de la figure, Young marque les lieux où les rayons ainsi réfléchis pourront rencontrer les rayons simplement déviés et où auront donc lieu les interférences : le deuxième rayon croisant le sixième, le troisième croisant le quatrième. L'état d'interférence au lieu du croisement dépend de la différence de temps de parcours des deux rayons qui se croisent relativement à la période d'ondulation de la lumière. Les lignes plus ou moins ombrées en bas de la figure schématisent la périodicité des franges colorées résultant de la différence de marche calculée pour les rayons se croisant sur ces lignes.

[203] Isaac Newton, *Opticks,* Livre III, Partie I, 1730, 292-313. Passage dédié à l'observation des franges colorées attribuées au phénomène que Newton appelle *inflexion* et à leur explication par la théorie des accès. Young affirme ici, sans pourtant le présenter, que le résultat numérique de son calcul correspond parfaitement aux mesures effectuées par Newton. André Chappert fait cependant remarquer que dans son article suivant (*An account of some Cases of the production of Colours not hitherto described*, Philosophical Transactions of the Royal Society of London, 1802, 388-389) Young examine le cas des franges colorées obtenues par diffraction de la lumière blanche par un fil de laine ou de soie selon le modèle présenté ici. Puis qu'il calcule une différence des chemins produisant la première frange en lumière rouge qui présente une différence sensible (11%) par rapport au résultat de Newton. Mais qui indique néanmoins selon lui une coïncidence « *suffisamment parfaite pour garantir complètement l'explication du phénomène* » (André Chappert, *Histoire de l'optique ondulatoire de Fresnel à Maxwell*, Belin, Paris, 2007, 43).

[204] Suite à l'abandon de l'hypothèse d'existence d'une atmosphère éthérée de densité décroissante entourant les corps matériels (Hypothèse IV), Young réédite une version de *la Théorie de la Lumière et des Couleurs* où l'interprétation présentée ici des couleurs par inflexion sera maintenue, mais essentiellement pour la forme (Thomas Young, *A Course of Lectures in Natural Philosophy and Mechanical Arts*, vol. II, London, 1807). Ce qui se manifestera par la suppression pure et simple de toute la partie qui suit du Corollaire V (p.629). Plus révélateur encore, en guise de conclusion au corollaire la phrase qui précède a été réécrite par Young, donnant à sa version originale une dimension presque prophétique : « *Mais je ne considère pas l'existence d'une telle atmosphère comme nécessaire à l'explication des phénomènes ; l'opinion simple de Grimaldi et Hooke, qui supposaient que l'inflexion émerge de la tendance naturelle de la lumière à diverger, paraissant également probable.* » Ainsi, si l'abandon de l'idée d'une atmosphère éthérée de densité décroissante force Young à abandonner son ambition de réunir l'explication de l'ensemble des phénomènes physiques par l'action d'un unique milieu éthéré, il retombera ingénieusement sur ses pieds dans le domaine restreint de l'optique en attribuant simplement à la divergence naturelle des ondes lumineuses (Proposition III) le rôle de déviation de la lumière à proximité des obstacles que le gradient d'atmosphère éthérée ne peut plus assurer (Geoffrey N. Cantor, *The changing role of Young's ether,* The British Journal for the History of Science, vol. 3, 17, 1970, 44-62).

[205] Thomas Young, *Outlines of Experiments and Inquiries respecting Sound and Light*, Philosophical Transactions of the Royal Society of London, vol. 90, 1800, 106-150. A la page 128 de ce texte, Young évoque l'inflexion de la lumière et l'efficace réfutation de la répulsion des rayons infléchis par « *l'ingénieux* » M. Jordan dans une récente publication (Gibbes Walker Jordan, *The observations of Newton concerning the inflections of light : accompanied by other observations differing from his ; and appearing to lead to a change of his theory of light and colours*, London, Strand, 1799). Young semble néanmoins se raviser ici, et retirer le soutien qu'il paraissait apporter aux conclusions de cet auteur un an auparavant.

*Scholie.* Pour construire le diagramme, il devient nécessaire de trouver le temps passé par chaque rayon lors de son passage[206]. {P.44} Puisque la vélocité a été assignée à $x^{-\frac{1}{r}}$, dans l'hypothèse d'un projectile, elle sera comme $x^{\frac{1}{r}}$ dans l'hypothèse contraire, (*Phil. Trans. for 1801*, p.27. Schol. 2. Prop. I.)[207] et la fluxion de la distance parcourue étant $\frac{\dot{x}}{\sqrt{1-yy}}$ [208], celle du temps sera $\frac{x^{-\frac{1}{r}}\dot{x}}{\sqrt{1-yy}}$ ou $\frac{rs}{1-r}\cdot\frac{\dot{y}}{yy\cdot\sqrt{1-yy}}$ [209], dont la fluente est $\frac{r}{1-r}\cdot\frac{s}{y}\cdot\sqrt{1-yy}$. Par

---

[206] La démonstration qui suit est difficile à suivre en bonne partie parce qu'elle repose sur des notations et résultats extraits du texte cité précédemment – lui-même supposant un modèle corpusculaire de la lumière et faisant référence au formalisme géométrique de la Proposition 41 du Livre 1 des *Principia* – mais que Young ne prend pas la peine de rappeler et contextualiser. Nous allons faire notre possible pour en clarifier progressivement la logique dans la série de notes de bas de page qui vont suivre.

[207] Thomas Young, *On the Mechanism of the Eye*, Proposition I, Scholie 2, Philosophical Transactions of the Royal Society of London, vol. 91, 1801, 27. Cette modélisation mathématique de l'évolution de la vitesse de la lumière dans un milieu de densité variable est utilisée par Young dans la résolution du problème de la propagation de la lumière dans un milieu inhomogène (*On the Mechanism of the Eye,* addendum Prop. VIII, 83, déjà mentionnée plus haut), mais dans le cadre d'un modèle corpusculaire de la lumière. La *densité réfractive* du milieu traversé y est supposée variable comme $x^{-\frac{1}{r}}$, où $x$ représente la distance au *centre* (zone la plus dense du milieu éthéré), et $r$ une constante caractérisant l'évolution spatiale de la densité. Dans ce même modèle, Young pose que la *densité réfractive* « *varie […] comme la vélocité de la lumière dans le milieu* », se rangeant alors à l'interprétation corpusculaire de la lumière. Mais la scholie 2 de la page 27 citée ici confronte succinctement l'opinion de Barrow, Huygens, Euler et Young selon laquelle la lumière accélère dans les milieux les plus rares, à l'opinion contraire défendue par la majorité. Young se sert donc de ce rappel pour amender doublement la proposition faite à la fin de sa conférence précédente : en associant tout d'abord la *densité réfractive* du milieu à l'inverse de la vitesse de la lumière ; puis en changeant le signe de l'exposant dans la formule donnant l'évolution spatiale de la vitesse, qui devient $x^{\frac{1}{r}}$. Cette dernière formule indique implicitement que dans le cadre de la résolution de ce problème, la vitesse de la lumière dans l'air sera supposée égale à 1 et non à *c*. Impliquant, par continuité de la variation de la vitesse à l'endroit où la densité d'éther devient constante, que le rayon CA de l'atmosphère éthérée doit lui aussi être pris égal à 1.

[208] Newton développe à la fin du XVIIe siècle ce qu'il appelle la méthode des *fluxions* et qui n'est autre que ce que l'on nomme aujourd'hui calcul différentiel. Il la baptise ainsi car il considère les variations infiniment faibles d'une quantité comme étant des fluxions ou des accroissements momentanés de celle-ci. Dans ce formalisme la quantité $x$ variable est appelée la *fluente*, et sa différentielle est appelée *fluxion* et notée $\dot{x}$. Ici il s'agit dans un premier temps d'évaluer par cette méthode la distance élémentaire $dl$ parcourue par la lumière dans une couche d'épaisseur $dx$ extrêmement fine d'atmosphère éthérée. Dans cette formule, l'écriture $yy$ signifie $y^2$. Où $y$ sera défini plus loin comme le sinus de l'angle d'incidence $i$ à la surface d'une sphère quelconque séparant deux milieux d'indices proches. Donc ici la grandeur $\sqrt{1-yy}$ peut être réécrite sous la forme plus moderne $\sqrt{1-\sin^2 i} = \cos i$. On a donc bien $dl = \frac{\dot{x}}{\sqrt{1-yy}} = \frac{dx}{\cos i}$. D'où Young déduit ensuite le temps élémentaire $dt$ nécessaire pour parcourir cette distance : $dt = \frac{dl}{v} = \frac{x^{-\frac{1}{r}}\dot{x}}{\sqrt{1-yy}}$.

[209] L'équivalence entre les deux formules est acquise dès lors que l'on pose $x^{-\frac{1}{r}+1} = \frac{s}{y}$, où $s$ est le Sinus de l'angle d'incidence $i_0$ du rayon lumineux à son entrée dans l'atmosphère éthérée : $s = CA \cdot \sin i_0 = \sin i_0$. Le changement de variable n'est pas explicitement justifié par Young, parce qu'il fait appel à un résultat analogue bien connu à l'époque, généralement appliqué au problème de la réfraction atmosphérique.

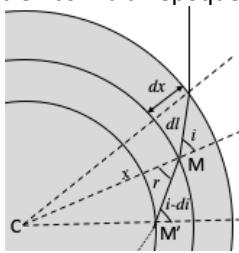

conséquent, avec le rayon <du cercle> $x^{1-\frac{1}{r}}$, décrivez un cercle concentrique avec les surfaces de l'atmosphère infléchissante, alors l'angle décrit par le rayon <lumineux> au cours de son passage à travers l'atmosphère sera toujours à l'angle sous-tendu par la ligne du rayon incident prolongé découpée par ce cercle, dans le rapport de $r$ à $r-1$; et le temps passé pour cette traversée, sera dans le même rapport au temps qui aurait été dépensé en parcourant cette portion interceptée avec la vélocité initiale[210]. Car $y$, qui est égal à $s.x^{\frac{1}{r}-1}$, est le <rapport du> Sinus de l'inclinaison du rayon <lumineux> incident au rayon <du cercle>,

---

En modélisant l'atmosphère comme un empilement de couches concentriques de densité croissante, on peut effectivement montrer que le produit $\frac{x.\sin i}{v}$ est constant le long de la trajectoire du rayon. Sur la Figure jointe, la lumière est réfractée au point M, située à la distance $x$ du centre C, sous l'effet d'un changement infinitésimal de sa vitesse passant de $v$ à $v\text{-}dv$. La loi de la réfraction au point M s'écrit donc : $\frac{\sin i}{\sin r} = \frac{v}{v-dv}$ (Proposition V). La formule des sinus appliquée au triangle CMM' permet par ailleurs d'écrire : $\frac{\sin r}{x-dx} = \frac{\sin[\pi-(i-di)]}{x} = \frac{\sin(i-di)}{x}$. Par conséquent : $\frac{x.\sin i}{v} = \frac{(x-dx).\sin(i-di)}{v-dv}$, ce qu'il fallait démontrer. Le changement de variable proposé par Young est la simple application de ce résultat au point d'entrée de la lumière dans le gradient d'atmosphère éthérée, puis à un point quelconque situé à une distance $x$ du centre : $\frac{1.\sin i_0}{1} = \frac{x.\sin i}{\frac{1}{x^{\frac{1}{r}}}}$, soit dans ses notations : $s = y.x^{1-\frac{1}{r}}$. Dès lors, on obtient effectivement en dérivant $\left(-\frac{1}{r}+1\right).x^{-\frac{1}{r}}.dx = -\frac{s.dy}{y^2}$, soit : $x^{-\frac{1}{r}}.\dot{x} = \frac{r}{r-1}.\frac{s.\dot{y}}{yy}$. Young obtient alors la durée totale $\Delta t = \left[\frac{r}{1-r}.\frac{s}{y}.\sqrt{1-yy}\right]_{entrée}^{sortie}$ du trajet de la lumière à travers le milieu à gradient de densité par intégration (ou calcul de la fluente) de la durée élémentaire $dt$. Ainsi le traversée complète de A à B du gradient d'atmosphère dure un temps $\Delta t = \left[\frac{r}{1-r}.\frac{s}{y}.\sqrt{1-yy}\right]_A^B = \frac{2r}{r-1}.\frac{s}{y_A}.\sqrt{1-y_A y_A}$, par symétrie du problème.

[210] Le résultat annoncé par Young est mieux compris avec la Figure 4bis. En s'appuyant sur le cas d'un cercle de rayon CA = $x^{1-\frac{1}{r}}$ choisi par Young pour des raisons pédagogique, car le résultat vaut pour tout cercle, on peut observer que l'angle A'CB' représentant « *l'angle décrit par le rayon lumineux au cours de son passage à travers l'atmosphère* », et l'angle A'CD' correspondant à « *l'angle sous-tendu par la ligne du rayon incident prolongé découpé par [le] cercle de rayon $x^{1-\frac{1}{r}}$* » sont dans un rapport $\frac{A'CB'}{A'CD'} = \frac{r}{r-1}$. Cette première affirmation n'est pas démontrée car il s'agit du résultat prouvé dans l'addendum à la Proposition VIII cité plus haut (*On the Mechanism of the Eye,* addendum Prop. VIII, 83-84).

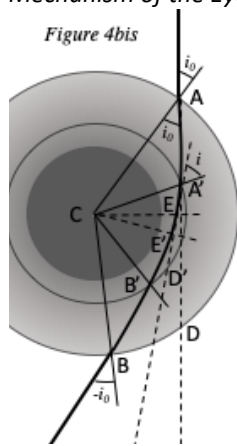

Figure 4bis

Cependant, le calcul mené jusqu'à présent est effectivement une démonstration de la seconde affirmation, selon laquelle : le temps total $\Delta t$ de traversée du cercle de rayon $x^{1-\frac{1}{r}}$ par le rayon lumineux est au temps $\Delta t'$ qu'aurait mis ce rayon pour parcourir A'D' « *la ligne du rayon incident prolongé découpée par* <le> cercle <de rayon $x^{1-\frac{1}{r}}$> » en conservant « *sa vitesse initiale* », $v_0 = 1$ dans l'air dans le rapport : $\frac{\Delta t}{\Delta t'} = \frac{r}{r-1}$. La démonstration de ce résultat suit.

là où il rencontre ce cercle ; par conséquent, par la proposition citée, l'angle décrit est dans un rapport donné à l'angle au centre, qui est la différence des inclinaisons. En faisant de $x^{1-\frac{1}{r}}$ ou $\frac{s}{y}$ le rayon <du cercle>, le Sinus au lieu de $y$ devient $s$, et le Cosinus $\sqrt{\frac{ss}{yy} - ss}$, ou $\frac{s}{y}\sqrt{1-yy}$, et quand $y = ss$, $\sqrt{1-ss}$ ; par conséquent la ligne interceptée est à la différence des fluentes comme $r$ est à $r - 1$.[211] (Voir aussi *Le Programme de* YOUNG, Art. 372.)[212]

Proposition IX
*La Lumière Radiante consiste en Ondulations de l'Ether luminifère.*

Cette proposition est la conclusion générale de toutes celles qui précèdent ; et il est estimé que celles-ci conspirent à la prouver de manière aussi satisfaisante qu'il peut possiblement être attendu de la {P.45} nature du sujet. Il est clairement concédé par NEWTON qu'il y a des ondulations, cependant il nie qu'elles constituent la lumière ; mais il est montré dans les trois premiers Corollaires de la dernière Proposition que tous les cas d'accroissement ou d'affaiblissement de la lumière sont soumis à un accroissement ou un affaiblissement de telles ondulations, et que toutes les affections auxquelles les ondulations seraient sujettes sont visibles distinctement dans les phénomènes de la lumière ; il peut par conséquent être très logiquement inféré que les ondulations sont la lumière.

Quelques remarques indépendantes serviront à écarter certaines objections qui pourront être élevées contre cette théorie.

1. NEWTON a avancé la réfraction singulière du cristal d'Islande comme argument du fait que les particules de lumière doivent être des corpuscules projetés ; puisqu'il pense probable que les différents côtés de ces particules doivent être attirés différemment par le cristal, et puisque Huygens a confessé son incapacité à rendre compte de tous les phénomènes de manière satisfaisante[213]. Mais, contrairement à ce que l'on aurait pu attendre des habituelles précision et franchise de Newton, il a établi une nouvelle loi pour la réfraction sans fournir de raison pour rejeter celle de Huygens, que M. Haüy[214] a découvert être plus précise que celle de Newton ; et, sans tenter de déduire de son propre système la moindre explication des effets les plus universels et frappants des spaths doubleurs, il a omis d'observer que la théorie des plus élégantes et ingénieuses de Huygens s'accorde parfaitement avec ces effets généraux dans tous les détails, et en tire forcémment des prétentions supplémentaires à la vérité : il omet ceci afin de pointer une difficulté pour laquelle on ne peut trouver qu'une solution verbale dans sa propre théorie, et qui restera probablement inexpliquée par toute autre pour longtemps.

---

[211] Ici se conclue enfin la démonstration du résultat : $\frac{\Delta t}{\Delta t'} = \frac{r}{r-1}$. Young a déjà démontré que pour tout cercle $\Delta t = \frac{2r}{r-1}.\frac{s}{y}.\sqrt{1-yy}$, il suffit donc pour conclure de montrer que $\Delta t' = \frac{2s}{y}.\sqrt{1-yy}$. Young rappelle d'abord que le rayon du cercle considéré vaut CA' $= x^{1-\frac{1}{r}} = \frac{s}{y}$. Dès lors « *le Sinus* » de l'angle CA'E' est bien : CA'.$\sin i = \frac{s}{y}.y = s$. Et « *le Cosinus* » de l'angle CA'E' est bien A'E' $= \sqrt{R^2 - R^2.\sin^2 i} = R.\sqrt{1-\sin^2 i} = \frac{s}{y}.\sqrt{1-yy}$. Le produit $\frac{2s}{y}.\sqrt{1-yy}$ correspond donc à la longueur totale de la ligne A'D', mais aussi au temps $\Delta t'$ mis pour parcourir cette ligne à la vitesse $v = 1$. L'égalité $\frac{\Delta t}{\Delta t'} = \frac{r}{r-1}$ est par conséquent vérifiée pour le cercle de rayon CA'. Elle est enfin généralisée à la traversée de l'ensemble du gradient d'atmosphère éthérée en posant « $y = ss$ ». Bien qu'il semble que ce soit « $y = s$ » que Young ait voulu écrire à cet endroit, il s'agit là en fait de remonter au point d'entrée A du rayon lumineux dans l'atmosphère avec un angle d'incidence $i_0$ tel que $y = \sin i_0 = s$. Dès lors, l'expression de la longueur AE devient effectivement $\sqrt{1-ss}$, et le temps mis pour parcourir AB est dans un rapport $\frac{r}{r-1}$ au temps qu'aurait mis le rayon pour parcourir AD à la vitesse qui est la sienne dans l'air. On notera donc enfin qu'en plus de sa démonstration particulièrement tortueuse, cette scholie n'est que l'une des étapes de la détermination mathématique des positions des franges colorées produites par inflexion, représentées par Young en bas de Figure 4. On ne peut donc effectivement pas estimer ici que le calcul de ces franges a été exposé jusqu'à son terme. Cette tendance à ne pas exposer intégralement le traitement mathématique des problèmes qu'il aborde est d'ailleurs l'une des raisons souvent attribuées au manque de résonance des théories de Young au début du XVIIIe siècle, en particulier lorsqu'on les compare aux travaux extrêmement détaillés d'Augustin Fresnel sur des sujets similaires.

[212] Thomas Young, *Syllabus of a course of Lectures on Natural and Experimental Philosophy*, Royal Institution, London, Art. 372, 1802, 111. Où Young évoque sans le démontrer un résultat semblable dans le cas de la réfraction atmosphérique, laquelle consiste aussi en la déviation progressive du rayon lumineux par un milieu à gradient d'indice.

2. M. Michell a réalisé certaines expériences qui paraissent montrer que les rayons de lumière possèdent un réel moment, au {P.46} moyen duquel un mouvement est produit lorsqu'ils tombent sur une fine lame de cuivre délicatement suspendue. (Optique de Priestley.) [215] Mais, en tenant pour acquis l'exacte perpendicularité de la lame et l'absence de tout courant d'air ascendant, toutefois dans toute expérience de la sorte, une plus grande quantité de chaleur doit être communiquée à l'air à la surface sur laquelle la lumière tombe qu'à la surface opposée, <et> l'excès d'expansion doit nécessairement produire un excès de pression sur la première surface et une récession très perceptible de la lame dans la direction de la lumière. M. BENNETT a reproduit l'expérience avec un appareillage beaucoup plus sensible et aussi en l'absence d'air ; et infère très justement de son échec total un argument en faveur du système ondulatoire de la lumière (*Phil. Trans. for 1792*, p. 87.)[216] Car, en concédant aux corpuscules de lumière la subtilité la plus extrême qui soit imaginable, on pourrait naturellement s'attendre à ce que leurs effets soient dans une certaine proportion aux effets des mouvements beaucoup moins rapides du fluide électrique, qui sont si facilement perceptibles même dans leurs états les plus faibles.

---

[213] Isaac Newton, *Opticks,* Livre III, Question 25 : « *N'y a-t-il pas d'autres propriétés originales des rayons de lumière que celles déjà décrites ?* » et Question 26 : « *Les rayons n'ont-ils pas plusieurs côtés doués de propriétés originales ?* », 1730, 328-336. Comme pour toutes les questions concluant l'*Opticks,* Newton répond implicitement par l'affirmative à celles-ci, en se basant spécifiquement dans ce cas précis sur une interprétation de la double réfraction des cristaux transparents de carbonate de calcium, ou spath d'Islande : les différents côtés du corpuscule lumineux étant responsables de chacun des deux types ordinaire et extraordinaire de réfraction dans ce cristal. Newton précise à cette occasion que la double réfraction dans le spath d'Islande a déjà été décrite par Bartholin (*Experimenta crystalli Islandi disdiaclastici quibus mira et insolita refractio deiegitur*, 1669) et Huygens (*Traité de la lumière*, Chapitre 5 : « De l'étrange réfraction du cristal d'Islande », 1690) mais estime qu'ils ne l'ont pas interprétée correctement. La lecture de ce long chapitre du *Traité de la lumière* révèle cependant une très minutieuse interprétation du phénomène par Huygens, qui introduit le concept d'ondes *sphéroïdes* (volume obtenu en faisant tourner une ellipse autour de son petit axe) et anticipe ainsi sur les interprétations modernes de la biréfringence. Huygens renonce seulement à interpréter dans sa conclusion les apparitions et disparitions alternées des deux rayons émergeants lorsqu'ils passent par un second cristal dont on s'autorise à modifier l'orientation ; phénomène sur lequel Newton décide précisément de développer son raisonnement. Reste que Newton jette un sérieux discrédit sur le travail de Huygens relatif à la double réfraction, à la fois par la nouvelle interprétation qu'il en propose et parce qu'il donne des valeurs sensiblement différentes des angles du cristal et de la double réfraction.

[214] René Just Haüy (1743-1822) est un minéralogiste français, qui à ce titre s'est intéressé aux propriétés du spath d'Islande. Il a donc mesuré très précisément à son tour les angles que font entre elles les faces du cristal, et fourni des valeurs compatibles avec celles proposées par Huygens ; ce que fait remarquer Young, en soulignant le manque de « *précision* » et de « *franchise* » de Newton dans le traitement de ce cas. Il n'est pas inintéressant de noter que quelques mois après la publication de la *Théorie de la lumière et des couleurs*, Wollaston (1766-1828), confirme à l'aide d'un instrument optique extrêmement précis que les mesures rapportées par Huygens sur les inclinaisons des faces de la calcite sont non seulement correctes, mais aussi plus justes que celles obtenues depuis par Haüy. Haüy est alors chargé par l'Institut de France de dire qui de Newton ou de Huygens a raison, et tranche définitivement en faveur de ce dernier. Dans la foulée, l'Institut annonce qu'un prix sera attribué à qui donnera « de la double réfraction que subit la lumière en traversant diverses substances cristallisées, une théorie mathématique vérifiée par l'expérience ». Et c'est Etienne Louis Malus (1755-1812) qui l'obtiendra grâce à sa *Théorie de la double réfraction de la lumière*, 1810, dans lequel il rendra compte pour la première fois de la *polarisation* par réflexion de la lumière.

[215] Joseph Priestley (1733-1804), *The History and Present State of Discoveries Relating to Vision, Light and Colours*, 1772. Dans cet ouvrage (p. 387) Priestley rapporte une expérience de John Michell ayant consistée à focaliser la lumière du soleil sur une fine lame de cuivre à l'aide d'un miroir concave. Selon Priestley le mouvement de la lame détecté dans cette expérience révèle le caractère corpusculaire de la lumière.

3. Il y a certains phénomènes de la lumière des phosphores solaires qui à première vue pourraient sembler être en faveur du système corpusculaire[217] ; par exemple, le fait qu'elle subsiste pendant de nombreux mois comme dans un état latent, et sa ré-émission ultérieure par l'action de la chaleur. Mais après considération plus poussée, il n'y a pas de difficulté à supposer que les particules des phosphores que l'action de la lumière a fait vibrer voient leur action abruptement suspendue par l'intervention du froid, comme par contraction du volume de la substance ou autrement ; et encore, après que la restriction soit supprimée, qu'elles reprennent leur mouvement comme le ferait un ressort qui aurait pour un temps été maintenu bloqué dans un état intermédiaire de sa vibration ; non plus qu'il est impossible que la chaleur elle-même puisse dans certains circonstances devenir latente d'une manière similaire. (NICHOLSON's {P.47} Journal. Vol. II. P.399.)[218] Mais les affections de la chaleur nous seront peut-être rendues plus intelligibles ci-après ; à présent, il semble hautement probable que la lumière diffère de la chaleur seulement par la fréquence de ses ondulations ou vibrations ; les ondulations qui sont comprises entre certaines limites, en termes de fréquence, étant capables d'affecter le nerf optique et constituant la lumière ; et celles qui sont plus lentes, et probablement plus fortes, constituant la chaleur seulement ; que la lumière et la chaleur se présentent à nous chacune selon deux états, l'état vibratoire ou permanent, et l'<état> ondulatoire ou transitoire ; la lumière vibratoire étant le mouvement infime des corps ignés[219], ou des phosphores solaires, et la lumière ondulatoire ou radiante le mouvement du milieu éthéré excité par ces vibrations ; la chaleur vibratoire étant un mouvement dont toutes les substances matérielles sont susceptibles, et qui est plus ou moins permanent ; la chaleur ondulatoire <étant> ce mouvement du même milieu éthéré, dont il a été démontré par M. KING, (*Morsels of Criticism*. 1786. p. 99,)[220] et M. PICTET, (*Essais de Physique.* 1790)[221] qu'il est aussi capable de réflexion que la lumière, et par le Dr. HERSCHEL qu'il est capable de réfraction séparée. (*Phil. Trans. for 1800*. p. 284.)[222] Combien la chaleur est communiquée avec plus d'empressement par l'accès libre aux substances plus froides que par radiation ou par transmission à travers un milieu au repos, a été démontré par les précieuses expériences du Comte RUMFORD[223]. Il est aisé de concevoir que certaines substances, perméables à la lumière, puissent être inaptes à la transmission de chaleur, de la même manière que des substances particulières peuvent transmettre certaines sortes de lumière alors qu'elles sont opaques à d'autres.

---

[216] Abraham Bennet, *A New Suspension of the Magnetic Needle, Intended for the Discovery of Minute Quantities of Magnetic Attraction : Also an Air Vane of Great Sensibility ; With New Experiments on the Magnetism of Iron Filings and Brass*. Dans lequel l'auteur ne semble pas observer de mise en mouvement de ses instruments sous une quelconque pression de la lumière, et y voir un argument en faveur d'une théorie ondulatoire. On peut noter que l'existence d'une pression de radiation, responsable notamment en partie de la queue des comètes, a été révélée est expliquée depuis (E. Nichols et G. F. Hull, *A Preliminary Communication on the Pressure of Heat and Light Radiation*, Physical Review, vol. 13, n°5, 1903, 307-320. E. Nichols et G. F. Hull, *The Pressure due to Radiation*, Physical Review, vol. 17, n°1, 1903, 26-50).

[217] Corps phosphorescents. Vers 1630, Vincenzo Cascariolo, cordonnier à Bologne, remarque qu'une substance baptisée plus tard phosphore de Bologne, ou pierre de Bologne, présente la propriété de briller dans le noir après avoir été exposée au soleil. Cet effet, plus facile à justifier par un modèle corpusculaire (la matière aurait stocké des corpuscules qu'elle libérerait lentement), a longtemps était considéré comme un argument crucial en faveur de celui-ci.

[218] Correspondant anonyme, *Observations on Electricity, Light and Caloric, chiefly directed to the results of Dr. Pearson's Experiment on Electric Discharges in Water*, A Journal of Natural Philosophy, Chemistry and the Arts, vol. 2, 1799, 396-400. En note de bas de la page 399 de cette lettre, son auteur anonyme défend la possibilité d'existence d'une chaleur latente. Tout en admettant que l'idée est « *absurde* » (puisque la chaleur est envisagée dans cette lettre comme un mouvement et qu'un mouvement latent constitue un paradoxe), il décrit néanmoins rapidement plusieurs moyens d'envisager la possibilité de celle-ci.

[219] Comprendre : « corps enflammés ».

  Dans l'ensemble il parait que les quelques phénomènes optiques qui admettent une explication par le système corpusculaire sont également cohérents avec cette théorie ; que beaucoup d'autres, qui sont connus depuis longtemps mais n'ont jamais été compris, deviennent par ces moyens parfaitement intelligibles ; et que plusieurs faits nouveaux se {P.48} trouvent ainsi être réductibles à une parfaite analogie avec d'autres faits, et aux principes simples du système ondulatoire. Il est supposé que dorénavant les deuxième et troisième livres de l'*Optics* de Newton seront considérés comme plus complètement compris que le premier ne l'a été jusqu'à présent[224] ; mais s'il devait paraître à des juges impartiaux que des preuves supplémentaires font encore défaut pour l'établissement de la théorie, il sera aisé d'entrer plus minutieusement dans les détails des diverses expériences et de montrer les insurmontables difficultés entourant les doctrines Newtoniennes, qu'il serait sans nécessité, fastidieux et désobligeant d'énumérer. Les qualités en philosophie naturelle de leur auteur sont d'une grandeur dépassant toute contestation ou comparaison ; sa découverte optique de la composition de la lumière blanche aurait à elle seule immortalisé son nom ; et les arguments mêmes qui tendent à renverser son système fournissent les preuves les plus solides de l'admirable précision de ses expériences.

---

[220] Edward King (1735-1807), *Morsels of Criticism : tending to illustrate some few passages in the Holy Scriptures upon the philosophical principles and an enlarged view of things*, 1800. Dans sa tentative d'analyse critique des Saintes Ecritures, King dédie quelques pages à « *une Note concernant le Fluide Élémentaire de Chaleur, ou le Feu* » dans laquelle il propose des clés d'analyse du sens des termes de Chaleur ou Feu qui y sont régulièrement employés. Il y défend l'idée que le fluide élémentaire de chaleur peut exister dans sept états différents (par analogie directe avec les sept couleurs du spectre de la lumière) parmi lesquelles il identifie la chaleur radiante, la flamme, le fluide électrique et la lumière.
[221] Marc-Auguste Pictet (1752-1825), *Essais de Physique, tome 1 : Essai sur le feu,* 1790. Le chapitre II de cet essai s'intitule « Le feu et la lumière ont quelqu'analogie », et le chapitre III propose des « Expériences diverses sur la chaleur », notamment sur la réfraction de la chaleur, sa vitesse de propagation, et la réflexion apparente du froid (p.74-85). Noter que dans la réédition de *On the Theory of Light and Colours* survenue en juillet 1802 dans le Nicholson's Journal, Young ajoute ici une référence bibliographique supplémentaire (proposée par Pictet lui-même dans ses *Essais*) : Horace-Benedict Saussure, *Le Voyage dans les Alpes,* 1786.
[222] William Herschel, *Experiments on the refrangibility of the invisible rays of the sun*, Philosophical Transactions of the Royal Society of London, vol. 90, 1800, 284-292. Voire note de bas de page déjà dédiée à cet article.
[223] Benjamin Thomson, Comte Rumford (1753-1814). Voir la note de bas de page qui lui est déjà dédiée.
[224] Les deuxième et troisième livres de l'*Opticks* sont dédiés, comme on l'a vu au fil de ce texte, à la description des couleurs des lames fines, des lames épaisses, des corps naturels et aux couleurs par inflexion, ainsi qu'à leur interprétation par la théorie des accès et aux *questions* laissées en suspens par Newton. Alors que le premier livre est consacré à une analyse et l'interprétation des couleurs obtenues avec un prisme. Si le premier livre donne donc l'impression d'être extrêmement complet et abouti, les deux suivants sont ceux dans lesquels on trouve les propositions de Newton les plus sujettes à discussion.

Aussi suffisants et décisifs ces arguments puissent-ils paraître, il ne peut être superflu d'en chercher une plus ample confirmation ; qui peut être espérée avec grande confiance d'une expérience très ingénieusement suggérée par le Professeur R‍obison sur la réfraction de la lumière nous revenant des extrémités opposées de l'anneau de Saturne[225] ; car, selon la théorie corpusculaire, l'anneau doit être considérablement distordu lorsqu'il est vu à travers un prisme achromatique : une distorsion similaire devrait aussi être observée dans le disque de Jupiter ; mais s'il est trouvé qu'une déviation égale de la totalité de la lumière réfléchie par ces planètes a lieu, il ne restera pratiquement plus le moindre espoir d'expliquer les affections de la lumière par une comparaison avec les mouvements des projectiles.

---

[225] John Robison (1739-1805) a rédigé plusieurs articles de la troisième édition de l'*Encyclopedia Britannica*, dont John Robison, *Theory of Optics*, Encyclopedia Britannica, 3ème Edition, 1797, 278-363. Fervent défenseur de la théorie corpusculaire, il évoque néanmoins la théorie ondulatoire dans la première page de cet article, tout en signalant « *qu'on ne doit pas désespérer de parvenir par l'expérience à déterminer laquelle de ces opinions est la plus proche de la vérité* ». Geoffrey N. Cantor (*Optics after Newton : Theories of Light in Britain and Ireland, 1704-1840*, Manchester University Press, Manchester, 1983, 87-88) liste une série de telles expériences envisagées par Robison, dont celle décrite ici par Young concernant les extrémités des anneaux de Saturne et évoquée dans une lettre à William Herschel non datée mais précédant la publication de *la Théorie de la Lumière et des Couleurs* (John Robison à William Herschel, Edinburgh University Library, La. II. 223, non daté). Cantor nous permet de remonter par ailleurs à une lettre ultérieure de John Robison à William Herschell (Royal Astronomy Society, Herschel W. Correspondence : 13.R.9, 14 avril 1804) dont le ton se fait beaucoup plus urgent, en réaction à la pression mise sur la théorie corpusculaire par les publications de Young qui se succèdent entre 1800 et 1804. Cette lettre précise le principe de cette expérience : « *si la lumière consiste en l'émission de matière depuis les corps lumineux, et obéit aux grandes lois du Mouvement, il semble suivre que la lumière réfléchie depuis l'extrémité Est de l'anneau de Saturne devrait venir à nous avec une vélocité excédent celle de la lumière réfléchie par l'extrémité Ouest d'environ 1/2500. Si c'est le cas, elle devrait être différemment réfractée.* » Robison suppose donc que ce ne devrait pas être le cas dans l'hypothèse ondulatoire, ce que semble confirmer Young par sa conclusion. Nous n'avons pas trouvé plus de détails sur cette expérience ou son éventuelle réalisation. Son évocation ici révèle néanmoins la très vive conviction que Young a développée au sujet du système ondulatoire, qu'il ne doute pas de voir confirmé par une expérience cruciale.

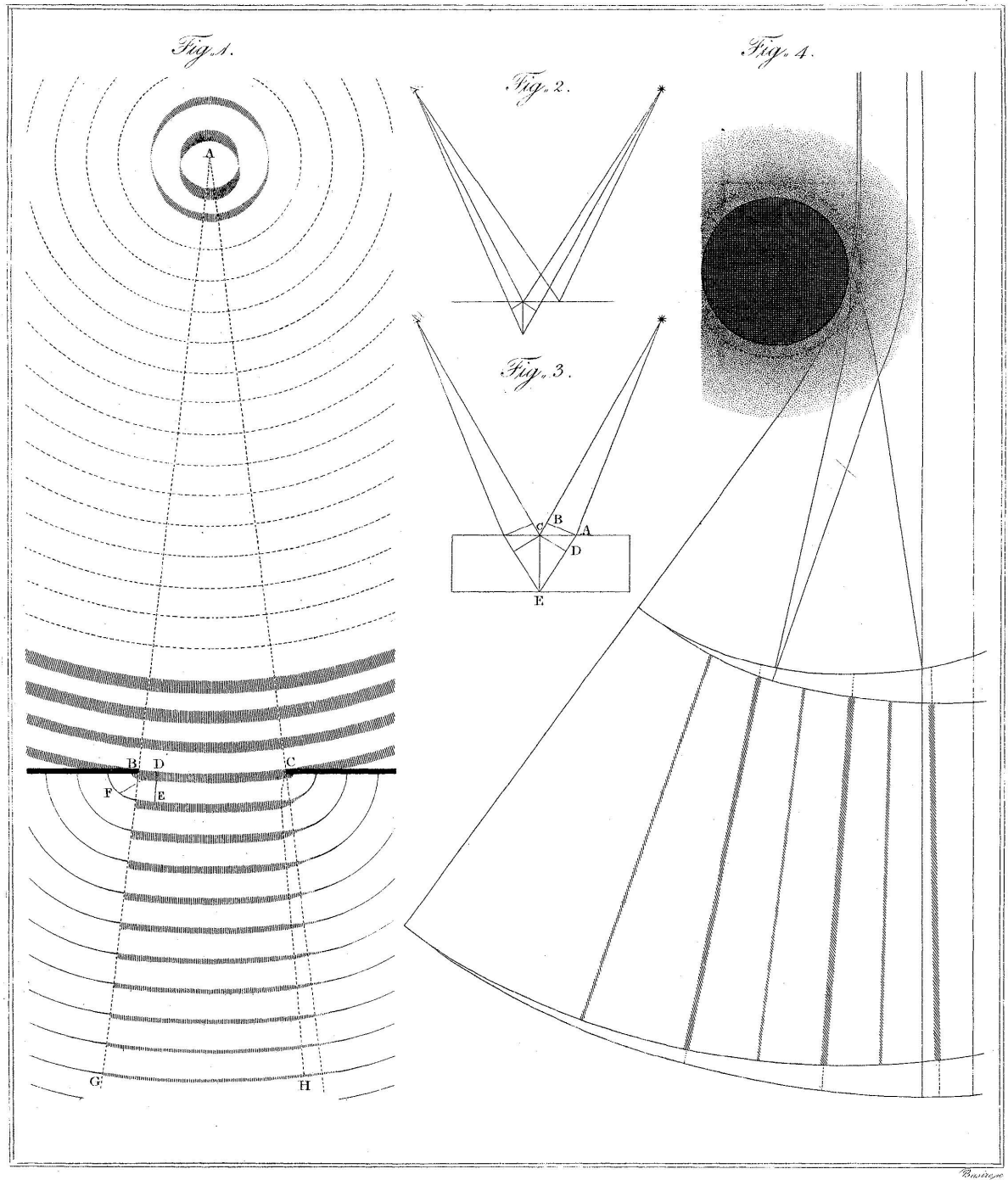

Source : Philosophical Transactions of the Royal Society of London, 1802, Vol. 92 (1802), Plate I, p. 48.
Published by: Royal Society.